\documentclass[floats,floatfix,showpacs,amssymb,prd,twocolumn,superscriptaddress,nofootinbib,nolongbibliography,reprint]{revtex4-2}

\usepackage{amssymb,amsmath,verbatim,mathtools,needspace,enumitem,etoolbox,graphicx,microtype,afterpage,xspace,tabularx,lmodern,multirow}
\usepackage{gensymb}
\usepackage[dvipsnames, usenames]{xcolor}
\definecolor{linkcolor}{rgb}{0.0,0.3,0.5}
\usepackage[unicode, colorlinks=true, linkcolor=linkcolor, citecolor=linkcolor, filecolor=linkcolor, urlcolor=linkcolor, linktocpage, breaklinks]{hyperref}
\usepackage[all]{hypcap}
\usepackage[T1]{fontenc}
\usepackage[utf8]{inputenc}
\usepackage[usenames,dvipsnames]{xcolor}
\definecolor{romared}{RGB}{142,0,28}
\hypersetup{colorlinks=true,citecolor=romared,linkcolor=romared,urlcolor=romared}
\usepackage{aas_macros}
\usepackage{booktabs}
\usepackage{bm}
\usepackage{multirow}
\usepackage{graphicx}
\usepackage{ulem}

\newcommand{\be}{\begin{equation}}
\newcommand{\ee}{\end{equation}}
\newcommand{\beqa}{\begin{eqnarray}}
\newcommand{\eeqa}{\end{eqnarray}}
\newcommand{\dd}{{\rm d}}

\begin{document}

\title{Hunting intermediate-mass black holes with LISA binary radial velocity measurements}

\author{Vladimir Strokov}
\email[]{vstroko1@jhu.edu}
\affiliation{Department of Physics \& Astronomy, Johns Hopkins University, Baltimore, MD 21218, USA}

\author{Giacomo Fragione}
\email[]{giacomo.fragione@northwestern.edu}
\affiliation{Department of Physics \& Astronomy, Northwestern University, Evanston, IL 60208, USA}
\affiliation{Center for Interdisciplinary Exploration \& Research in Astrophysics (CIERA), Northwestern University, Evanston, IL 60208, USA}

\author{Kaze W. K. Wong}
\email[]{kwong@flatironinstitute.org}
\affiliation{Center for Computational Astrophysics, Flatiron Institute, New York, NY 10010, USA}

\author{Thomas Helfer}
\email[]{thelfer1@jhu.edu}
\author{Emanuele Berti}
\email[]{berti@jhu.edu}
\affiliation{Department of Physics \& Astronomy, Johns Hopkins University, Baltimore, MD 21218, USA}

\date{\today}

\begin{abstract}
Despite their potential role as massive seeds for quasars, in dwarf galaxy feedback, and in tidal disruption events, the observational evidence for intermediate-mass black holes (IMBHs) is scarce. LISA may observe stellar-mass black hole binaries orbiting Galactic IMBHs, and reveal the presence of the IMBH by measuring the Doppler shift in the gravitational waveform induced by the binary's radial velocity. We estimate the number of detectable Doppler shift events from the Milky Way globular clusters (assuming they host IMBHs) and we find that it decreases with the IMBH mass. A few Galactic globular clusters (including M22 and $\omega$~Centauri) may produce at least one event detectable by LISA. Even in more pessimistic scenarios, one could still expect~$\sim$ 1 event overall in the Milky Way. We also estimate the number of Doppler shift events for IMBHs wandering in the Milky Way as a result of the disruption of their parent clusters. If there is at least one binary black hole orbiting around each wandering IMBH, LISA may detect up to a few tens of Doppler shift events from this elusive IMBH population. Under more pessimistic assumptions, LISA may still detect~$\sim 1$ wandering IMBH that would hardly be observable otherwise.
\end{abstract}

\maketitle

\section{Introduction\label{intro}} 

The existence of intermediate-mass black holes (IMBHs) is still controversial. While the boundaries between different classes of black holes are largely a matter of convention, IMBHs are usually assumed to have masses in the range $\sim 10^2-10^{5}\,M_\odot$, filling the gap between stellar black holes (SBHs, with mass $\lesssim 100\,M_\odot$) and supermassive black holes (with mass $\gtrsim 10^6\,M_\odot$). Unlike their lighter and heavier counterparts, IMBHs remain elusive~\cite{2018ApJ...868..152B,2018ApJ...863....1C,2018NatAs...2..656L}. Finding IMBHs would have important implications for a wide range of phenomena, including the seeding of supermassive black holes, galaxy evolution, accretion, tidal disruption events, and gravitational waves (GWs)~\cite{2020ARA&A..58..257G}.

There are three main classes of proposed formation mechanisms for IMBHs~\cite{1978Obs....98..210R}. The first predicts that IMBHs of $\sim 10^4-10^5\,M_\odot$ are produced from the direct collapse of a metal-poor gas cloud, without passing through all the phases of stellar evolution~\cite{1994ApJ...432...52L,2003ApJ...596...34B,2013MNRAS.433.1607L,2017NatAs...1E..75R}. The second requires massive Population III stars, which can collapse to IMBHs of $\sim 100\,M_\odot$ as a result of inefficient cooling~\cite{2001ApJ...551L..27M,2002ApJ...564...23B,2004ARA&A..42...79B,2001ApJ...550..372F}. The third involves dense star clusters, where an IMBH of mass $\sim 10^2-10^4\,M_\odot$ can form either as a result of repeated mergers of SBHs, or from the collapse of a very massive star formed via stellar mergers~\cite{2002ApJ...576..899P,2016MNRAS.459.3432M,2019MNRAS.487.2947D,2019PhRvD.100d3027R,2021MNRAS.501.5257R,2021MNRAS.507.5132D,2004ApJ...604..632G,2015MNRAS.454.3150G,2019arXiv190500902A,2021ApJ...908L..29G,2021arXiv211109223M,2002MNRAS.330..232C,2004IJMPD..13....1M,2019MNRAS.486.5008A,2020MNRAS.498.4591F,2020ApJ...902L..26F,2021MNRAS.505..339M,2021MNRAS.505..339M}.

\begin{figure*}
\centering
\includegraphics[width=1.6\columnwidth]{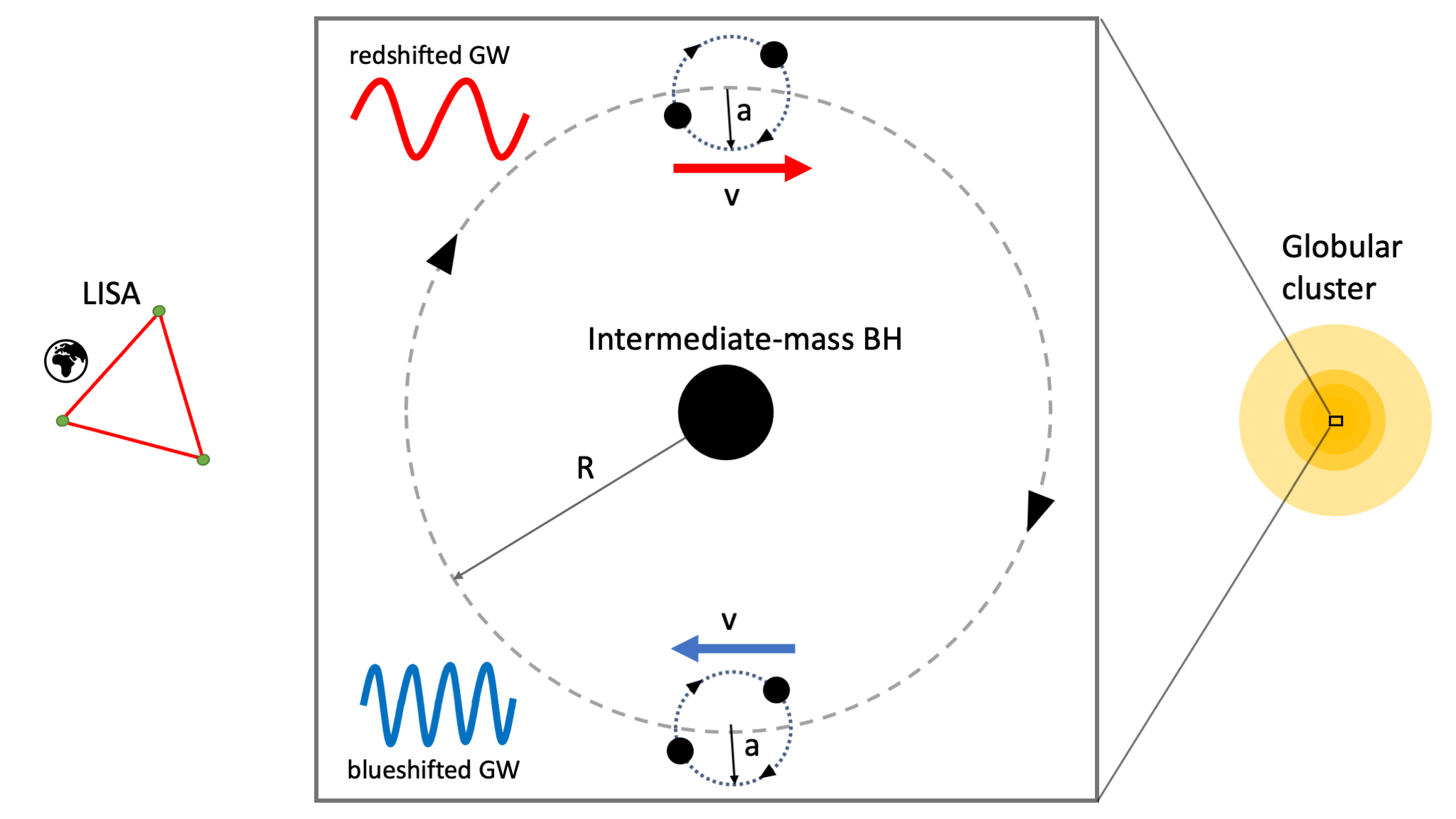}
\caption{Schematic illustration of the IMBH detection method explored in this paper. The radial motion of a BBH orbiting an IMBH located in a GC causes a periodic Doppler shift -- either a redshift or a blueshift -- in the GW signal observed by LISA. The detection of the Doppler shift can be used to infer the presence of the IMBH.
\label{illustration}}%
\end{figure*}

Observational signatures of IMBHs are under intense scrutiny. Accreting IMBHs could be detected in galactic nuclei using observations ranging from the radio to the X-ray band, or as ultraluminous X-ray sources in the field~\cite{2007ApJ...670...92G,2017ARA&A..55..303K,2018ApJ...868..152B}. The existence of nonaccreting IMBHs can be inferred by tracking the orbits of stars and gas in their vicinity~\cite{2010MNRAS.409.1146G,2019MNRAS.482.3669G}. However, stellar crowding makes these detections very challenging~\cite{2019MNRAS.488.5340B,2010ApJ...710.1063V,2010ApJ...719L..60N,2019MNRAS.488.5340B,2011A&A...533A..36L,2013ApJ...769..107L,2013A&A...552A..49L}. IMBHs that lurk in the centers of dense star clusters can interact and disrupt stars, resulting in detectable tidal disruption events~\cite{2014ApJ...784...87S,2016ApJ...821...25L,2018NatAs...2..656L,2018ApJ...867...20C,2018ApJ...867..119F,2019MNRAS.484.4665S,2021ApJ...918...46W}, similar to those observed in galactic nuclei harboring supermassive black holes (see e.g.~\cite{2015ApJ...808L..11C,2021MNRAS.504.4885B,2021ARA&A..59...21G,2021MNRAS.tmp.3135A}). Close to the Galactic center, electromagnetically quiet IMBHs can be also detected by pulsar timing with a timing accuracy of $100$~ns--$10\,\mu\mbox{s}$ using pulsars at distances~$0.1$--$1$~pc from the center~\cite{2012ApJ...752...67K}. Finally, the inspiral of a SBH into an IMBH and the merger of an IMBH binary could be detected with present and upcoming GW interferometers such as LIGO/Virgo/KAGRA, LISA, Cosmic Explorer and the Einstein Telescope. These systems are particularly interesting for multiband GW astronomy~\cite{2007CQGra..24R.113A,2019BAAS...51c.109C,2020CQGra..37u5011A}. In particular, LISA will be able to observe IMBH--SBH binaries and IMBH binaries up to redshifts $z\sim 1$--$2$~\cite{2002MNRAS.330..232C,2007CQGra..24R.113A,2008ApJ...681.1431M,2014MNRAS.444...29L,2016ApJ...832..192H,2018ApJ...856...92F,2020NatAs...4..260J,2020ApJ...899..149R,2021A&A...652A..54A}. The LIGO/Virgo Collaboration (LVC) recently detected the first IMBH ever, the $\sim 150\,M_\odot$ remnant from the GW190521 binary black hole (BBH) merger event~\cite{2020PhRvL.125j1102A,2021PhRvX..11b1053A}, thus validating the expectation that GWs are uniquely well suited to finding IMBHs. The second part of LIGO/Virgo's third observing run (O3b) revealed another significant, somewhat less massive IMBH candidate, GW200220\_061928, and a few other events on the lower end of the IMBH mass range~\cite{2021arXiv211103606T}.    

In this paper we explore the possibility of an {\it indirect} detection of IMBHs through LISA observations of GWs from stellar-mass BBHs. As pointed out in Ref.~\cite{2019MNRAS.488.5665W}, and schematically illustrated in Fig.~\ref{illustration}, the radial motion of a binary system orbiting a third, more massive body can produce Doppler shifts in the GW signal from the stellar-mass binary. These Doppler shift may be detectable by LISA (see also \cite{2017PhRvD..96f3014I,2017PhRvD..95d4029B,2020PhRvD.101f3002T,2021MNRAS.502.4199X,2011PhRvD..83d4030Y} for other applications of Doppler shift observations and~\cite{2017ApJ...834..200M} for Doppler shift detectability in LIGO).  Consider for example a BBH with component masses of $30\,M_\odot$ orbiting a $10^3\,M_\odot$ IMBH at a distance $R = 20$~AU: a few years before merger, LISA could measure the resulting Doppler shift in the GW signal as far as $16$~Mpc at signal-to-noise ratio (SNR) of 100, and as far as 162.6~Mpc at SNR of 10, with a relative error~$\sim 0.001\%$ and~$\sim 0.01\%$, respectively\,\footnote{Note that these numbers reproduce the SOBH case from Table~1 of Ref.~\cite{2019MNRAS.488.5665W} and have been updated by using the LISA power spectral density adopted in this work (see Sec.~\ref{waveform} for details).}.

Our main goal is to estimate the rates at which LISA could find IMBHs lurking in globular clusters (GCs) by measuring the radial velocity modulations in the GW signal of BBHs orbiting the IMBH. Most such BBHs are far from merger, and LISA can only measure the Doppler modulation if they are within $\sim 100$~kpc (that is, within the Milky Way). In this paper, we first estimate the rates of IMBH detections via LISA radial velocity measurements by using catalogs of the observed population of GCs in the  Milky Way. Then we ask whether the radial velocity method could spot IMBHs that are wandering in the Galaxy, being left behind as a result of the disruption of their parent clusters.

The paper is organized as follows. In Sec.~\ref{waveform} we present our waveform model and parameter estimation method, and in Sec.~\ref{astro} we describe our astrophysical models. In Sec.~\ref{results} we report our results for the number of detectable Doppler shift events, both in Galactic GCs and in the neighborhoods of wandering IMBHs. Finally, in Sec.~\ref{discuss} we summarize our conclusions and discuss possible directions for future work. Throughout this paper we use geometrical units ($G=c=1$).

\section{Waveform model and parameter estimation\label{waveform}}

In order to take into account the Doppler shifts in a GW signal, we start off with the frequency-domain expression for the gravitational waveform from a BBH of component masses $m_1$ and $m_2$:
\be
\label{wf0}
h^{(0)}_{\alpha}(f) = \frac{\sqrt{3}}{2}A(t)\mathcal{A}f^{-7/6}e^{i\left[\Psi(f)-\varphi_p(t)-\varphi_D(t)\right]}\,,
\ee
where
\be
\label{amp0}
\mathcal{A} = \sqrt{\frac{5}{96}}\frac{\mathcal{M}^{5/6}}{\pi^{2/3}D_L}\,,
\ee
the index $\alpha=I,II$ denotes the two independent interferometric responses in the LISA arms~\cite{1998PhRvD..57.7089C}, $f$ is the GW frequency, $\mathcal{M}=\eta^{3/5}(m_1 + m_2)$ is the chirp mass, $\eta=m_1 m_2/(m_1+m_2)^2$ is the symmetric mass ratio, and $D_L$ is the luminosity distance. The amplitude $A(t)$, polarization phase $\varphi_p(t)$ and Doppler phase $\varphi_D(t)$, where $t=t(f)$, arise from the geometry of the interferometer and from its motion around the Sun~\cite{1998PhRvD..57.7089C}. They can be expressed in terms of the sky position angles~$(\bar{\theta}_S$, $\bar{\phi}_S)$ of the BBH with respect to the Solar System, and the orientation angles $(\bar{\theta}_L$, $\bar{\phi}_L)$ of the BBH's orbital angular momentum. For the functions $\Psi(f)$ and $t=t(f)$, which encode the inspiral dynamics of the BBH under gravitational radiation reaction, we use expansions up to second post-Newtonian (2PN) order, assuming for simplicity that the binary components are nonspinning~\cite{2005PhRvD..71h4025B}. The inclusion of spins would have a mild effect on our final results. Including aligned spins adds two parameters to the waveform and gives rise to degeneracies, thus increasing Fisher matrix errors~\cite{2005PhRvD..71h4025B}. However, taking into account waveform modulations due to spin precession effectively breaks these degeneracies, making the errors comparable to the nonspinning case~\cite{2009PhRvD..80d4002S} (see Appendix~\ref{subsec:BBHspin} and~\ref{subsec:IMBHspin} for precession timescales). Without loss of generality, we set the coalescence time~$t_c$ and the coalescence phase~$\phi_c$ to zero. We collectively denote all parameters entering Eqs.~(\ref{wf0}) and~(\ref{amp0}) as the components of a vector $\theta_a \equiv \left\{\mathcal{M},\eta,D_{\rm L}, t_{\rm c}, \phi_{\rm c}, \bar{\theta}_S, \bar{\theta}_L, \bar{\phi}_S, \bar{\phi}_L\right\}$.

Let us now consider the case where the BBH orbits an IMBH of mass $M_{\rm IMBH}$ on a circular orbit of semimajor axis $R$ and inclination $I$ with respect to the plane of sky. The Doppler shift due to the motion of the binary gives rise to a correction to the phase. As long as the frequency of the GW signal changes slowly compared to the orbital period $P$ of the binary around the IMBH, the resulting waveform reads~\cite{2019MNRAS.488.5665W}
\be
\label{wf}
h_{\alpha}(f;\tilde{\theta}_1) = h^{(0)}_{\alpha}(f;\theta_a)\exp{\left[ifv_{||}P\sin{\left(\frac{2\pi t(f)}{P}\right)}\right]}\,.
\ee
Here $P$ is the BBH orbital period around the IMBH,
\be
\label{period}
P = 2\pi\sqrt{\frac{R^3}{M_{\rm tot}}}=1\,\mbox{yr}\;\left(\frac{R}{10\,\mbox{AU}}\right)^{3/2}\left(\frac{M_{\rm tot}}{10^3\,M_\odot}\right)^{-1/2}\,,
\ee
with $M_{\rm tot} = M_{\rm IMBH} + m_1 + m_2$, and $v_{||}$ is the amplitude of the radial velocity
\beqa
\label{velocity}
v_{||} &=& \frac{2\pi R\sin{I}}{P}\frac{M_{\rm IMBH}}{M_{\rm tot}}\nonumber\\
&=&300\,{\rm km\,s}^{-1}\,\left(\frac{R\sin{I}}{10\,\mbox{AU}}\right)\left(\frac{P}{1\,\mbox{yr}}\right)^{-1}\left(\frac{M_{\rm IMBH}}{M_{\rm tot}}\right)\,. 
\eeqa
The extended set of waveform parameters $\tilde{\theta}_a$ now includes both $v_{||}$ and $P$, i.e. $\tilde{\theta}_a = \theta_a \cup\left\{v_{||},P\right\}$.
The choice of the initial orbital phase is irrelevant as long as the observation time $T_{\rm obs} \gtrsim P$.

To estimate the relative errors $\Delta v_{||}/v_{||}$ and $\Delta P/P$ we use the Fisher matrix method (see e.g.~\cite{2005PhRvD..71h4025B,2008PhRvD..77d2001V}). In our particular case, the Fisher matrix~$\Gamma_{ab}$ and SNR read
\be
\label{inner}
\Gamma_{ab}  \equiv 4\,{\rm Re}\sum\limits_{\alpha=I,II}\int\limits_{f_0}^{f_0+\delta f}{\frac{\partial h_\alpha^*}{\partial\tilde{\theta}_a}\frac{\partial h_\alpha}{\partial\tilde{\theta}_b}\frac{df}{S_{\rm n}(f)}}\,,
\ee
\be
\label{SNR}
{\rm SNR} = \left[4\sum\limits_{\alpha=I,II}\int\limits_{f_0}^{f_0+\delta f}{\frac{\left|h_\alpha\right|^2 df}{S_{\rm n}(f)}}\right]^{1/2}\,,
\ee
where~$S_{\rm n}(f)$ denotes the LISA noise power spectral density~\cite{2021arXiv210801167B}. Here we use the noise power spectral density $S_{\rm n}(f)$ corresponding to the LISA Science Requirements Document (SciRD), corrected for the ``de-averaging'' factor of $3/20$, and including the foreground of Galactic white dwarf binaries corresponding to $4$ years of observation. The errors in the parameters are given by the diagonal terms of the correlation matrix (the inverse of the Fisher matrix), i.e. $\Delta\tilde{\theta}_a = \left(\Gamma^{-1}\right)_{aa}$ (no summation implied). In the equations above, $f_0$ is the GW frequency at the beginning of the BBH observation, and $\delta f$ is the change in frequency during the observation time. Given an expression for~$t(f)$, as discussed below Eqs.~(\ref{wf0}) and~(\ref{amp0}), the change in frequency can be found from $t(f_0+\delta f)-t(f_0)=T_{\rm obs}$, where we assume $T_{\rm obs}=4$~yr to be the nominal observation time for LISA~\cite{2017arXiv170200786A} (see \cite{2021arXiv210709665S} for a discussion of different options for the mission duration).

Since the BBHs under consideration are usually observed long before merger, typically $\delta f/f_0\ll 1$. This means that we must be careful to evaluate the integral in Eq.~(\ref{inner}) with sufficient accuracy. Indeed, the integrand~$\gamma_{ab}$ is a sum of two direct vector products: $\gamma_{ab}\propto \partial_a\hat{h}\,\partial_b\hat{h} + \hat{h}^2\partial_a\phi\,\partial_b\phi$, where $\hat{h}\equiv |h_\alpha|/\sqrt{S_{\rm n}}$ and $\phi = \arg{h_\alpha}$. Therefore, if we expand $\Gamma_{ab}$ to first order in~$\delta f$, $\det{\Gamma_{ab}} \propto (\delta f)^{N}\,\det{\gamma_{ab}} = 0$, where $N$ is the number of parameters ($N=11$ in our case). This makes the determinant at least~$\mathcal{O}\left((\delta f)^{N+1}\right)$, and special care is required when inverting the matrix. We use the Python library \texttt{mpmath}~\cite{mpmath} for arbitrary precision arithmetic, setting the number of significant digits equal to $75$.

\section{Astrophysical scenarios\label{astro}}

In this section we introduce some astrophysical models to estimate the number of systems that yield detectable Doppler shifts. We consider two scenarios. In the first (Sec.~\ref{cluster}), BBHs orbit IMBHs located in Galactic GCs. In the second (Sec.~\ref{evolve_GC}), they orbit wandering IMBHs left behind when clusters dissolve by losing their mass due to tidal stripping by the Galaxy, stellar evolution, and star ejections.

\subsection{Intermediate-mass black holes in Milky Way globular clusters\label{cluster}}

In our first scenario, we consider IMBHs that may be located at the centers of Galactic GCs. We extract GC parameters (luminosities, angular positions, distances, and metallicities) from the 2010 edition of the Harris catalog\footnote{Available at~\url{https://physics.mcmaster.ca/~harris/mwgc.dat}.}~\cite{1996AJ....112.1487H}. We only exclude the cluster GLIMPSE02, since its absolute magnitude (as well as many other parameters) is not reported. In order to convert absolute visual magnitudes to cluster masses, we assume a mass-to-light ratio of $1.5\,M_\odot/L_\odot$~\cite{2017ApJ...836...67H}. Note that this value is close to the typical value derived from a sample of the Milky Way GCs~\cite{2000ApJ...539..618M,2007A&A...469..147R}, although some models could yield higher mass-to-light ratios~\cite{2017MNRAS.464.2174B}.

For simplicity, we assume that every cluster hosts an IMBH in its center, with a mass making up a fixed fraction
\begin{equation}
 f_{\rm IMBH}=\frac{M_{\rm IMBH}}{M_{\rm GC}}
\label{eq:fimbh}
\end{equation}
of the cluster mass $M_{\rm GC}$. Motivated by observational constraints on the masses of IMBH candidates, and in order to bracket the uncertainties on estimated IMBH masses in GCs (see e.g. Table~3 in~\cite{2020ARA&A..58..257G}), we explore the range of mass fractions $f_{\rm IMBH}=10^{-3}-10^{-1.5}$, which corresponds to $0.1\%$--$3.2\%$ of the host GC mass. In that range, we consider $10$~values of $f_{\rm IMBH}$ equally spaced on a log scale. As long as calculating this fraction results in a black hole of $>100\,M_\odot$, we consider it to be an IMBH and use it in the rest of the simulation.

To obtain the masses $m_1$ and~$m_2$ of the BBH components orbiting the IMBH, we first sample the masses of their stellar progenitors from a Kroupa initial mass function~\cite{2001MNRAS.322..231K}
\begin{equation}
\xi(m_*)=k_1
\begin{cases}
\left(\frac{m_*}{0.5}\right)^{-1.3}& \text{$0.08\le m_*/\mathrm{M}_\odot\leq 0.50$},\\
\left(\frac{m_*}{0.5}\right)^{-2.3}& \text{$0.50\le m_*/\mathrm{M}_\odot\leq 100.0$},
\end{cases}
\label{eqn:imf}
\end{equation}
where $k_1\approx 0.62$. We evolve stars with mass $m>20\mathrm{M}_\odot$ using the latest version of \textsc{sse}~\cite{2000MNRAS.315..543H, 2020A&A...639A..41B}, updated with the most up-to-date prescriptions for stellar winds and remnant formation, until they form an SBH. Stellar tracks are computed using the metallicity appropriate to each GC in the Milky Way. For clusters with no estimated value of the metallicity, we set it equal to the average catalog metallicity $\overline{Z}=0.05\ Z_\odot$, where $Z_\odot\approx 0.02$ is the solar metallicity~\cite{1989GeCoA..53..197A}. For metalicities $Z\lesssim 0.1Z_\odot$ (which includes $2/3$ of GCs in the Harris catalog), the masses of the SBH remnants predicted with SSE are about $ 5\,M_\odot$--$45\,M_\odot$ (see Fig.~1 of Ref.~\cite{2020ApJ...902L..26F}). For comparison, for the solar metallicity the mass range reduces to about $ 5\,M_\odot$--$15\,M_\odot$.

Next, we randomly combine pairs of SBH remnants to form BBHs. For the majority of the Milky Way GCs the BBH masses will be approximately in the range $10\,M_\odot$--$90\,M_\odot$. For each BBH we draw its semimajor axis $a$ from a log-uniform distribution between $a_{\rm min}=0.01$~AU and $a_{\rm max}=100$~AU, while the semimajor axis $R$ of its orbit around the central IMBH is drawn from a uniform distribution in the range $[0,r_{\rm infl}]$. Here $r_{\rm infl}$ is the influence radius of the IMBH, related to the the GC half-mass radius $r_{\rm h}$ by
\be
\label{influence_rad}
r_{\rm infl} = f_{\rm IMBH}\,r_{\rm h}\,.
\ee
In turn, the half-mass radius is computed using the following expression for the half-mass density~\cite{2014ApJ...785...71G}:
\beqa
\rho_{\rm h}&=& 10^3\,\frac{M_\odot}{\mbox{pc}^3}\,\min\left\{100, \max\left[1,\left(\frac{M_{\rm GC}}{2\times 10^5\,M_\odot}\right)^{2}\right] \right\}\,. \nonumber \\
& & \label{rhohalf}
\eeqa
For $10^4\,M_\odot<M_{\rm GC}<10^7\,M_\odot$, this results in a half-mass radius $1.5\,{\rm pc}<r_{\rm h}< 3.5\,{\rm pc}$.
Now, when we sample $a$ and $R$, we check that the triple system (BBH+IMBH) is stable under the tidal disruption condition~\cite{2001MNRAS.321..398M, 2012ApJ...757...27A,2019MNRAS.488...47F}
\be
\label{stability_crit}
\frac{R}{4r_{\rm t}} > 1\,,
\ee
where
\beqa
\label{disruption_radius}
r_{\rm t} &=& 0.05\,\mbox{AU}\,\left(\frac{a}{0.01\,\mbox{AU}}\right)\nonumber\\
&\times & \left(\frac{M_{\rm IMBH}}{10^3\,M_\odot}\right)^{1/3}\left(\frac{m_1 + m_2}{20\,M_\odot}\right)^{-1/3}
\eeqa
is the tidal disruption radius. Finally, the orientations of both the BBH's orbital angular momentum and of the BBH orbit around the IMBH are assumed to be distributed isotropically: the direction of the orbital angular momentum $\mathbf{n}_L\equiv (\sin{\bar{\theta}_L}\cos{\bar{\phi}_L}, \sin{\bar{\theta}_L}\sin{\bar{\phi}_L}, \cos{\bar{\theta}_L})$ points to a random direction on the sphere, and $\cos{I}$ is distributed uniformly in the range $[-1,1]$.

Typically, $\mathcal{O}(10)$ BBHs lurk in a star cluster at any given time~\cite{2015ApJ...800....9M,2020ApJ...898..162W}. However, this number could significantly change under the assumption of high primordial binary fractions and/or for a top-heavy initial mass function -- in particular, in the case where massive black holes are formed~\cite{2021ApJ...908L..29G,2021ApJ...907L..25W} -- or in the case of core-collapse star clusters~\cite{2020ApJ...903...45K}. To bracket these uncertainties, we consider two possibilities for the number of BBHs in a catalog. In Sec.~\ref{results} we refer to these possibilities as the cases of ``many'' and ``few'' BBHs, by which we mean that a star cluster hosts $\mathcal{O}(100)$ or $\mathcal{O}(10)$ BBHs. If there is an IMBH in a cluster, SBHs may be less abundant, e.g. because the IMBH forms from the reservoir of SBHs or because the SBHs are ejected from the cluster as a result of scatterings~\cite{2004ApJ...613.1143B,2014MNRAS.444...29L,2015MNRAS.454.3150G,2016ApJ...819...70M,2019arXiv190500902A}. Therefore, the case of ``many'' BBHs represents an optimistic upper limit and helps highlight the most promising candidates with $\sim\mbox{a few}$ Doppler shift events (see Fig.~\ref{MWindnums} below). The case of ``few'' BBHs in turn provides an estimate for the number of events if the IMBH is light ($\sim 100\,M_\odot$) and less likely to affect the BBH abundance (for example, an IMBH forming from stellar collisions~\cite{2021ApJ...908L..29G,2021MNRAS.501.5257R,2021MNRAS.507.5132D}). This case also serves as a consistency check, where we should expect the ``many'' BBH number to scale down by about a factor of $10$. If a cluster model yields a different number of BBHs per cluster (e.g.~\cite{2018PhRvL.120s1103K,2019PhRvD..99f3003K}), it is also easy to adjust our results proportionately.

In our models, we sample the respective number of BBHs using the procedure outlined above and select only those emitting in the LISA band, i.e. those with GW frequency $2/P_{12} > 10^{-5}$~Hz, where $P_{12}$ is the period of the binary. The results for the number of BBHs with detectable Doppler shifts are reported in Sec.~\ref{subsec:MW_results} below.

\subsection{Wandering intermediate-mass black holes\label{evolve_GC}}

In the second scenario, we explore the possibility that BBHs may orbit around wandering IMBHs. During their evolution, some of the clusters dissolve by losing their mass due to tidal stripping by the Galaxy, stellar evolution, and star ejections~\cite{2014ApJ...785...71G}, in which case they may leave behind the IMBHs that were hosted in their centers~\cite{2018ApJ...856...92F,2018ApJ...867..119F}.

To estimate how many of these wandering IMBHs could be revealed by the method of radial velocities, we start off by simulating the evolution of GCs in the Milky Way following the prescription from~\cite{2014ApJ...785...71G} (for other applications of that prescription see, for example, \cite{2018ApJ...856...92F,2018ApJ...867..119F,2018PhRvL.121p1103F,2020ApJ...899..149R}). Initially, GCs are assumed to comprise a fraction $f_{\rm GC}=0.01$ of the Milky Way's mass $M_{\rm gal} = 5\times10^{10}\,M_\odot$ (approximately equal to the estimated stellar mass~\cite{2015ApJ...806...96L}), and their masses are sampled from the distribution
\be
\label{IMF_GC}
F(M_{\rm GC}) \propto M_{\rm GC}^{-2}\exp{\left(-\frac{M_{\rm GC}}{10^6\,M_\odot}\right)}
\ee
in the range $[10^4\,M_\odot, 10^7\,M_\odot]$. The initial distribution of GC distances to the Galactic center is assumed to follow a spherical S{\'{e}}rsic profile~\cite{1963BAAA....6...41S} with total stellar mass~$M_{\rm gal}$, effective radius~$r_{\rm e}=4$~kpc, and concentration index~$n_{\rm s}=2.2$\,. When computing GC orbits in the Galaxy, we also include the contribution of the gravitational potential of dark matter, described by a Navarro--Frenk--White profile~\cite{1997ApJ...490..493N} with total mass~$M_{\rm h}=10^{12}\,M_\odot$, scale radius $r_{\rm s} = 20$~kpc, and virial radius~$R_{\rm vir}=10\;r_{\rm s}$\,.

The initial cluster positions change in time because of dynamical friction, so that their distance $r$ to the Galactic center decreases in accordance with equation
\be
\frac{d r^2}{d t} = -\frac{r^2}{t_{\rm df}(r)}\,,
\ee
where~\cite{1943ApJ....97..255C,2008gady.book.....B}
\be
\label{friction}
t_{\rm df} = 0.45\,\mbox{Gyr}\,\left(\frac{r}{\mbox{kpc}}\right)^2\left(\frac{V_{\rm c}(r)}{\mbox{km\,s}^{-1}}\right)\left(\frac{M_{\rm gc}}{10^5\,M_\odot}\right)^{-1}f_{\rm e}\,.
\ee
Here $V_{\rm c}$ is the circular velocity in the Galaxy, and $f_{\rm e}=0.5$ is a correction for eccentric GC orbits~\cite{2014ApJ...785...71G}.

\begin{figure*}
  \centering
  \includegraphics[width=0.49\textwidth]{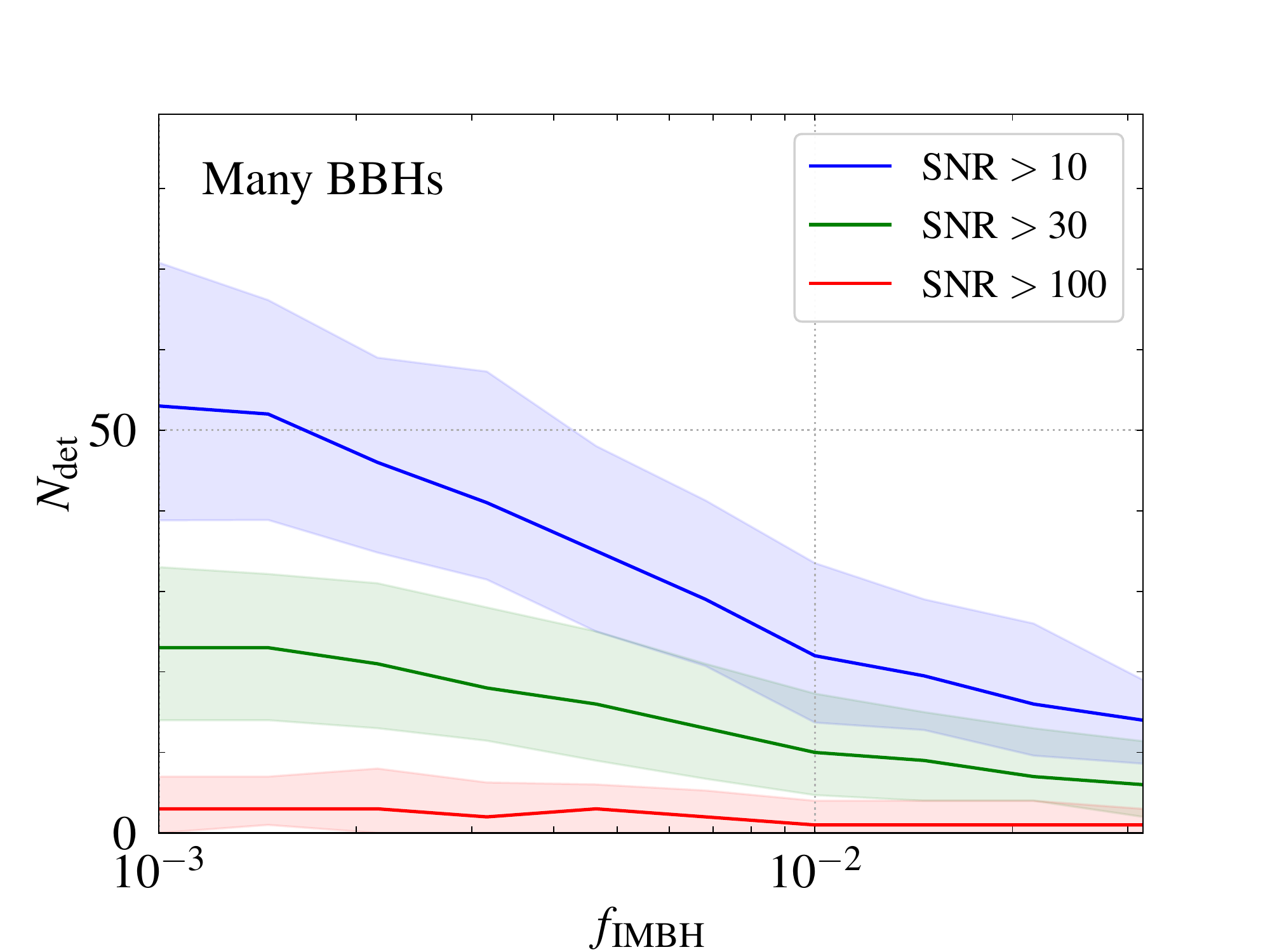}
  \includegraphics[width=0.49\textwidth]{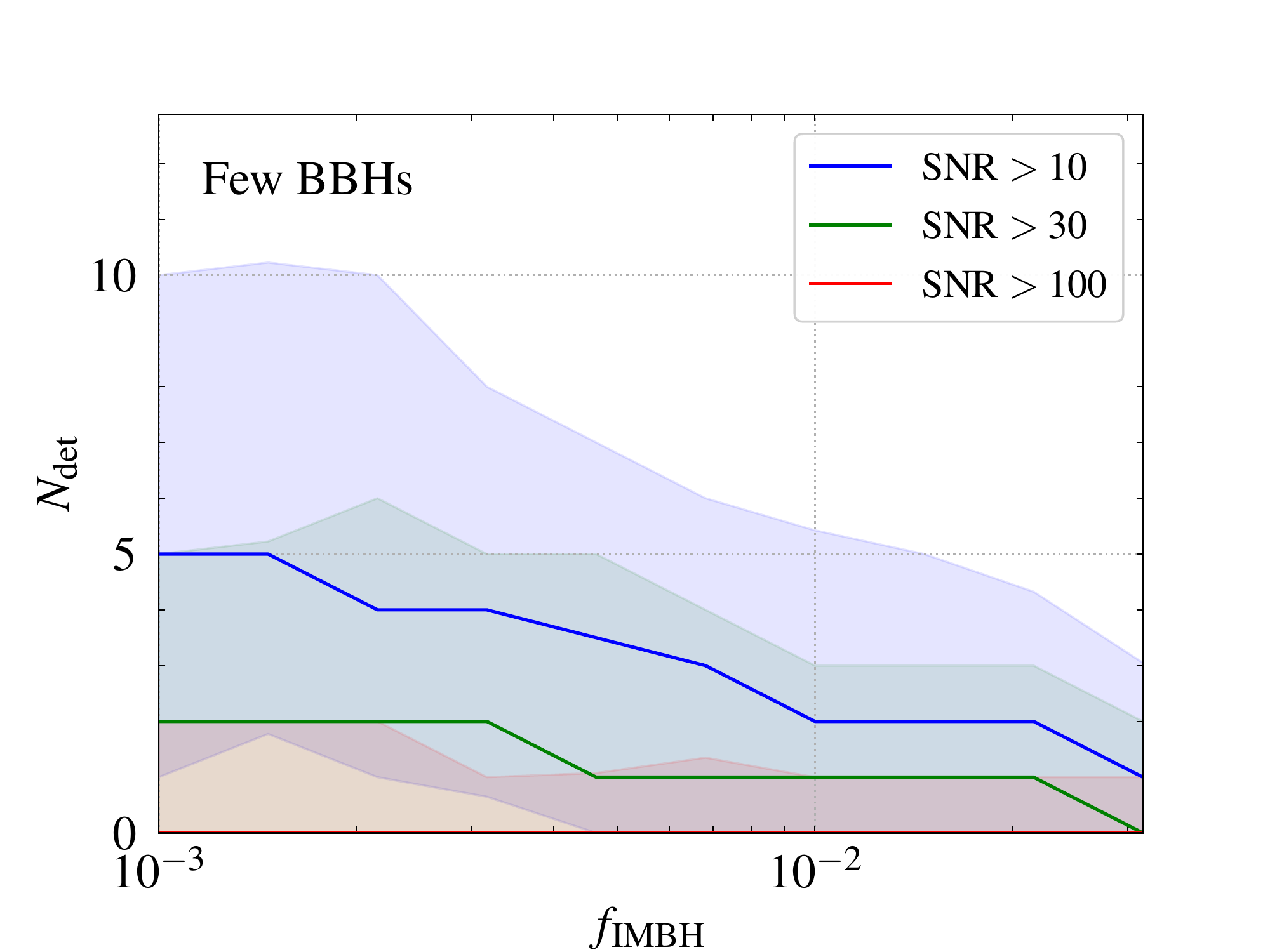}
  \caption{\label{MW_total}Total number of detectable Doppler shift events in the Milky Way's GCs, $N_{\rm det}$, as a function of the ratio of IMBH mass to the mass of its host cluster, $f_{\rm IMBH}$, in the case of many BBHs (left panel) and few BBHs (right panel). Different colors represent different values of the SNR threshold (blue: $\mbox{SNR}>10$; green: $\mbox{SNR}>30$; red: $\mbox{SNR}>100$). Shaded areas indicate statistical uncertainties ($95$-percent quantiles for 112 simulation runs).}%
\end{figure*}

As mentioned above, clusters can lose their mass due to stellar evolution (winds), star ejections after close encounters, and tidal stripping in the Galaxy. To model the mass loss due to stellar evolution, we sample stellar masses from the Kroupa initial mass function of Eq.~(\ref{eqn:imf}), which we evolve using \textsc{sse} assuming that $\overline{Z}=0.05Z_\odot$ (the average metallicity in the Harris catalog). This allows us to obtain remnant masses as a function of initial main-sequence masses (see also~\cite{2021arXiv210600699M}). Thus, the mass lost via stellar evolution
is simply due to the mass of the stars that evolved out of their main sequence to form compact remnants.
We parametrize the typical timescale of the mass loss due to star ejections as~\cite{2014ApJ...785...71G}
\be
\label{eq:ejections}
t_{\rm ej} \approx 17\,\mbox{Gyr}\,\left(\frac{M_{\rm gc}}{2\times 10^5\,M_\odot}\right)
\ee
and the timescale of tidal stripping as~\cite{2001MNRAS.325.1323B,2008MNRAS.389L..28G,2008ApJ...689..919P,2001ApJ...561..751F}
\beqa
\label{eq:tidal}
t_{\rm tid} &\approx& 10\,\mbox{Gyr}\, \left(\frac{M_{\rm gc}}{2\times 10^5\,M_\odot}\right)^{2/3}P(r)\,, \nonumber \\
P(r)&=& 100\left(\frac{r}{\mbox{kpc}}\right)\left(\frac{V_{\rm c}(r)}{\mbox{km\,s}^{-1}}\right)^{-1}\,.
\eeqa
To evolve the mass of a given GC, we first calculate the time steps at which stars in the cluster leave the main sequence. At each time step, the mass decreases by an amount equal to the difference between the mass of the star which is leaving the main sequence and the corresponding remnant mass. In between steps, the evolution of the cluster mass is governed by the equation
\be
\frac{d M_{\rm GC}}{d t} = -M_{\rm GC}\left(\frac{1}{t_{\rm ej}} + \frac{1}{t_{\rm tid}}\right)\,.
\ee
If a cluster approaches the Galactic center, we assume that it is disrupted as soon as the local Galactic density~$\rho(r)$ exceeds the average density of the cluster at the half-mass radius~\cite{2014ApJ...785...71G}, $\rho(r)>\rho_{\rm h}$, eventually leaving behind the central IMBH and any BBHs orbiting it. We evolve the clusters for $10$~Gyr and find that about $200$~clusters survive, $\sim 10{,}000$ clusters get disrupted, and a few dozens of the disrupted clusters end up within $10$~pc of the Galactic center. The mass distribution and Galactic density of surviving clusters is consistent with the Harris catalog~\cite{2014ApJ...785...71G,2019ApJ...871L...8F}.

The number and masses of the wandering IMBHs left behind by the disrupted clusters are uncertain. For an upper bound, we assume that every GC hosted an IMBH, whose mass is a fraction of the initial mass of the cluster. As in the previous section, we consider a range of fractions $f_{\rm IMBH}=10^{-3}-10^{-1.5}$, as long as the resulting IMBH mass is greater than $100\,M_\odot$. This prescription results in $\approx 800$--$10{,}000$ wandering IMBHs in the Milky Way, with the number and mass distribution of these IMBHs depending on the assumed initial mass fraction. Since we use constant ratios of the mass of an IMBH to the mass of its host/parent cluster, the mass spectrum of wandering IMBHs is roughly proportional to the IMF of GCs, Eq.~(\ref{IMF_GC}). It is somewhat biased towards lighter IMBHs, because light GCs are more likely to be disrupted.
For $f_{\rm IMBH}\sim 0.01$ this bias is negligible, because the number of disrupted GCs is not very different
from the initial number of GCs. At higher values of $f_{\rm IMBH}$, all GCs leave an IMBH (in the sense that
$f_{\rm IMBH}M_{\rm GC}>100M_\odot$ for all of them). For $f_{\rm IMBH}\sim 0.001$ this bias is more pronounced,
but not sufficient to make the mass function top-heavy.

The number of BBHs that orbit a wandering IMBH is uncertain. To estimate the number of detectable Doppler shift events, we simply assume that there is exactly one equal-mass BBH around each wandering IMBH. We consider two values of the BBH masses, motivated by the least and most massive SBHs produced from the stellar initial mass function of Eq.~(\ref{eqn:imf}), namely $m_1=m_2=10\,M_\odot$ and $m_1=m_2=50\,M_\odot$\footnote{These limiting masses may correspond to different metallicity; see for example Fig.~1 in Ref.~\cite{2020ApJ...902L..26F}.}. We sample the orbital semimajor axis of the BBH
from a uniform distribution with upper limit set by the influence radius of the wandering IMBH $R_{\rm hs}$, determined from
\be
\label{dispersion}
\frac{GM_{\rm IMBH}(m_1 + m_2)}{R_{\rm hs}} = 0.5\langle m_\star\rangle\sigma^2(r)\,,
\ee
where $\langle m_\star\rangle\approx 0.5\,M_\odot$ is the average stellar mass (assuming a canonical stellar initial mass function) and ~$\sigma(r)$ is the local Galactic velocity dispersion. To calculate the velocity dispersion, we use the Galactic mass profile described above and solve the Jeans equation. We have checked that the profile of $\sigma(r)$ obtained in this way is consistent with current data (see e.g.~\cite{2006MNRAS.369.1688D,2010AJ....139...59B}). The other parameters that describe the BBHs are generated as described in Sec.~\ref{cluster}.

Our estimates for the number of BBHs orbiting wandering IMBHs and producing detectable Doppler shifts are reported in Sec.~\ref {subsec:wandering_results} below.

\section{Results\label{results}}

In this section we present the results from our simulations. We first discuss the distribution of IMBHs in Galactic GCs that can be detected via Doppler shift (Sec.~\ref{subsec:MW_results}) and then we estimate the number of detectable IMBHs wandering in the Galaxy, left behind as a result of the disruption of their parent clusters (Sec.~\ref{subsec:wandering_results}).

In both cases we define a detectable Doppler shift event as a LISA observation such that the relative errors $\Delta v_{||}/v_{||}$ and $\Delta P/P$ are both smaller than $0.1$ (this value is somewhat arbitrary, but it was chosen as a proxy for sufficiently precise measurements) for three selected values of the SNR threshold, namely $10$, $30$, and $100$. Note, however, that in most cases $\Delta v_{||}/v_{||}\gg \Delta P/P$, so the velocity measurement is usually the limiting factor.

\subsection{Events from the Milky Way globular clusters\label{subsec:MW_results}}

In Fig.~\ref{MW_total} we show the total number of detectable Doppler shift events $N_{\rm det}$ as a function of $f_{\rm IMBH}$ for the cases of many BBHs (left panel) and few BBHs (right panel). We report $N_{\rm det}$ for three selected values of the SNR threshold, namely $10$, $30$, and $100$. We find that $N_{\rm det}$ decreases as a function of $f_{\rm IMBH}$, i.e., $N_{\rm det}$ is smaller for larger IMBH masses. This can be explained considering that more massive IMBHs have larger influence radii, which leads to longer periods of BBHs around the IMBHs: see Eqs.~(\ref{influence_rad}) and~(\ref{period}). As the periods exceed the LISA mission duration, it is increasingly difficult to measure the velocity and period, and the measurement errors grow (see Fig.~2 of Ref.~\cite{2019MNRAS.488.5665W} for the dependence of the errors on the period).

In Table~\ref{tab:Nev} we list $N_{\rm det}$ for the same three values of the SNR threshold listed above ($10$, $30$, and $100$) and for two selected values of $f_{\rm IMBH}=0.001,\,0.01$. The entries in the Table can be thought of as two vertical ``cuts'' in the left and right panels of Fig.~\ref{MW_total}.

\begin{table}
  \caption{Number of detectable events $N_{\rm det}$ in the case of IMBHs at the center of GCs for selected values of $f_{\rm IMBH}$, as defined in Eq.~(\ref{eq:fimbh}), and of the SNR threshold.\label{tab:Nev}}
\begin{center}
\begin{tabular}{c | c c c }\hline\hline
 \multicolumn{4}{c}{Many BBHs }\\ \hline
 \\[-0.8em]
 $f_{\rm IMBH}$ & SNR $>$ 10 & SNR $>$ 30  &  SNR $>$ 100 \\[1ex] \hline
 \\[-0.8em]
 0.001 & $53^{+18}_{-14}$ & $23^{+10}_{-9}$ & $3^{+4}_{-3}$  \\[1ex]   
 0.01 & $22^{+12}_{-8}$& $10^{+7}_{-5}$& $1^{+3}_{-1}$     \\[1ex] \hline\hline
 \multicolumn{4}{c}{Few BBHs }\\\hline
 \\[-0.8em]
 $f_{\rm IMBH}$ & SNR $>$ 10 & SNR $>$ 30  &  SNR $>$ 100 \\[1ex] \hline
 \\[-0.8em]
 0.001 & $5^{+5}_{-4}$& $2^{+3}_{-2}$& $0^{+2}_{-0}$ \\ [1ex]  
 0.01 & $2^{+4}_{-2}$& $1^{+2}_{-1}$& $0^{+1}_{-0}$ \\ [1ex]  
\hline
\end{tabular}
\end{center}
\end{table}

\begin{figure}
  \centering
  \includegraphics[width=\columnwidth]{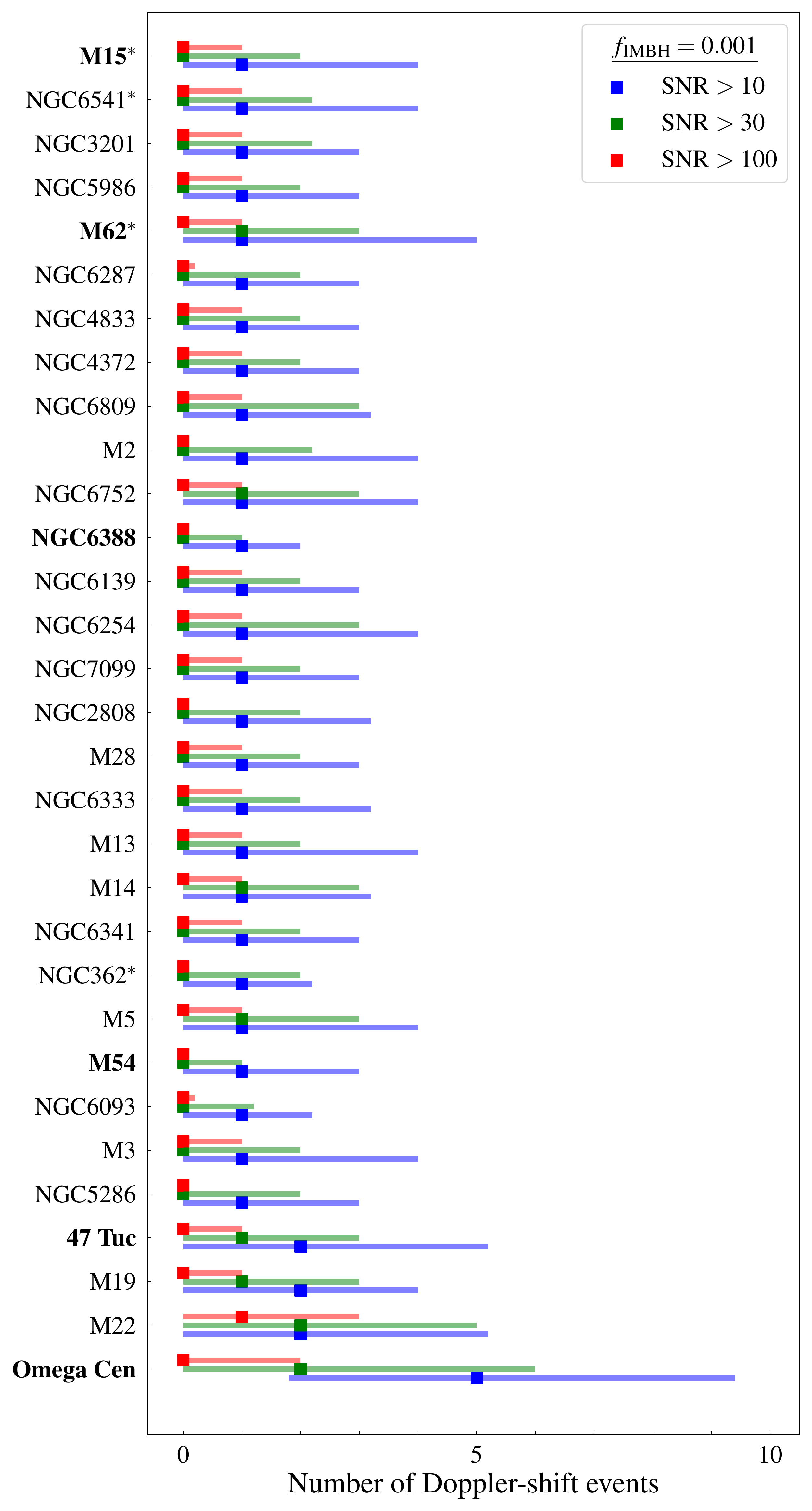}
  \includegraphics[width=\columnwidth]{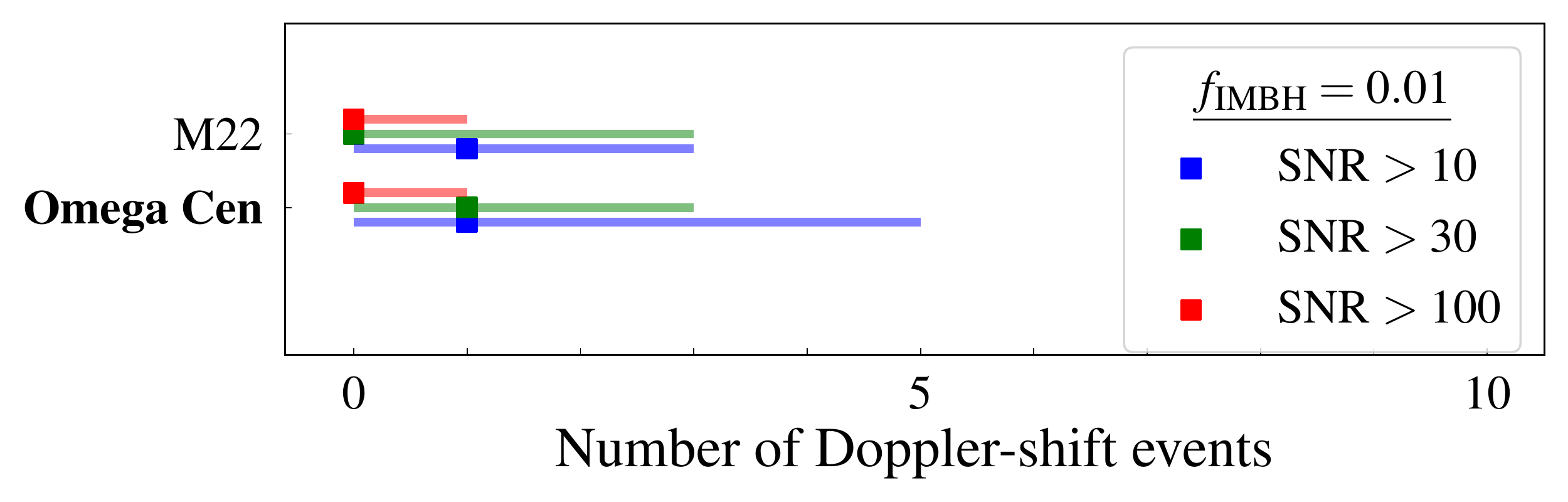}
  \caption{Total number of Doppler shift events in individual clusters for $f_{\rm IMBH}=0.001$ (top) and $f_{\rm IMBH}=0.01$ (bottom) for three different values of the SNR threshold (blue: $\mbox{SNR}>10$; green: $\mbox{SNR}>30$; red: $\mbox{SNR}>100$) in the case of many BBHs (see Sec.~\ref{cluster}). The plots include only GCs with a median value of one or more events with $\mbox{SNR}>10$. The horizontal bars indicate statistical uncertainties (95-percent quantiles for 112 simulation runs) around the median. Clusters with observational constraints on the IMBH mass (see Table~3 in Ref.~\cite{2020ARA&A..58..257G}) are highlighted in bold. Asterisks mark GCs that are core-collapsed according to the 2010 edition of the Harris catalog~\cite{1996AJ....112.1487H}. %
    \label{MWindnums}}%
\end{figure}

In Fig.~\ref{MWindnums} we further break down the results of our simulations by listing the most promising Galactic GCs that could yield detectable Doppler shift events. We focus, for concreteness, on the case of many BBHs. The top and bottom panels refer to $f_{\rm IMBH}=0.001$ and $f_{\rm IMBH}=0.01$, respectively. Both panels include only GCs yielding a median of one or more events with $\mbox{SNR}>10$. The median values, as well as the statistical uncertainties, are based on an ensemble of $112$~simulation runs\,\footnote{The uneven number of the runs stems from efficient usage of computing resources.} for each value of $f_{\rm IMBH}$. 

Let us focus on $f_{\rm IMBH}=0.001$ first (top panel). We find in total $31$~Galactic GCs yielding at least a median of one event at $\mbox{SNR}>10$. Out of these GCs, $4$ (47~Tucanae, M19, M22, and $\omega$~Centauri) yield $2$~events or more.

Some of the GCs in the list (NGC6388, M15, M62, M54, 47~Tucanae, and $\omega$~Centauri) could harbor IMBHs with estimated masses of
\mbox{$\lesssim 10^3\,M_\odot$}--$10^4\,M_\odot$, %
\mbox{$\lesssim 1,500\,M_\odot$}--$3,000\,M_\odot$, %
\mbox{$\lesssim 1,000\,M_\odot$}--$3,000\,M_\odot$, %
\mbox{$\lesssim 10,000\,M_\odot$}, %
\mbox{$\lesssim 2,000\,M_\odot$}, and %
\mbox{$\lesssim 10^3\,M_\odot$}--$10^4\,M_\odot$, %
respectively (see Table~3 in Ref.~\cite{2020ARA&A..58..257G}).
The putative IMBH in these GCs can account for at most $\sim 0.1\%$--$1\%$ of the cluster mass, making these clusters good candidates for Doppler shift events. Note however that M15 and M62 (marked with an asterisk in Fig.~\ref{MWindnums}) are known to be core-collapsed clusters, thus they are likely to contain few SBHs (see e.g.~\cite{2020ApJS..247...48K}), rendering the detection of a Doppler shift event less likely.

The best candidate is represented by $\omega$~Centauri, the most massive of the Milky Way's GCs. For this GC, we predict $5^{+5}_{-3}$ and $2^{+4}_{-2}$ events at $\mbox{SNR}>10$ and $\mbox{SNR}>30$, respectively. Note that $\omega$~Centauri could harbor an IMBH~\cite{2010ApJ...719L..60N,2010ApJ...710.1032A,2010ApJ...710.1063V} with mass $\lesssim 10^3\,M_\odot$--$10^4\,M_\odot$ (corresponding to $f_{\rm IMBH}\lesssim 10^{-3}$--$10^{-2}$) and it is not core-collapsed, so it may host a relatively abundant population of SBHs (see e.g.~\cite{2020ApJ...898..162W,2020ApJ...904..198C}). A more massive IMBH~$\sim 10^4\,M_\odot$ appears to be excluded by the absence of its dynamical influence in the core~\cite{2019MNRAS.488.5340B} and accretion~\cite{2018ApJ...862...16T}. However, the method of GW Doppler shift measurements studied in this paper is more favorable for searches of lighter IMBHs with $\lesssim 10^3\,M_\odot$ ($f_{\rm IMBH}\lesssim 10^{-3}$).

We now turn to $f_{\rm IMBH}=0.01$ (Fig.~\ref{MWindnums}, bottom panel), where we are still considering the case of many BBHs. There are now only $2$~GCs with $1+$ Doppler shift events observable at $\mbox{SNR}>10$. Both of these GCs ($\omega$~Centauri and M22) also appear on the top panel of Fig.~\ref{MWindnums}, and they include $1$~GC with an IMBH candidate: $\omega$~Centauri, our most promising target, which now yields $1^{+4}_{-1}$ and $1^{+2}_{-1}$ at $\mbox{SNR}>10$ and $\mbox{SNR}>30$, respectively.

In both the top and bottom panels, the median values for $\mbox{SNR}>100$ are mostly zero. However, all of our results should be understood as statistical averages (recall that each simulation ensemble comprises $112$~runs). The nonvanishing error bars indicate that events with such high SNRs could still be observed.
The median values fluctuate between different simulation ensembles. Due to these statistical fluctuations, at least one GC (M22)
can yield a number of detections compatible with one, and in some of our runs two of the GCs (M19 and 47~Tucanae) happened to appear also in the bottom panel of Fig.~\ref{MWindnums}.  

The case of few BBHs (not shown in Fig.~\ref{MWindnums}) yields more pessimistic results. We find only a handful of GCs having $\sim 1$ event with SNR $>10$: only a few GCs, if any, would have detectable Doppler shift events. The formal median values are now zero for all GCs considered, but the error bars may allow for up to $2$ events. A comparison of the two cases shows that the number of BBHs in a GC directly affects the number of detectable Doppler shift events.

We now discuss the impact of our assumptions on our results. To begin with, we parametrize the distribution of BBH separations, $a$, and the distribution of distances to IMBHs, $R$, with power-law functions   
\be
\label{eq:SMA_distribution}
f(a) \propto \frac{1}{a^\alpha}\,, \quad g(R) \propto\frac{1}{R^{\beta-2}}\,,
\ee
which reduce to our fiducial model when $\alpha=1$ and $\beta=2$. To test these changes to the distributions, we consider $\alpha\in[0,1.5]$ and $\beta\in[1.5,2.5]$. The function $g(R)$ is directly related to the density profile~$n(R)$, $g(R)\propto n(R)R^2\propto R^{-\beta+2}$, so that our fiducial model correspond to a cuspy profile $R^{-2}$. Power-law cusps in the proximity of a massive black hole are a general prediction of stellar dynamics models~\cite{2004ApJ...613.1143B,2018A&A...609A..28B} (see also the review~\cite{2017ARA&A..55...17A} and references therein).

\begin{figure}[!htbp]
  \centering
  \includegraphics[width=\columnwidth]{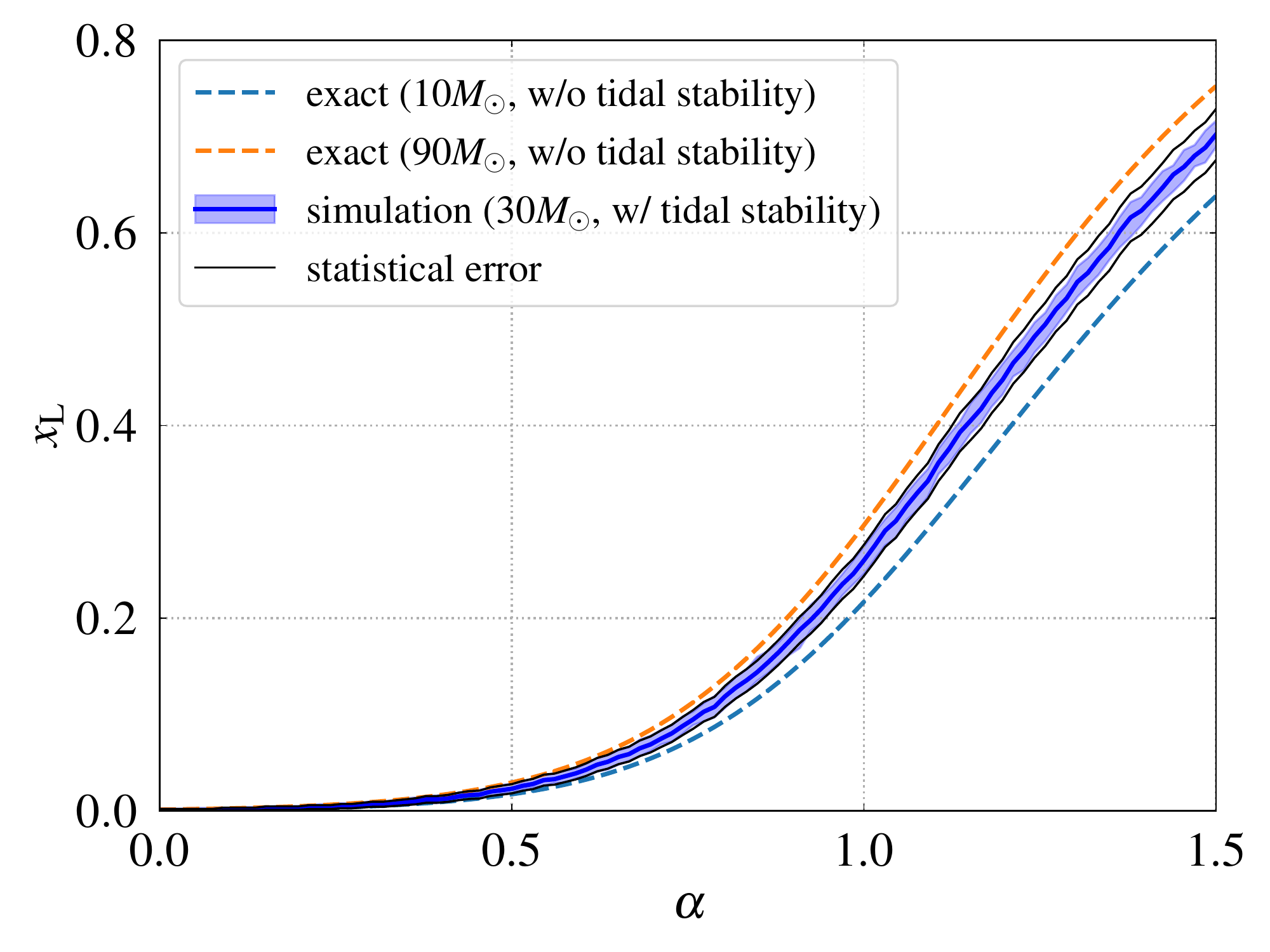}
  \caption{Fraction of BBHs in the LISA band vs. the power-law exponent~$\alpha$ of the distribution of semimajor axes defined in~Eq.~(\ref{eq:SMA_distribution}). Dashed lines show the dependence for a fixed total mass $m=m_1+m_2=10\,M_\odot$ (cyan line) and $90\,M_\odot$ (orange line) as given by Eq.~(\ref{eq:fraction_LISA}), which does not take into account the stability criterion~(\ref{stability_crit}). The shaded region with a blue line in the middle results from a simulation for a total mass of~$30\,M_\odot$, for $100$~values of $\alpha\in[0,1.5]$ and $100$~values of $\beta\in[1.5,2.5]$ at each~$\alpha$ where we do apply the stability criterion. The blue line represents the median over different values of~$\beta$, and the shaded region is the $68$-percent quantile. The fraction~$x_{\rm L}$ is evaluated by counting how many of $n_{\rm samples}=1,000$ BBHs with randomly drawn~$a$ and~$R$ end up in the LISA band. Thin lines show the statistical Poisson error estimate $\pm\sqrt{x_{\rm L}/n_{\rm samples}}$.\label{fig:fraction_LISA}}%
\end{figure}
\begin{figure*}[!htbp]
  \centering
  \includegraphics[width=0.325\textwidth]{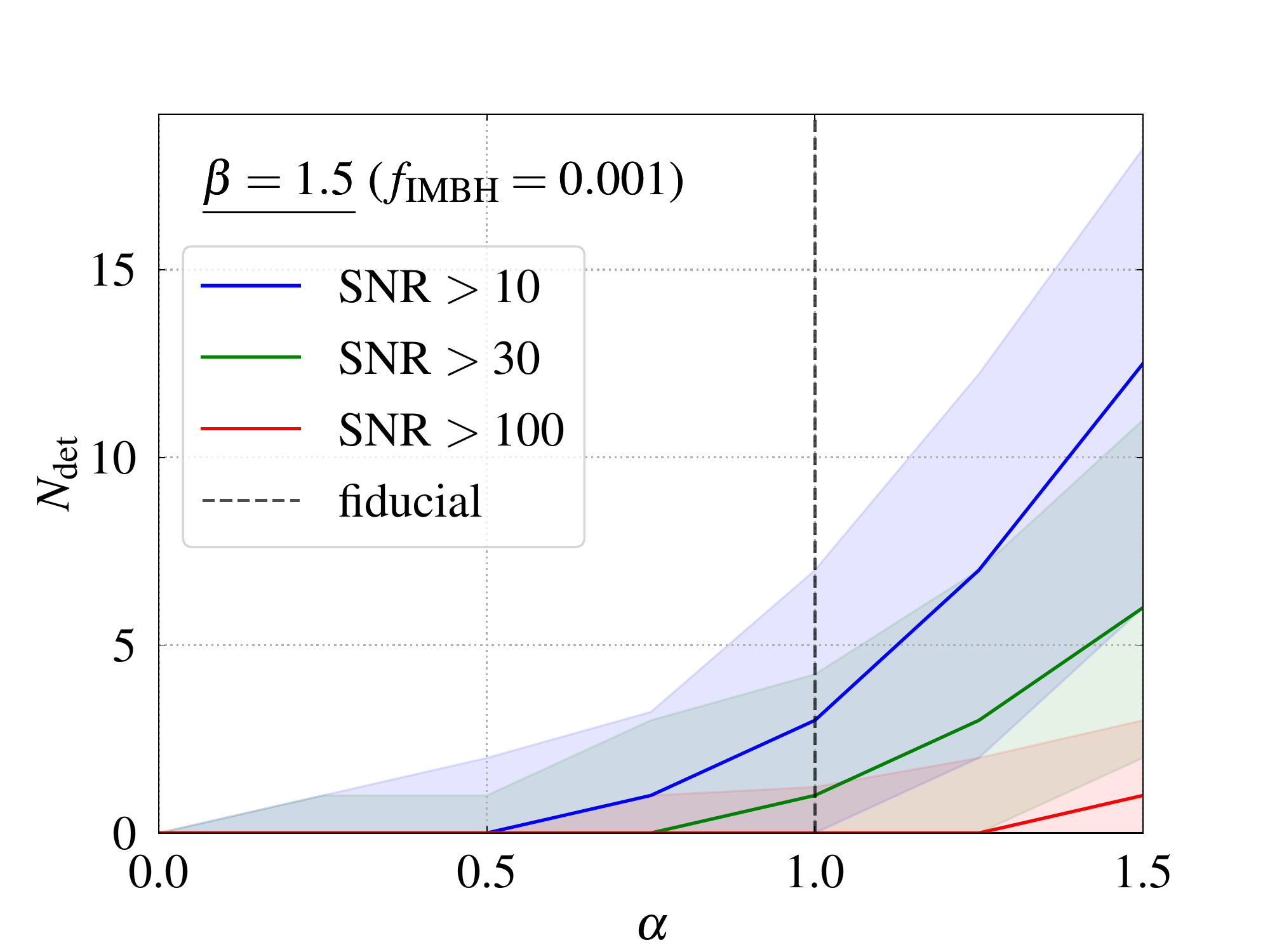}
  \includegraphics[width=0.325\textwidth]{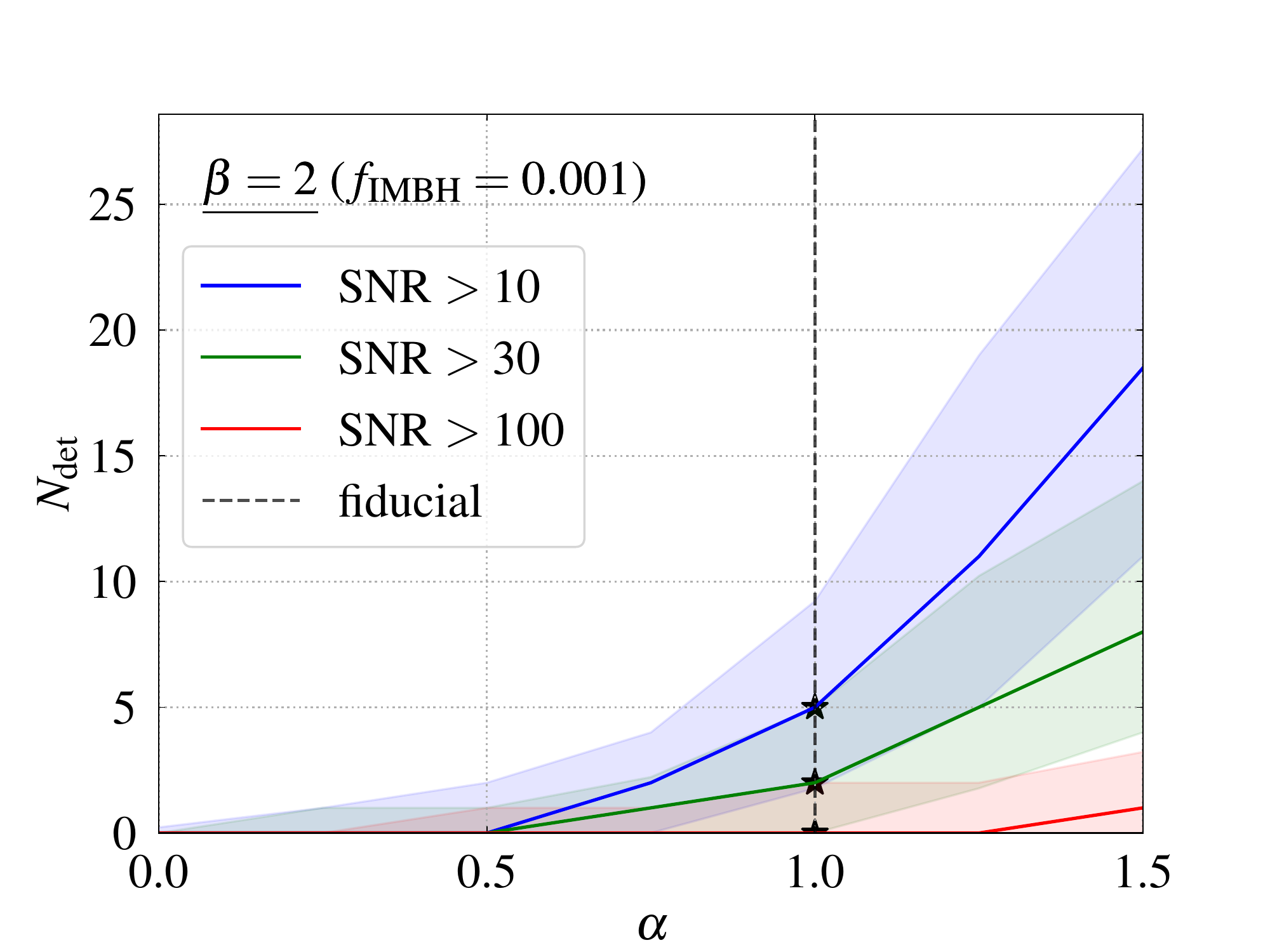}
  \includegraphics[width=0.325\textwidth]{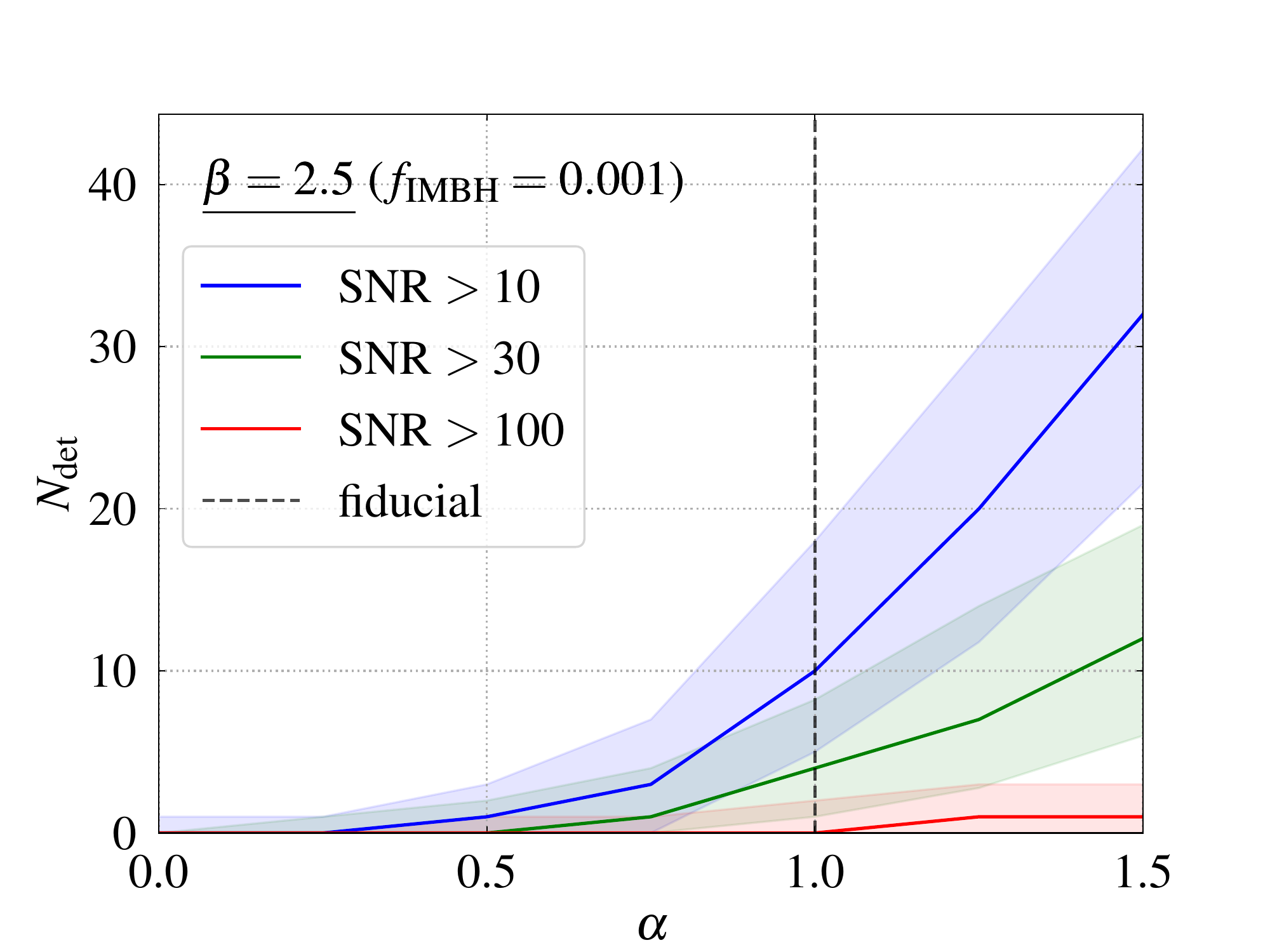}
  \includegraphics[width=0.325\textwidth]{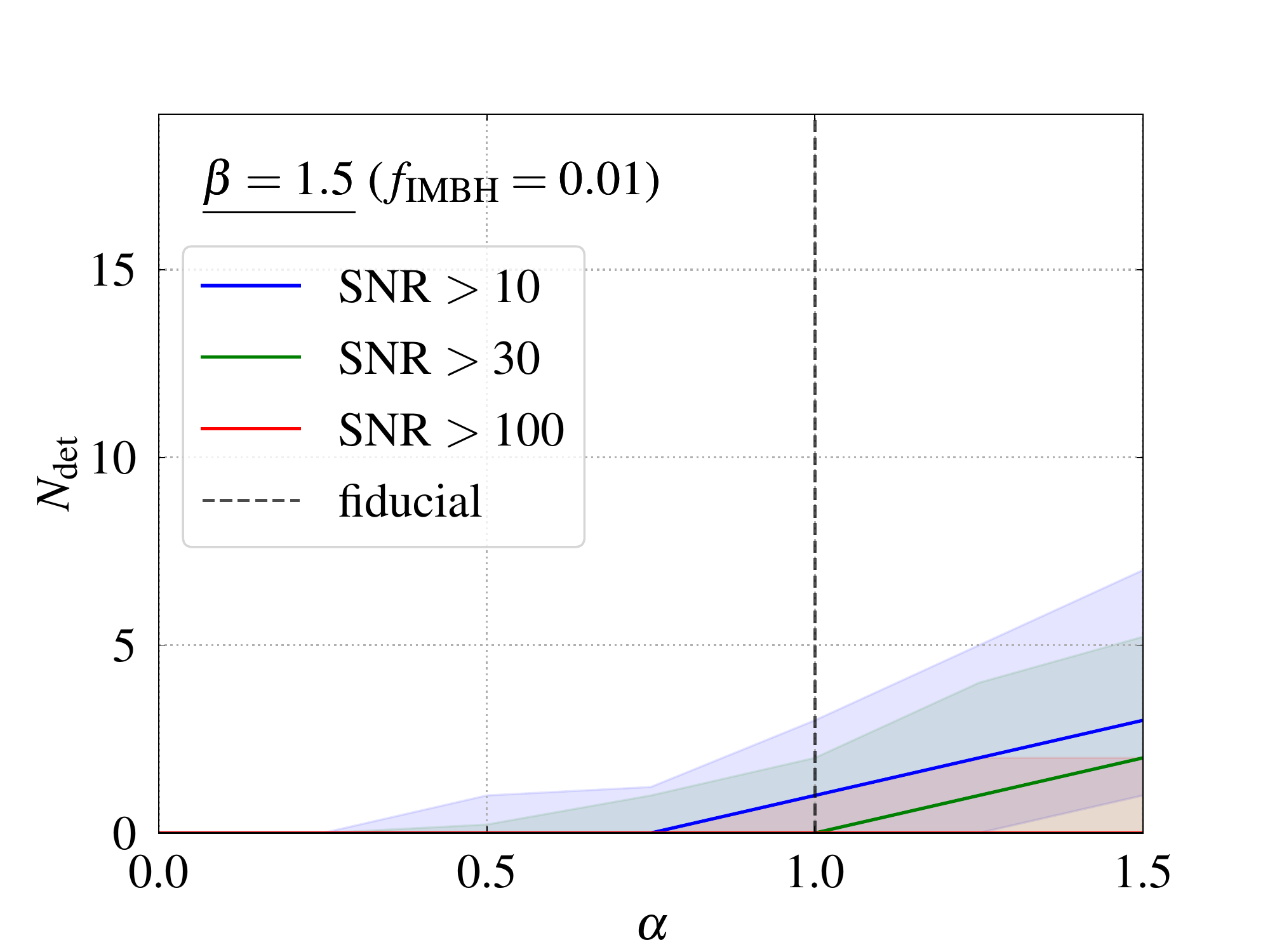}
  \includegraphics[width=0.325\textwidth]{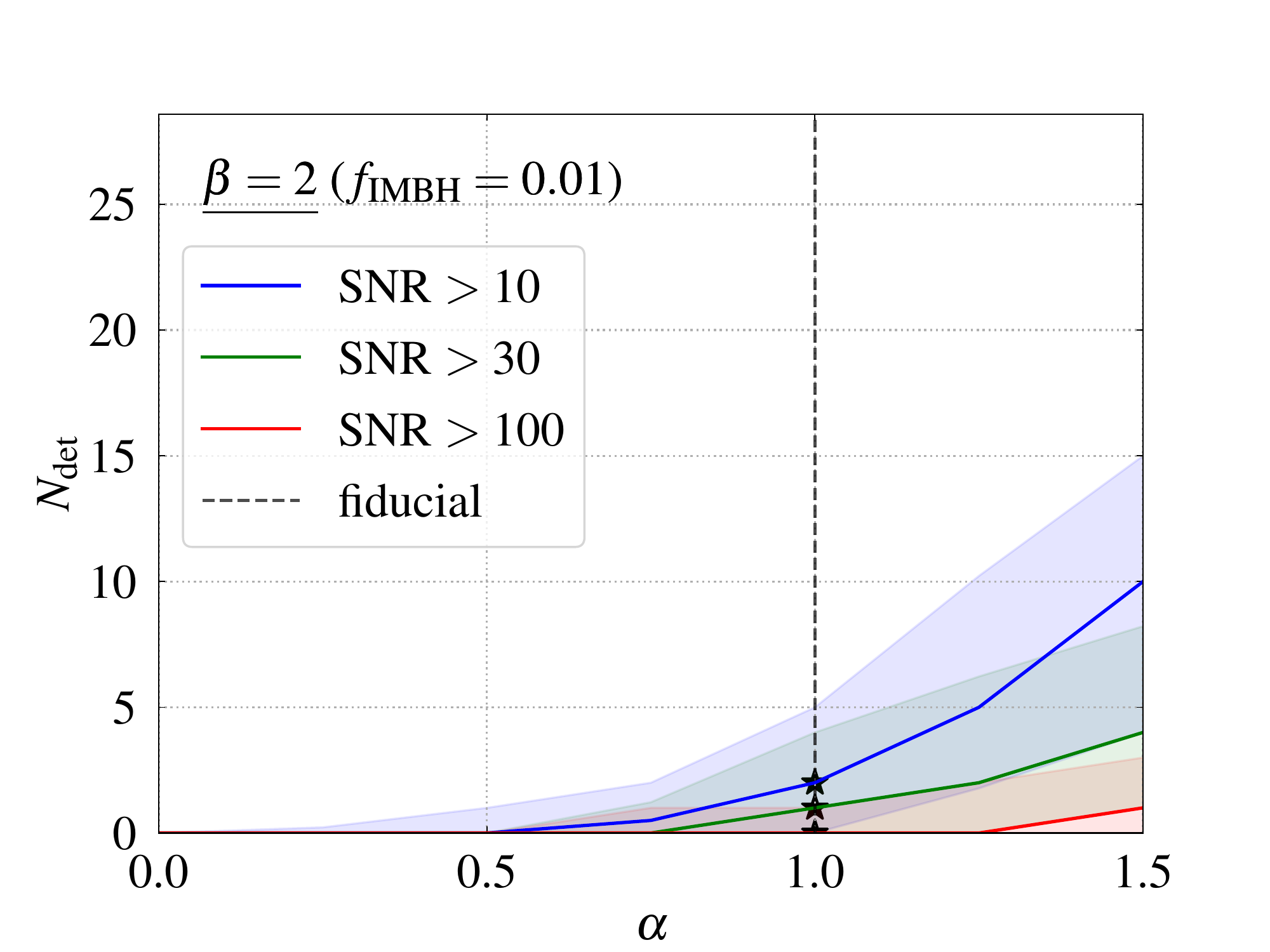}
  \includegraphics[width=0.325\textwidth]{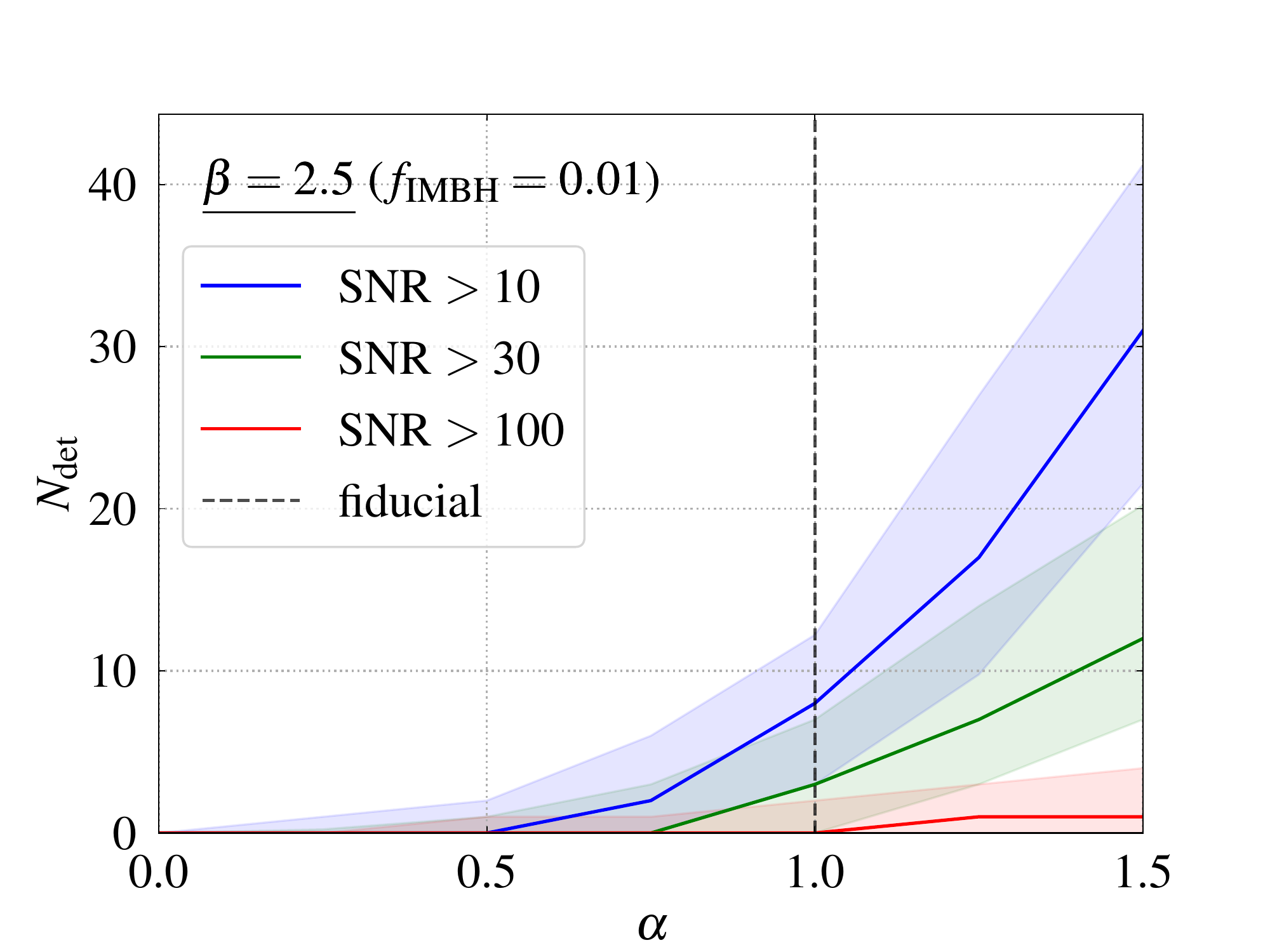}
  \caption{Total number of the Doppler shift events in the Milky Way GCs as a function of the slope of a power-law distribution of BBH separations. Three columns correspond to three values of the slope~$\beta$ in the distribution of IMBH--BBH distances, $\beta=1.5$ (left), $\beta=2$ (middle), and $\beta=2.5$ (right). Two rows show the number of detections for two values of the ratio of the mass of an IMBH to the mass of its host cluster, $f_{\rm IMBH}=0.001$ (top) and $f_{\rm IMBH}=0.01$ (bottom). The fiducial value of~$\alpha$ is indicated with vertical dashed lines, and black stars mark the results for our fiducial model.\label{fig:slopes}}%
\end{figure*}

First, let us consider how the fraction of BBHs in the LISA band, $f_{\rm min}\equiv 10^{-5}\,\mbox{Hz}<2/P_{12}<1\,\mbox{Hz}$, changes as a function of~$\alpha$ (without applying the stability criterion~(\ref{stability_crit})). If $a_{\rm L}=m^{1/3}/(\pi f_{\rm min})^{2/3}$, with $m=m_1+m_2$, is the separation of a BBH entering the LISA band, the fraction of BBHs in band is
\beqa
\label{eq:fraction_LISA}
x_{\rm L} &=& \left\langle\int\limits_{a_{\rm min}}^{a_{\rm L}(m)}{f(a)\,\dd a}\right\rangle \nonumber \\
&=& 
\begin{cases}
\frac{\left\langle\left(m/m_0\right)^{\frac{1-\alpha}{3}}\right\rangle - 1}{\left(a_{\rm max}/a_{\rm min}\right)^{1-\alpha} - 1} & \quad \alpha\neq 1\,,\\
\frac 13\frac{\left\langle\ln\left(m/m_0\right)\right\rangle}{\ln{\left(a_{\rm max}/a_{\rm min}\right)}} & \quad \alpha=1\,, \\
\end{cases}
\eeqa
where
\beqa
m_0 & \equiv & 0.025\,M_\odot\,\left(\frac{f_{\rm min}}{10^{-5}\,\mbox{Hz}}\right)^2\left(\frac{a_{\rm min}}{0.01\,\mbox{AU}}\right)^{3}\,. 
\eeqa
The angle brackets stand for averaging over the BBH mass spectrum. We find that the dependence on the total mass~$m$ is rather weak, as illustrated in Fig.~\ref{fig:fraction_LISA}. Here, we show how the estimate of $x_{\rm L}(\alpha)$ for fixed total masses of~$10\,M_\odot$ and~$90\,M_\odot$ can closely bracket an estimate of $x_{\rm L}(\alpha)$ for a typical mass spectrum. Figure~\ref{fig:fraction_LISA} also reports the estimate of $x_{\rm L}(\alpha)$ in the case when we do apply the stability criterion, which implicitly introduces a dependence of~$x_{\rm L}$ on~$\beta$. In this case, we compute $x_{\rm L}(\alpha)$ numerically for $100$ equally spaced values of $\beta\in[1.5,2.5]$, drawing each time $n_{\rm samples}=1,000$ BBH whose semimajor axes~$a$ and orbital radii~$R$ around IMBHs are distributed according to Eq.~(\ref{eq:SMA_distribution}) and subject to the stability criterion~(\ref{stability_crit}). The fraction $x_{\rm L}$ is obtained by counting how many of them end up in the LISA band. We compare the scatter due to the dependence on~$\beta$ (the blue median line surrounded by a shaded 68-percentile region) to statistical scatter (delineated by two thin black lines) and find good agreement, which implies no strong dependence on~$\beta$.

We also investigate how the number of detectable Doppler shift events changes as a function of the power-law exponents~$\alpha$ and~$\beta$. Figure~\ref{fig:slopes} shows the dependence of the total number of Doppler shift detections~$N_{\rm det}$ in the Milky Way as a function of~$\alpha$ for the case of few BBHs, for two values of the ratio of the mass of an IMBH to the mass of its host cluster, $f_{\rm IMBH}=0.001$ (top row) and $f_{\rm IMBH}=0.01$ (bottom row), and for three values of~$\beta$: $\beta=1.5$ (left column), $\beta=2$ (middle column), and $\beta=2.5$ (right column). The fiducial value~$\alpha=1$ is indicated with vertical dashed lines and the results for the fiducial model are marked with black stars in the middle column. We find that changing~$\alpha$ has a major effect on the number of detections, such that $N_{\rm det}$ almost vanishes for~$\alpha=0$ (uniform distribution of the semimajor axes) and grows by a factor of three for~$\alpha=1.5$. Still, the region indicating the 95-percent quantile on~$N_{\rm det}$ has nonzero measure even for~$\alpha=0$ and $\mbox{SNR}>10$, meaning that a detection might still be possible even in our worst-case scenario. Note that relaxing the assumption on~$\beta$ changes the number of events within a factor of $\approx 2$. For a steeper cusp with~$\beta=2.5$, the number of events increases, while it decreases for a shallower profile. The fact that $\beta$ has a weaker effect on our results than~$\alpha$ is consistent with its weaker effect on~$x_{\rm L}$, as previously discussed.

\begin{figure}[!htbp]
  \centering
  \includegraphics[width=\columnwidth]{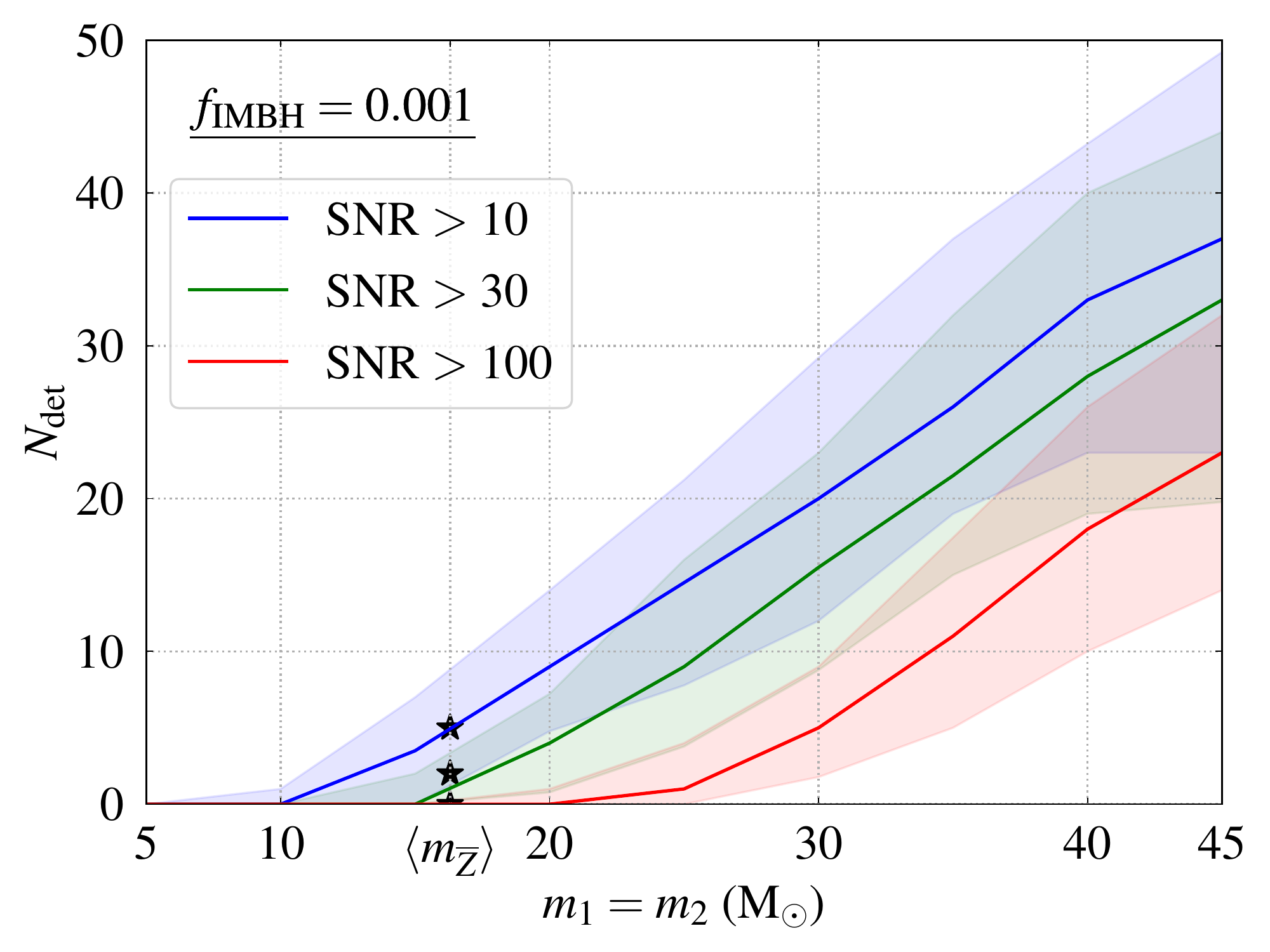}
  \includegraphics[width=\columnwidth]{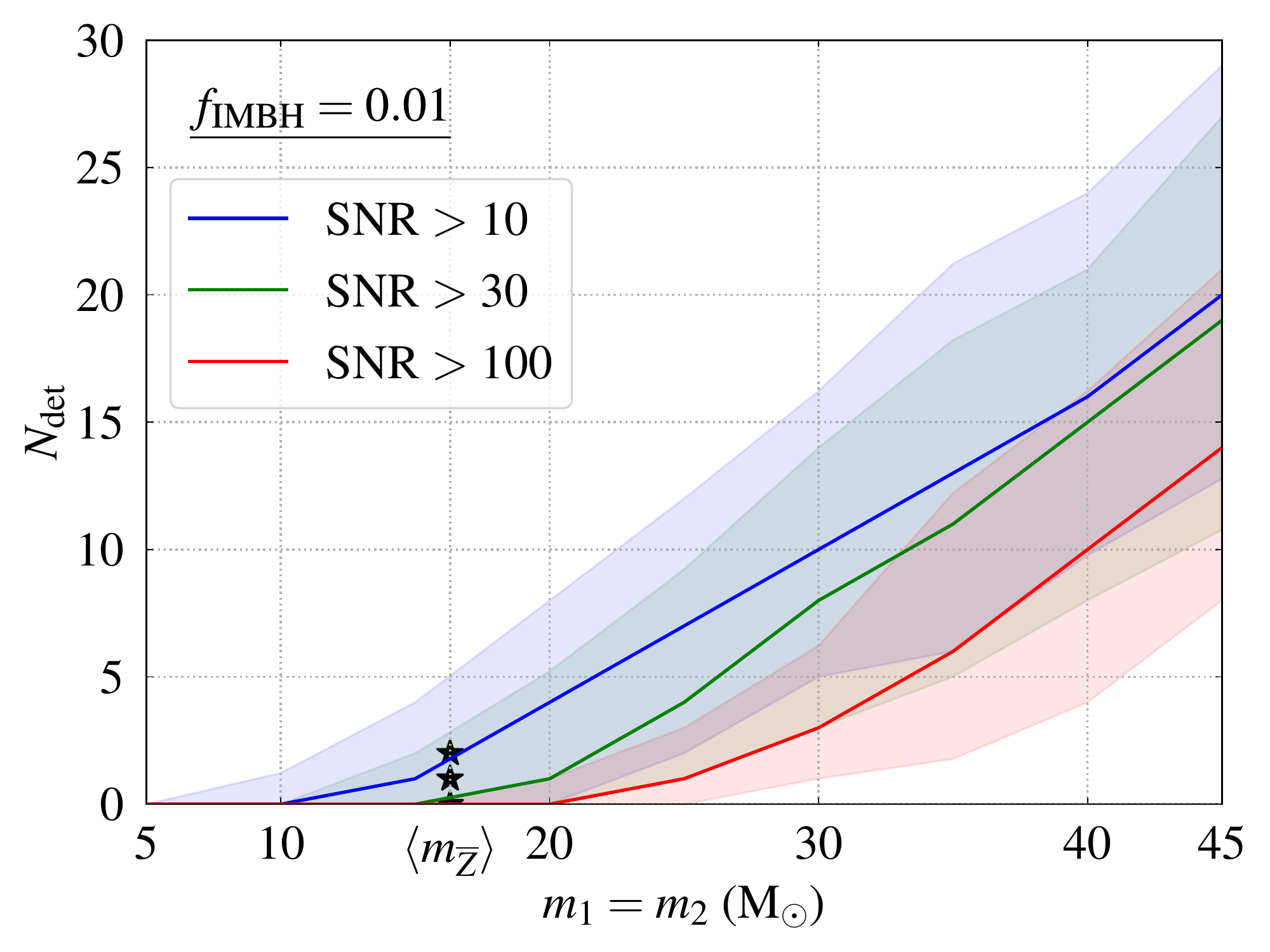}
  \caption{Total number of the Doppler shift events in the Milky Way GCs as a function of the components' mass in an equal-mass BBH for $f_{\rm IMBH}=0.001$ (top) and $f_{\rm IMBH}=0.01$ (bottom) in the case of few BBHs. Black stars mark the results from Table~\ref{tab:Nev} and are plotted against the mean mass $\langle m_{\overline{Z}}\rangle\approx 16\,M_\odot$ at $\overline{Z}=0.05Z_\odot$, the average metallicity in the Harris catalog.\label{fig:masses}}%
\end{figure}
\begin{figure*}
  \centering
  \includegraphics[width=0.49\textwidth]{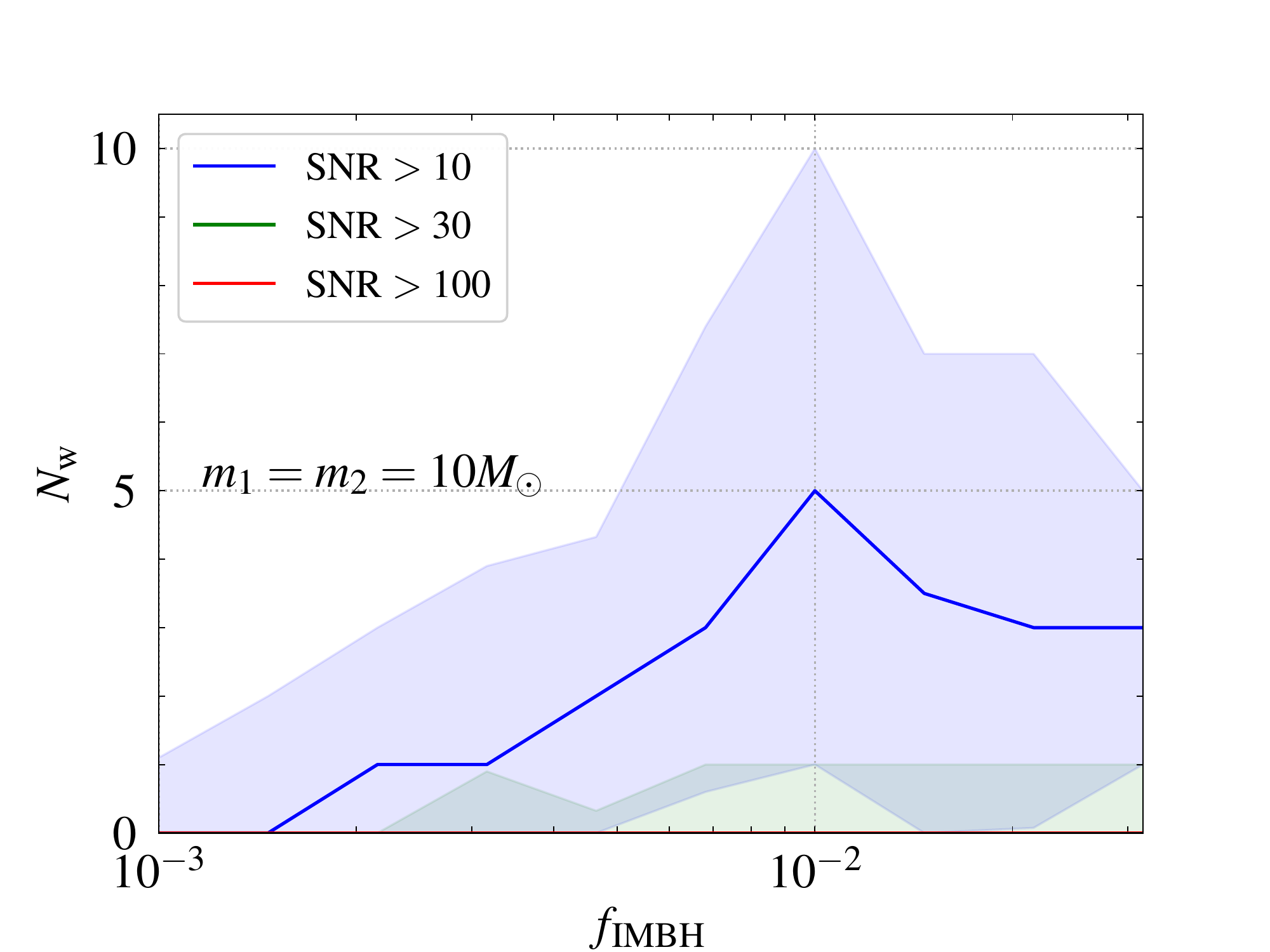}
  \includegraphics[width=0.49\textwidth]{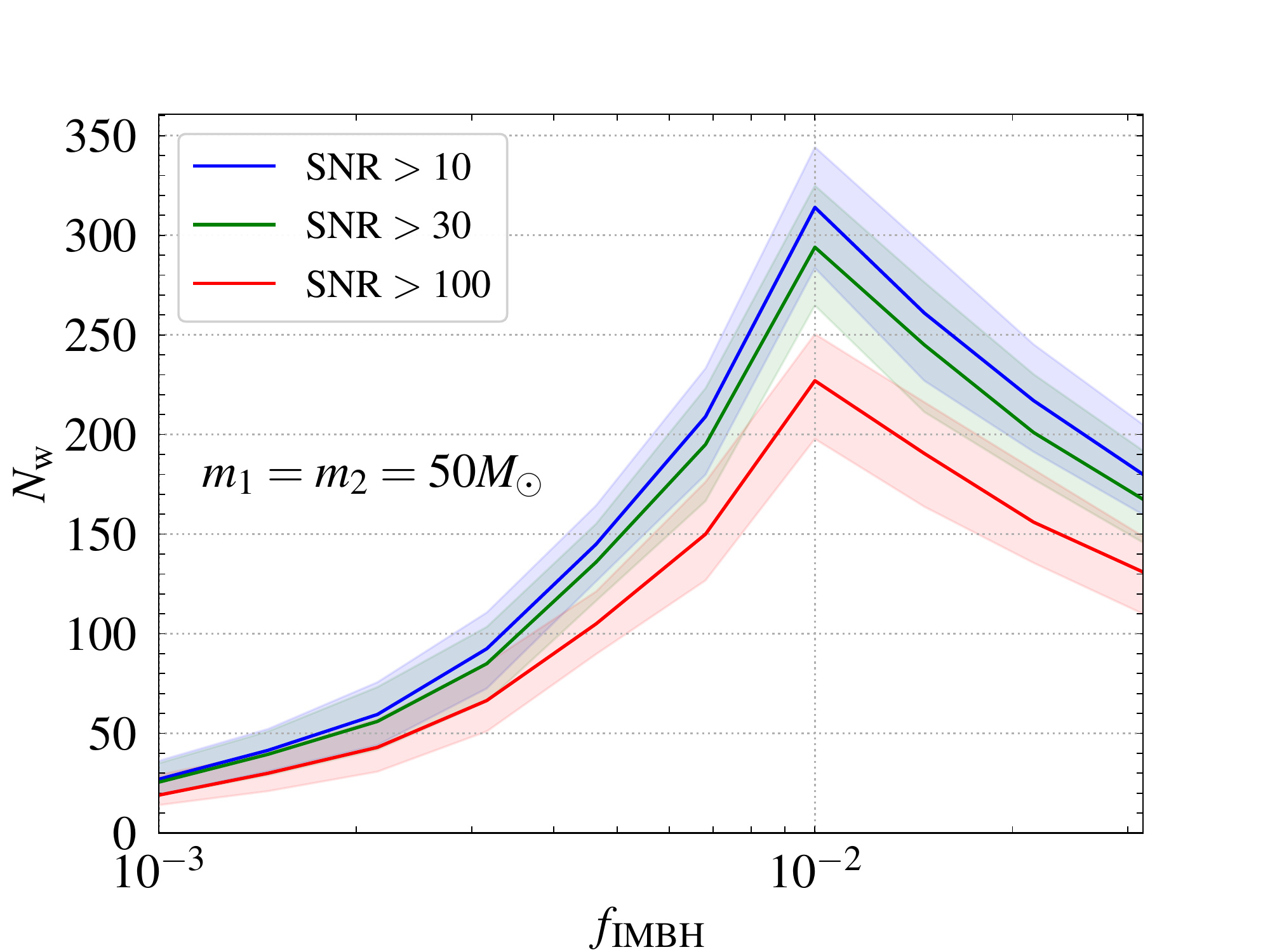}
  \caption{Total number of Doppler shift events around wandering IMBHs as a function of $f_{\rm IMBH}$ for equal-mass BBHs with individual component masses of $10\,M_\odot$ (left) and $50\,M_\odot$ (right) and three different SNR thresholds (blue: $\mbox{SNR}>10$; green: $\mbox{SNR}>30$; red: $\mbox{SNR}>100$). Shaded areas indicate statistical uncertainties ($95$-percent quantiles for 112 simulation runs). Note that $f_{\rm IMBH}$ is the ratio of IMBH mass to the initial mass of its destroyed parent cluster.\label{wandering_total}} 
\end{figure*}

Another factor that can play a role in determining the number of detectable events is the mass spectrum of BBHs, which could be affected by the presence of an IMBH and/or dynamical evolution. To check how our assumptions affect our results, we run a simulation for the case of few BBHs where all BBHs are assumed to be equal-mass. Figure~\ref{fig:masses} shows the number of events that can be detected as a function of the mass of either BBH component for $f_{\rm IMBH}=0.001$ (top panel) and $f_{\rm IMBH}=0.01$ (bottom panel). Our results from Table~\ref{tab:Nev} are marked with black stars which are plotted against the mean value of BH mass for $\overline{Z}=0.05Z_\odot$, the average metallicity in the Harris catalog. In the scenario where all BBHs are about $5\,M_\odot+5\,M_\odot$, there are no detections at all, while in the case of BBH components more massive than $\gtrsim 10\,M_\odot$, we expect $\gtrsim 1$ events.

Finally, note that a mass-to-light ratio different from the value of $1.5\,M_\odot/L_\odot$ used in our calculation (to convert luminosities to masses in the Harris catalog) would have little impact on our results. The values of the mass-to-light ratio inferred for our Galaxy are approximately within the range $1$--$5\,M_\odot/L_\odot$, and extreme values close to the upper limit are rare~\cite{2000ApJ...539..618M,2007A&A...469..147R,2017ApJ...836...67H,2017MNRAS.464.2174B}. To estimate the effect on the number of Doppler shift events, we ran the original simulation for the case of few BBHs assuming a doubled ratio of~$3\,M_\odot/L_\odot$. We find that there are $\approx 1.5$ more events, with the increase being more notable at low $f_{\rm IMBH}$ and for lower SNR thresholds.

\begin{figure*}
  \centering
  \includegraphics[width=0.49\textwidth]{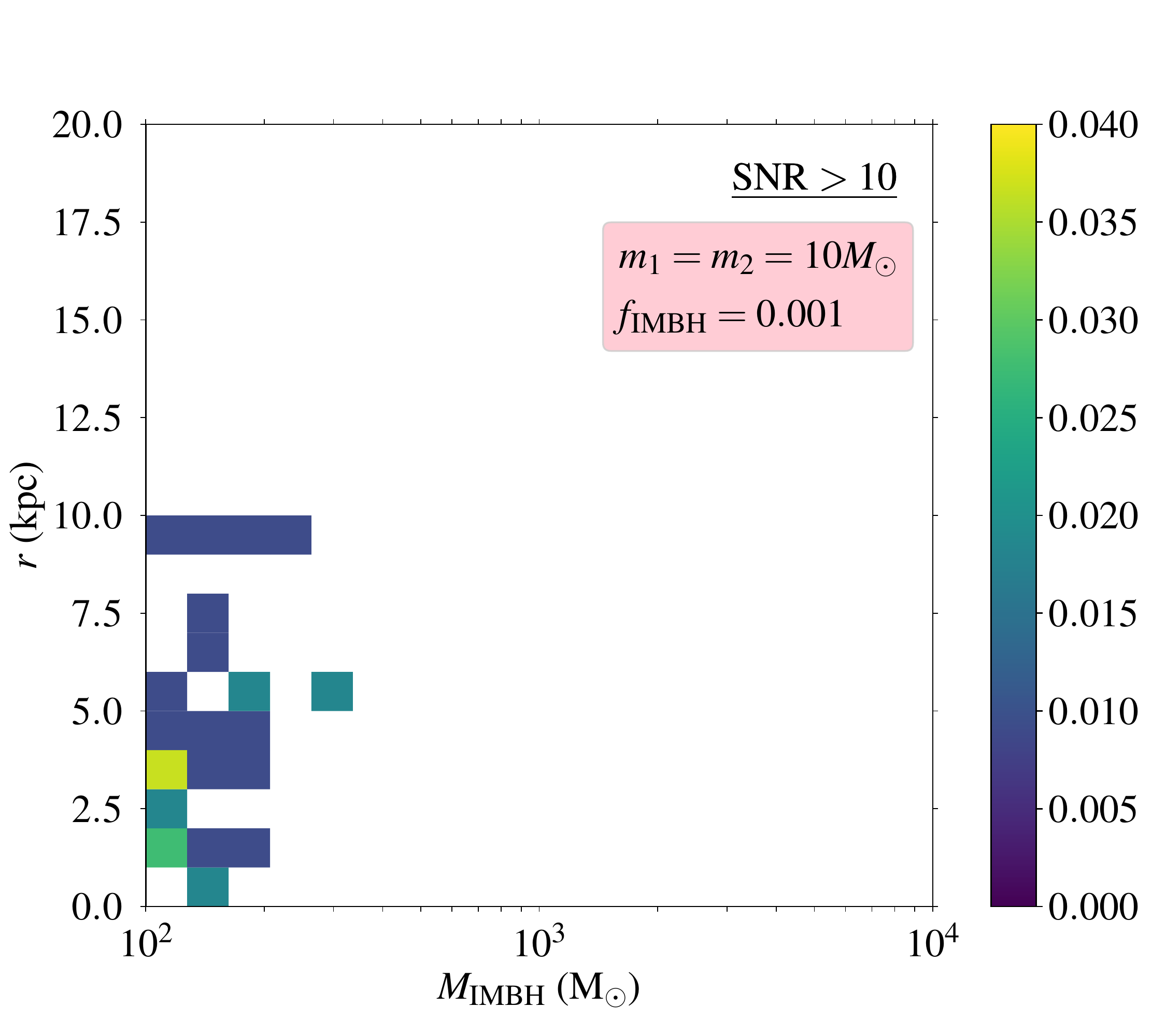}
  \includegraphics[width=0.49\textwidth]{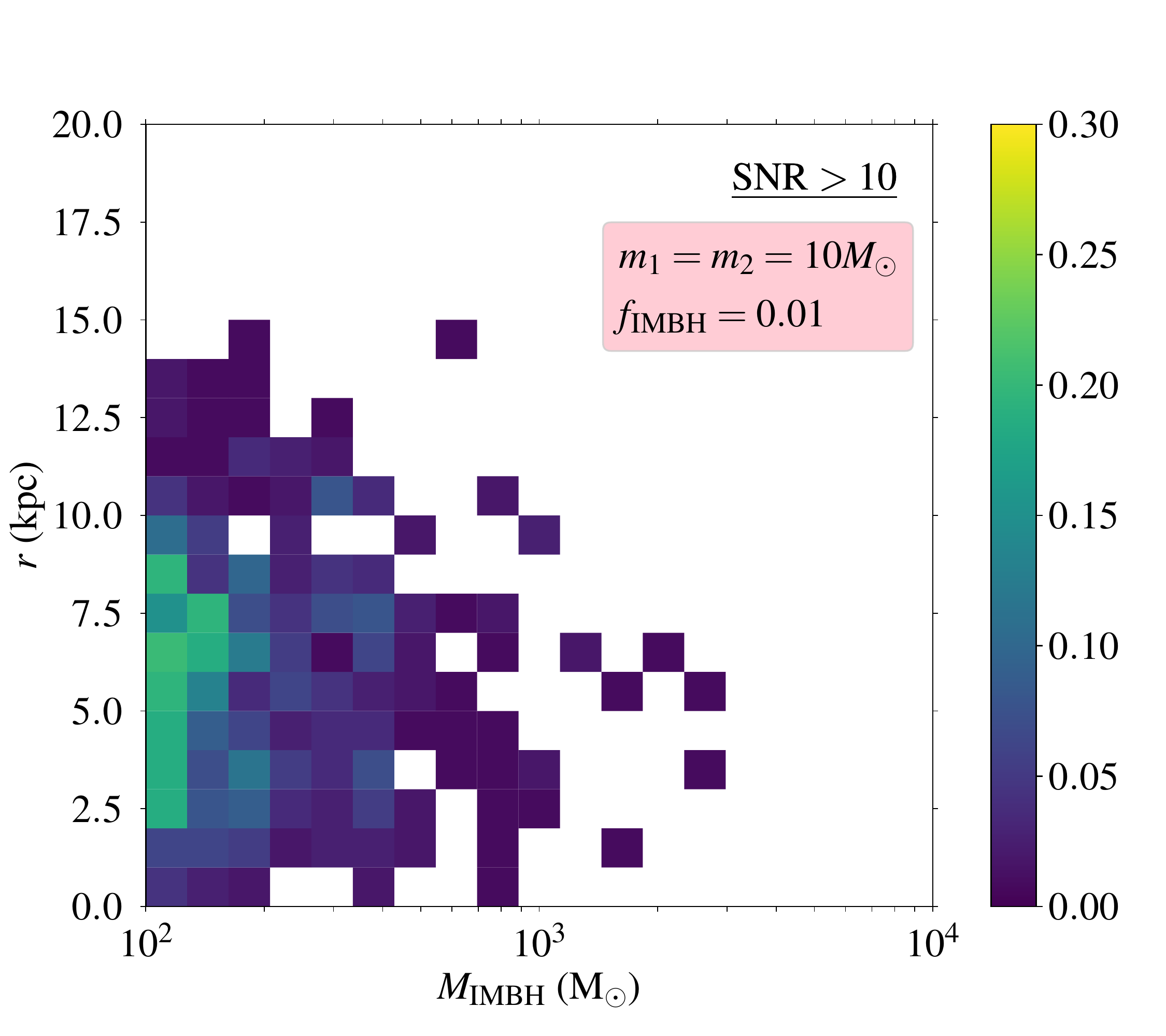}
  \includegraphics[width=0.49\textwidth]{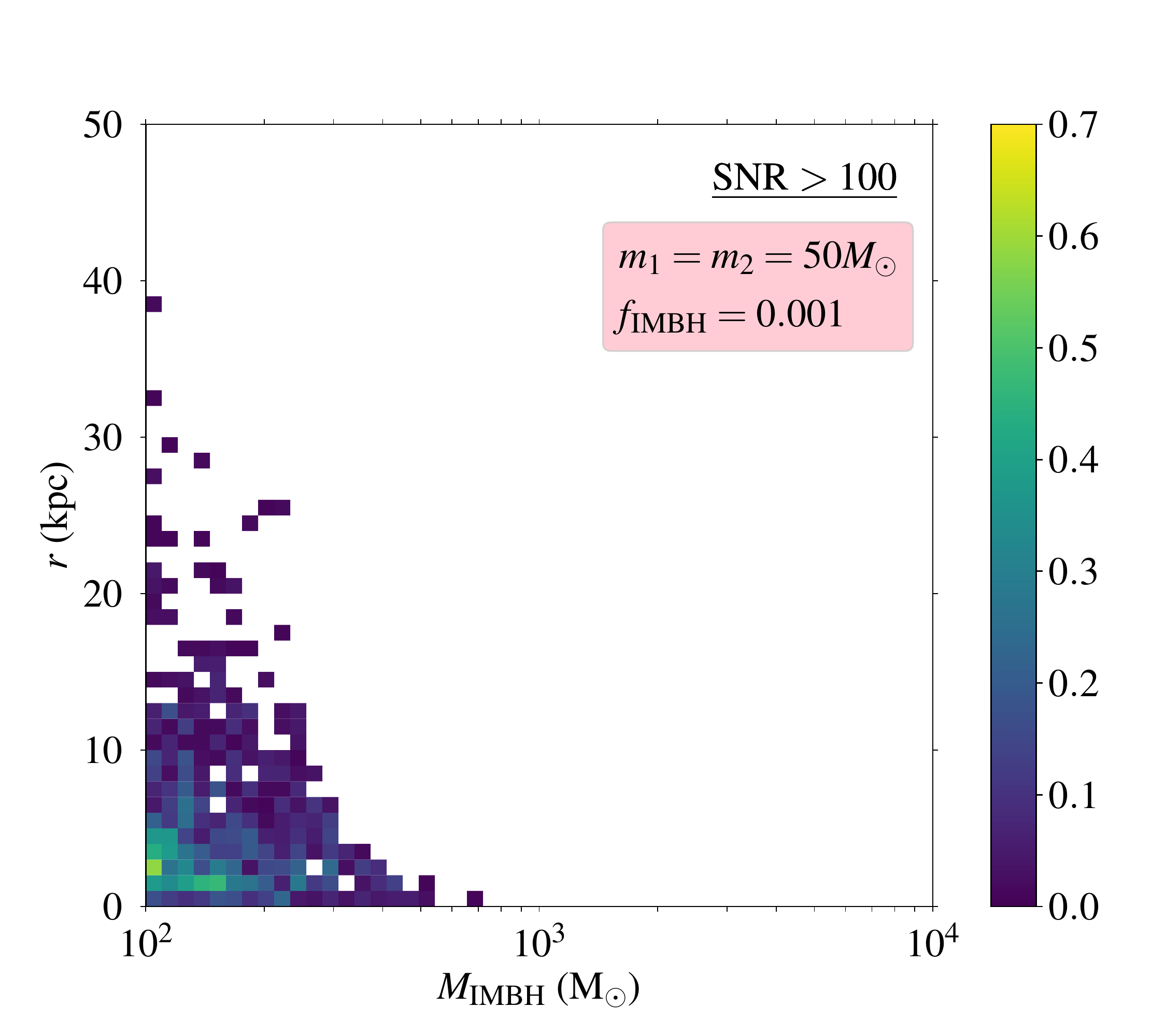}
  \includegraphics[width=0.49\textwidth]{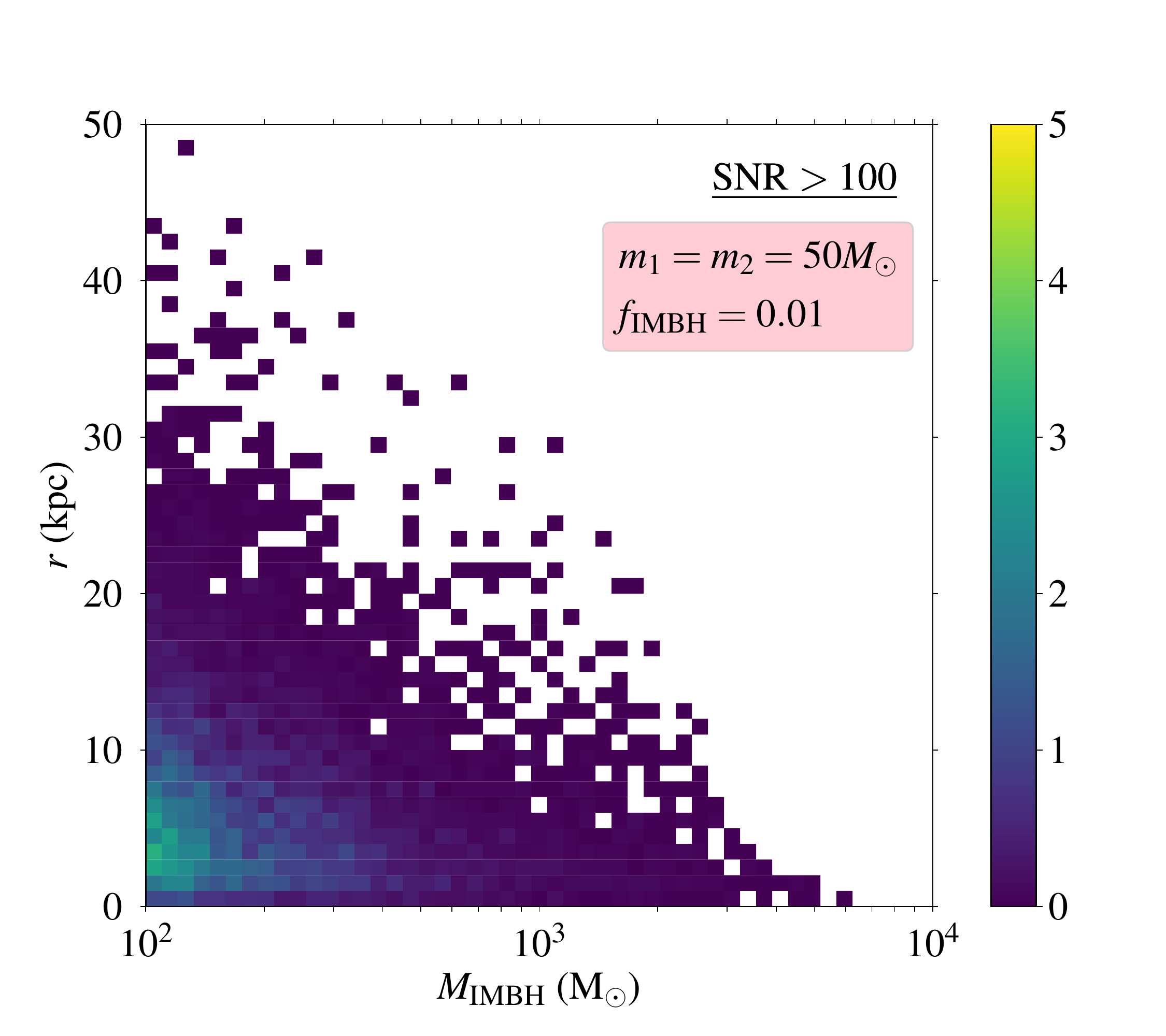}
  \caption{
  Distribution of the number of Doppler shift events around individual wandering IMBHs as a function of the mass of the wandering IMBH and the distance from the Galactic center for $f_{\rm IMBH}=0.001$ (left) and $f_{\rm IMBH}=0.01$ (right). The number of events is calculated over $112$~simulation runs. The top panels refer to light BBHs ($m_1=m_2=10\,M_\odot$) and a threshold $\mbox{SNR}>10$, because higher SNR~thresholds yield almost no events (see Table~\ref{tab:Nw}). The bottom panels refer to heavy BBHs ($m_1=m_2=50\,M_\odot$) and $\mbox{SNR}>100$, because the distributions for lower SNR~thresholds look qualitatively similar. Note also the different ranges of the color-coded scales.
    \label{isolated_hist_15}
}%
\end{figure*}

\subsection{Events from wandering intermediate-mass black holes\label{subsec:wandering_results}}

In Fig.~\ref{wandering_total} we plot estimates for the total number of Doppler shift events~$N_{\rm w}$ for wandering IMBHs as a function of $f_{\rm IMBH}$. Note that in this case $f_{\rm IMBH}$ is the ratio of the IMBH mass to the initial mass of its destroyed parent cluster. We assume that exactly one BBH orbits around each wandering IMBH at any given time (see Sec.~\ref{evolve_GC}).  The two panels refer to the cases of light ($m_1=m_2=10\,M_\odot$, left) and heavy ($m_1=m_2=50\,M_\odot$, right) BBHs. As before, we report $N_{\rm det}$ for three selected SNR thresholds: $10$, $30$, and~$100$.

The trend is different from the case of the Milky Way's GCs: now $N_{\rm w}$ increases for $10^{-3}<f_{\rm IMBH}<10^{-2}$, and it decreases for $f_{\rm IMBH} > 10^{-2}$. This can be explained as follows. The majority of primordial GCs have initial masses $\sim 10^4$\,M$_\odot$ as a result of the negative slope of the GC initial mass function: see Eq.~(\ref{IMF_GC}). Since they are not massive and dense enough to survive until now, most of these clusters are disrupted by the Galactic tidal field. However, no IMBHs will be left behind for $f_{\rm IMBH}<10^{-2}$, since this would correspond to IMBH masses $< 100$\,M$_\odot$. Therefore the number of events increases for $10^{-3}<f_{\rm IMBH}<10^{-2}$, since more and more disrupted GCs leave behind IMBHs with masses $>100$\,M$_\odot$. For example, the disruption of GCs produces $\sim 100$ wandering IMBHs heavier than $100\,M_\odot$ in the Galaxy for $f_{\rm IMBH}\approx 10^{-3}$, while almost every disrupted GC will leave behind an IMBH for \mbox{$f_{\rm IMBH}\approx 10^{-2}$}. Beyond the peak at \mbox{$f_{\rm IMBH}>10^{-2}$} the number of Doppler shift events decreases, since more massive IMBHs have both larger tidal radii and larger influence radii (see the discussion in Sec.~\ref{subsec:MW_results}).

In Table~\ref{tab:Nw} we report the total number of Doppler shift events~$N_{\rm w}$ around wandering IMBHs in the Galaxy for the three values of SNR (10, 30, and 100) and two values of~$f_{\rm IMBH}=0.001, 0.01$. As explained above, the number of events is largest for $f_{\rm IMBH}=0.01$. It is also larger (as expected) for the case of BBHs with heavy components $m_1=m_2=50\,M_\odot$, which generate a stronger GW signal. In this case, we find that $\sim 10$ and $\sim 100$ Doppler shift events could be detectable for $f_{\rm IMBH}=0.001$ and $0.01$, respectively.

\begin{figure*}
  \centering
  \includegraphics[width=0.8\textwidth]{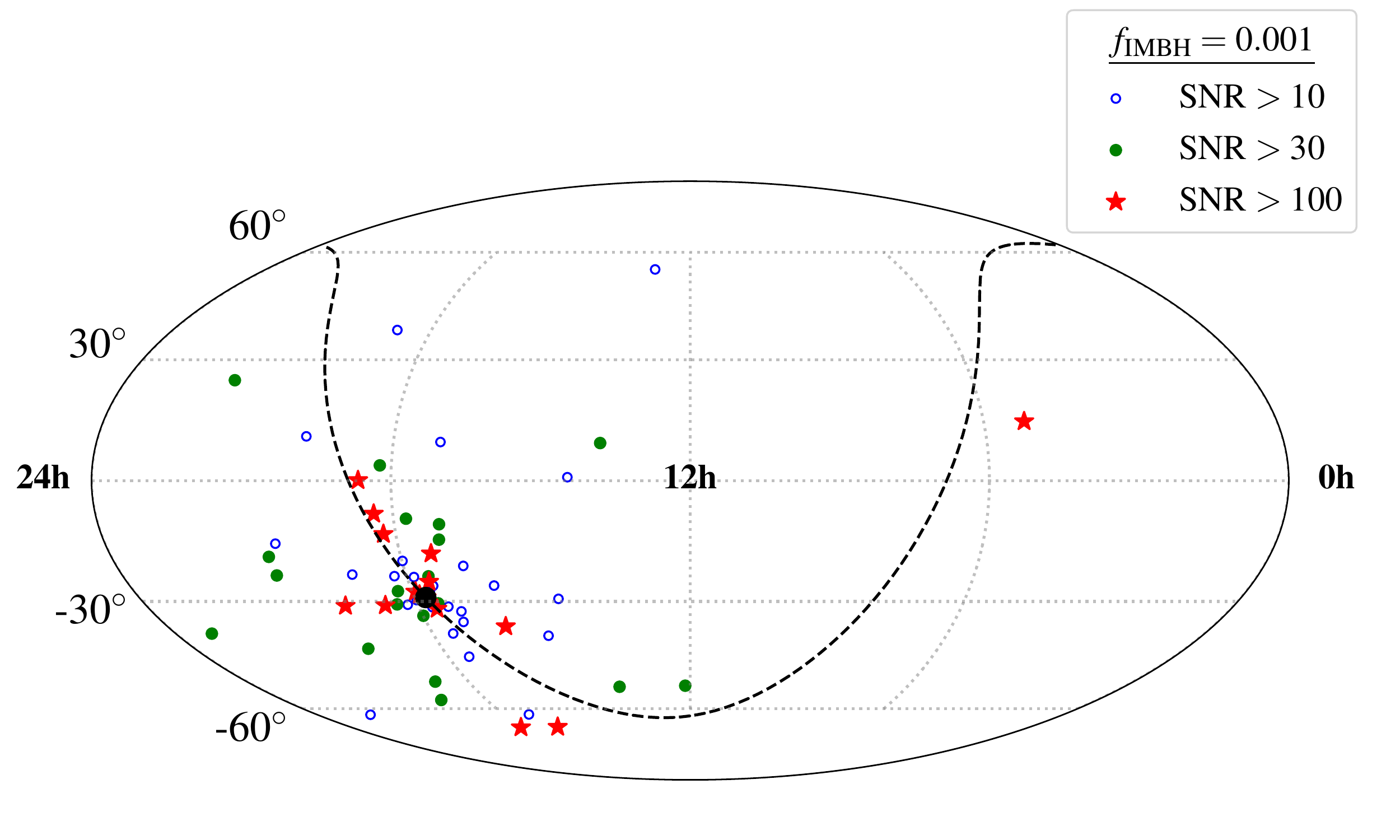}
  \caption{Sample distribution of Doppler shift events around wandering IMBHs on the sky in equatorial coordinates. Here we consider BBHs with component masses $m_1=m_2=50\,M_\odot$, $f_{\rm IMBH}=0.001$, and three values of the SNR threshold (blue empty dots: $\mbox{SNR}>10$; green filled dots: $\mbox{SNR}>30$; red stars: $\mbox{SNR}>100$). The distribution of events on the sky for other values of the component masses and $f_{\rm IMBH}$  is qualitatively the same. The Galactic plane is shown with a dashed line, while the filled black circle marks the Galactic center.\label{skymap}}%
\end{figure*}

Fig.~\ref{isolated_hist_15} shows the distribution of the number of Doppler shift events as a function of the masses of wandering IMBHs and their distances from the Galactic center. The top and bottom rows correspond to the light and heavy BBH cases, while the left and right columns refer to $f_{\rm IMBH}=0.001$ and $f_{\rm IMBH}=0.01$, respectively. In the case of light BBHs we only show the distribution for $\mbox{SNR}>10$, because the number of events practically vanishes at the higher SNR thresholds. In the case of heavy BBHs we only show the distribution for $\mbox{SNR}>100$, because the distributions for different SNR thresholds are qualitatively similar. The heat maps are weighted by the probability for an event to occur. More precisely, we run our simulation $112$~times for each wandering IMBH and each combination of parameters (SNR, $f_{\rm IMBH}$, and the component masses). Then we divide the number of simulation runs that resulted in an event by~$112$ to obtain the probability for an event with the given parameters to occur. Thus, all numbers smaller than unity in the heat maps can be interpreted as the probability for an event to occur. The number of simulation runs also gives a lower bound of $\approx 1\%$ on this probability, since the heat maps are based on 112~runs. Typical probabilities are in the range $1\%$--$10\%$.

\begin{table}
\caption{Approximate numbers of detectable events $N_{\rm w}$ around wandering IMBHs within a distance~$r$ from the Galactic center for selected values of $f_{\rm IMBH}$, as defined in Eq.~(\ref{eq:fimbh}). A fractional number of events should be interpreted as a probability of detection during the LISA mission.\label{tab:Nr}}
\begin{center}
\begin{tabular}{c | c c c }\hline\hline
 \multicolumn{4}{c}{$m_1=m_2=10\,M_\odot$}\\ \hline
 \\[-0.8em]
 $f_{\rm IMBH}$ & $r<5$~kpc & $r<10$~kpc & $r<20$~kpc \\[1ex] \hline
 \\[-0.8em]
 0.001 & 0.2 & 0.3 & 0.3  \\[1ex]   
 0.01 & 2 & 4 & 5     \\[1ex] \hline\hline
 \multicolumn{4}{c}{$m_1=m_2=50\,M_\odot$}\\\hline
 \\[-0.8em]
 $f_{\rm IMBH}$ & $r<5$~kpc & $r<10$~kpc & $r<20$~kpc \\[1ex] \hline
 \\[-0.8em]
 0.001 & 12 & 17 & 19 \\ [1ex]  
 0.01 & 110 & 180 & 210 \\ [1ex]  
\hline
\end{tabular}
\end{center}
\end{table}

The heat maps illustrate three trends.

First, the events clearly cluster at a distance of a few kpc from the Galactic center. The overall clustering of the wandering IMBHs towards the center is not unexpected: the primordial distribution of Galactic GCs follows the Galactic density profile, and GCs are more likely to be disrupted near the center of the Milky Way, leaving behind wandering IMBHs. The events cluster at some finite distance from the center, approximately corresponding to the distance at which the Milky Way velocity dispersion has a maximum. This is because larger dispersion results in smaller influence radii for the wandering IMBHs [cf. Eq.~(\ref{dispersion})], which in turn leads to tighter orbits of BBHs around the IMBHs and to shorter orbital periods, comparable to the observation time $T_{\rm obs}=4$~years. This makes it easier to measure Doppler shift modulations.
In Table~\ref{tab:Nr} we provide specific approximate numbers of events within distance $r=5$~kpc, $10$~kpc, and $20$~kpc from the Galactic center. In addition, Fig.~\ref{skymap} illustrates the distribution of the events on the sky in equatorial coordinates. The dashed line corresponds to the Galactic plane, and the filled black circle marks the Galactic center. Note that in our simulations the primordial GCs which host the IMBHs before being disrupted are isotropically distributed in the Galaxy. The map shows a random sample of events for the case of heavy BBHs, $f_{\rm IMBH}=0.001$, and at the three SNR thresholds.

\begin{table}
\caption{Number of detectable events $N_{\rm w}$ around wandering IMBHs for selected values of $f_{\rm IMBH}$, as defined in Eq.~(\ref{eq:fimbh}), and of the SNR threshold.\label{tab:Nw}}
\begin{center}
\begin{tabular}{c | c c c }\hline\hline
 \multicolumn{4}{c}{$m_1=m_2=10\,M_\odot$}\\ \hline
 \\[-0.8em]
 $f_{\rm IMBH}$ & SNR $>$ 10 & SNR $>$ 30  &  SNR $>$ 100 \\[1ex] \hline
 \\[-0.8em]
 0.001 & $0^{+2}_{-0}$ & $0$ & $0$  \\[1ex]   
 0.01 & $5^{+5}_{-3}$& $0^{+1}_{-0}$& $0$     \\[1ex] \hline\hline
 \multicolumn{4}{c}{$m_1=m_2=50\,M_\odot$}\\\hline
 \\[-0.8em]
 $f_{\rm IMBH}$ & SNR $>$ 10 & SNR $>$ 30  &  SNR $>$ 100 \\[1ex] \hline
 \\[-0.8em]
0.001 & $26^{+9}_{-9}$& $25^{+7}_{-8}$& $19^{+8}_{-7}$ \\ [1ex]  
0.01 & $297^{+32}_{-30}$& $276^{+32}_{-29}$& $214^{+27}_{-24}$ \\ [1ex]
\hline
\end{tabular}
\end{center}
\end{table}

\begin{table}
\caption{Approximate numbers of detectable events $N_{\rm w}$ around wandering IMBHs in three IMBH mass ranges and for selected values of $f_{\rm IMBH}$, as defined in Eq.~(\ref{eq:fimbh}). A fractional number of events should be interpreted as a probability of detection during the LISA mission.\label{tab:Nm}}
\begin{center}
\begin{tabular}{c | c c c }\hline\hline
 \multicolumn{4}{c}{$m_1=m_2=10\,M_\odot$}\\ \hline
 \\[-0.8em]
 $f_{\rm IMBH}$ & $100$--$300\,M_\odot$ & $300$--$1,000\,M_\odot$ & $1,000$--$3,000\,M_\odot$ \\[1ex] \hline
 \\[-0.8em]
 0.001 & 0.3 & 0. & 0.  \\[1ex]   
 0.01 & 4 & 0.8 & 0.1     \\[1ex] \hline\hline
 \multicolumn{4}{c}{$m_1=m_2=50\,M_\odot$}\\\hline
 \\[-0.8em]
 $f_{\rm IMBH}$ & $100$--$300\,M_\odot$ & $300$--$1,000\,M_\odot$ & $1,000$--$3,000\,M_\odot$ \\[1ex] \hline
 \\[-0.8em]
 0.001 & 18 & 1 & 0 \\ [1ex]  
 0.01 & 170 & 35 & 5 \\ [1ex]    
\hline
\end{tabular}
\end{center}
\end{table}

The second trend visible in Fig.~\ref{isolated_hist_15} is that there are more events around lighter IMBHs. This is purely due the large number of light IMBHs left behind by relatively light GCs, which are more prone to destruction and also more abundant, as a consequence of the bottom-heavy initial mass function of Eq.~(\ref{IMF_GC}).

Finally, Fig.~\ref{isolated_hist_15} demonstrates that events around heavier IMBHs are more likely to occur closer to the Galactic center. This is because heavier GCs are more likely to be destroyed and leave behind heavier IMBHs when they are close to the center. To be more quantitative, in Table~\ref{tab:Nm} we report estimates for the number of events around IMBHs in specific mass ranges: $100$--$300\,M_\odot$, $300$--$1,000\,M_\odot$, and $1,000$--$3,000\,M_\odot$.

We conclude with a few remarks about our assumptions in the wandering IMBH case. When considering the evolution of GCs in the Milky Way, we use an approximate formula for the evaporation timescale, Eq.~(\ref{eq:ejections}). The evaporation timescale is typically proportional to the half-mass relaxation time of a GC, that is $\propto M_{\rm GC}^{1/2}r_{\rm h}^{3/2}$ (see see Eq.~(7.108) in Ref.~\cite{2008gady.book.....B}). Taking into account the prescription for the half-mass density, Eq.~(\ref{rhohalf}), our equation overestimates the timescale for light GCs~$\sim 10^4\,M_\odot$. Moreover, this difference in timescales is only relevant when~$t_{\rm ej}<t_{\rm tid}$ (see Eq.~(\ref{eq:tidal})) which happens at $r\gtrsim 1$~kpc. Thus, for the light GCs located further than this distance one can expect faster evaporation due to ejections and, hence, a larger abundance of light wandering IMBHs as long as $f_{\rm IMBH}M_{\rm GC}>100\,M_\odot$. This would increase the number of detectable Doppler shift events at~$f_{\rm IMBH}\sim 0.01$, while keeping that number approximately constant at~$f_{\rm IMBH}\sim 0.001$\,.

We have assumed that every disrupted GC leaves behind an IMBH. However, the timescale for IMBH formation may be longer than that for GC disruption. It appears that at least lighter IMBHs form fast enough~\cite{2021ApJ...908L..29G,2021MNRAS.501.5257R,2021MNRAS.507.5132D}, although it might take up to several Gyrs to form a $\sim 10^4\,M_\odot$ IMBH (see Fig.~15 in Ref.~\cite{2015MNRAS.454.3150G}). Another possibility is that some GCs do not form an IMBH at all. The fraction of IMBH-forming GCs does not appear to exceed~$20\%$ in simulations~\cite{2015MNRAS.454.3150G,2021MNRAS.501.5257R} (a few percent more may come from young stellar clusters~\cite{2019MNRAS.487.2947D,2021MNRAS.507.5132D}). If only a fraction of GCs leaves IMBHs behind, the results presented in this subsection should be scaled down proportionately. Whenever this scaled number happens to be less than unity, it can be interpreted as the detection probability, rather than the number of events. We leave a detailed exploration of these effects for future work.

\section{Discussion\label{discuss}}

Finding IMBHs and characterizing their properties is of crucial importance since they play an important role in a wide range of phenomena, including seeds of massive black holes, accretion, tidal disruption events, and GWs. Despite major observational efforts IMBHs still remain elusive, and new methods are needed to detect them.

We have investigated the possibility that LISA may find IMBHs lurking in Galactic GCs by measuring the radial velocity modulations in the GW signal of BBHs orbiting around them. We have found that the number of Doppler shift events decreases for larger IMBH masses, because more massive IMBHs have a larger influence radius and larger tidal radii. Since tidal stability requires BBHs to orbit far away from the IMBH, the Doppler modulations in their gravitational waveforms are harder to detect. We have also estimated that $\sim 30$ Galactic GCs could produce at least one Doppler event detectable by LISA if an IMBH lurks in their center. Among these candidate Galactic GCs, M15, M62, NGC6388, M54, and 47~Tucanae may harbor an IMBH. The best candidate in our analysis is $\omega$~Centauri, if indeed it hosts an IMBH with mass $\lesssim 10^4\,M_\odot$, as suggested by dynamical measurements~\cite{2020ARA&A..58..257G}. 

We have also considered the possibility of hunting for wandering Galactic IMBHs left behind by the disruption of the parent cluster.
Assuming that each of these wandering IMBHs has at least one BBH orbiting around it, LISA could detect tens of Doppler events with SNR\,$>10$.

The numbers just cited represent optimistic scenarios and should serve as reference points. For the Milky Way GCs, we have run additional simulations to estimate the role played by the distributions of BBH semimajor axes and their distances from the IMBH, the BBH mass spectrum, as well as mass-to-light ratios for GCs. In doing so, we assumed more realistically~$\mathcal{O}(10)$ BBHs per cluster (``few'' BBHs). The results, part of which are presented in Figs.~\ref{fig:fraction_LISA},~\ref{fig:slopes} and~\ref{fig:masses}, show that, in more pessimistic scenarios, one can still expect~$\approx 1$--$5$ detections from GCs. This is consistent with scaling down the ``many'' BBHs case number by~$10$. Another factor that could affect the number of detections is the probability for a GC to form an IMBH. According to the literature (e.g.~\cite{2015MNRAS.454.3150G,2021MNRAS.501.5257R,2019MNRAS.487.2947D,2021MNRAS.507.5132D}) only~$\sim 20\%$ of GCs may form IMBHs. This would scale down our estimates by~$5$. Below we also discuss eccentricity, which makes the GW signal louder, and thus may increase the estimate. Whenever new insights are learned about any of these factors, our results can be scaled up or down to obtain less uncertain estimates. All in all, under moderately pessimistic assumptions one should expect $\sim 1$ event.

The same scaling considerations apply to the wandering IMBH simulations. These simulations assumed exactly one BBH per wandering IMBH, and bracketed the uncertainty about the BBH mass as presented in Fig.~\ref{wandering_total}. Major factors which can modify the numbers for this case are the probability for a GC to get disrupted, and for a disrupted GC to form and leave behind an IMBH. As we have discussed (see the last two paragraphs of Sec.~\ref{subsec:wandering_results}), by making an assumption about the evaporation time scale we may be underestimating the number of light wandering IMBHs for~$f_{\rm IMBH}=0.01$, thus underestimating the number of detections. Besides, only~$\sim 20\%$ of GCs may form IMBHs, which would scale down the numbers by a factor~$5$. With this scaling, it is realistic to expect $\sim 1$~detection from the wandering IMBH population. Even one such event would be of major importance, allowing us to spot an IMBH that would hardly be observable otherwise.

\begin{figure}
  \centering
  \includegraphics[width=\columnwidth]{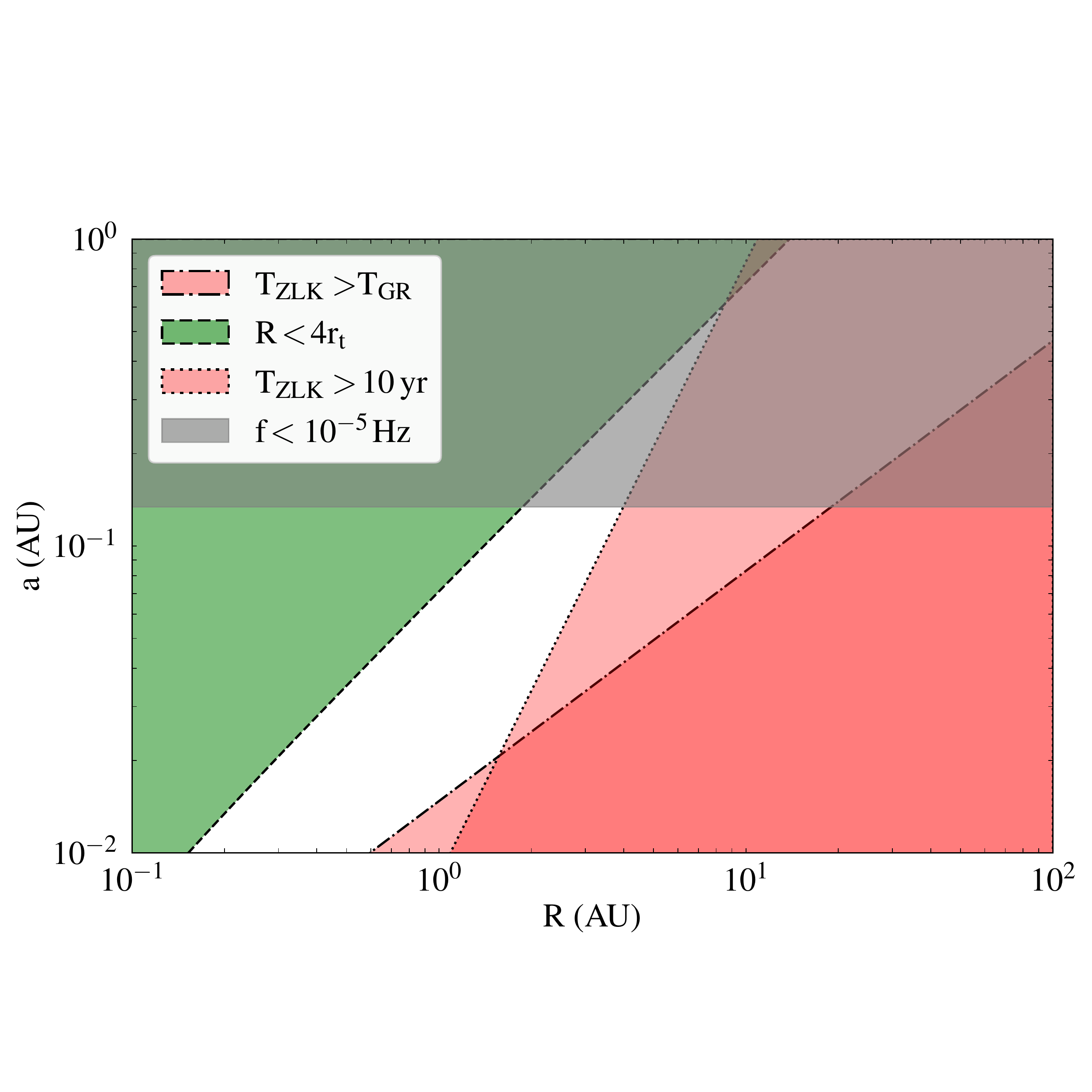}
  \caption{Constraints on the orbital radius~$a$ of a BBH with $m_1=m_2=30\,M_\odot$ and on the radius~$R$ of the BBH's orbit around an IMBH with $M_{\rm IMBH}=10^3\,M_\odot$. Red regions show the portion of the parameter space where the ZKL effect does not contribute to the binary's evolution during the nominal LISA observation time of $T_{\rm obs}=4$~yr. This happens because either the ZKL timescale is longer than the observation time, $T_{\rm ZKL}\gtrsim T_{\rm obs}$ (dotted line), or because the ZKL oscillations are quenched by the GR periapsis precession, $T_{\rm ZKL}\gtrsim T_{\rm GR}$ (dash-dotted line). The green region is excluded due to the tidal stability criterion, Eq.~(\ref{stability_crit}). In the gray shaded region, GW frequencies are outside of the LISA band.
    \label{fig:constraints}}%
\end{figure}
Note also that observational and theoretical evidence makes some of the GCs more promising targets than others for our present purpose. The modeling of stellar orbits close to the cluster center suggests that some Galactic GCs could indeed host an IMBH (see Table~3 in Ref.~\cite{2020ARA&A..58..257G} for a summary). Moreover, recent numerical models have shown that a higher degree of mass segregation in a cluster is linked to a less abundant population of SBHs~\cite{2020ApJ...898..162W}. This implies that GCs observationally classified as core-collapsed are likely to contain either fewer BBHs than those that have not undergone the core collapse yet~\cite{2020ApJS..247...48K,2020ApJ...898..162W}, or no BBHs at all.

Let us conclude by mentioning other uncertain assumptions that affect our astrophysical models. One of the most notable uncertainties is 
the minimum semimajor axis for BBHs ($a_{\rm min}=0.01$~AU in this work), which plays an important role in our estimates. 

The waveform model of Eq.~(\ref{wf}) is based on two main assumptions: (i) the frequency of the GW signal changes slowly compared to the orbital period~$P$, and (ii) the initial orbital phase is irrelevant. The first assumption is well justified, because the relative change in the GW frequency of detectable events is much smaller than that due to the Doppler shift, even for the BBHs with the smallest separation~$a_{\rm min}$. Regarding the second assumption, about half of the BBHs which contribute to the final count satisfy it within an order of magnitude. In fact, one can weaken the assumption to~$T_{\rm}\gtrsim P/4$, because only $1/4$ of a period is required for the Doppler shift modulation to develop. This weaker condition is generally satisfied by $\approx 80\%$ of the BBHs. Note also that our choice of phase is conservative: the observation is assumed to begin when the radial velocity is maximal (and its rate of change is minimal). We leave a detailed study of the effect of the phase to future work. Such a study should inevitably deal with eccentric orbits, where the initial phase plays a more important role.

We have assumed for simplicity that BBH have circular orbits, but nonzero eccentricities could increase the GW amplitude of the signals and yield more optimistic rate estimates~\cite{1963PhRv..131..435P}. In particular, BBHs sufficiently close to a central IMBH ($R\lesssim 1$~AU) may experience von Zeipel--Kozai--Lidov (ZKL) eccentricity oscillations (see Appendix~\ref{subsec:ZKL} for timescales and references). Figure~\ref{fig:constraints} shows the values of~$R$ and the BBH's orbital radius~$a$ allowing for the ZKL effect for an equal-mass BBH with total mass $m=60\,M_\odot$ and an IMBH of $10^3\,M_\odot$. Other values of~$m$ and~$M_{\rm IMBH}$ considered in this work give approximately the same estimate~$R\lesssim 1$~AU. If there are such relatively tight IMBH--BBH triples, our estimates for the number of Doppler shift events may increase, as the eccentricity oscillations enhance the SNR of perturbed BBHs~\cite{2019ApJ...875L..31H,2020ApJ...901..125D}.

Finally, we have considered only BBHs, neglecting black hole-neutron stars binaries and binary neutron stars as possible sources of Doppler shift events. This assumption is quite well justified: these two populations of compact object binaries would likely contribute much less than BBHs, because most of the neutron stars should be ejected as a result of natal kicks and do not efficiently segregate in the cluster center, close to the central IMBHs \cite{2018MNRAS.480.4955F,2020ApJ...888L..10Y,2020ApJ...901L..16F}. These uncertainties should be better quantified through further work. Our initial estimates suggest that LISA may detect tens of Doppler shift events, thus mapping the elusive population of IMBHs in the Milky Way.

\begin{acknowledgments}
V.S., T.H. K.W.K.W. and E.B. were supported by NSF Grants No. PHY-1912550 and AST-2006538, NASA ATP Grants No. 17-ATP17-0225 and 19-ATP19-0051, NSF-XSEDE Grant No. PHY-090003, and NSF Grant PHY-20043. G.F., V.S., T.H., and E.B. acknowledge support from NASA Grant 80NSSC21K1722.
G.F. acknowledges support from NSF Grant~AST-1716762 at Northwestern University. K.W.K.W. is supported by the Simons Foundation. The authors are very grateful to the anonymous referee for a detailed and constructive report which helped us significantly improve the paper. We would also like to thank Bence Kocsis and Kyle Kremer for helpful comments on our assumptions. G.F. is grateful to Sambaran Banerjee for insightful discussions on stellar evolution and for updating \textsc{sse}, and to Oleg Gnedin for useful discussions on star cluster evolution.
This research project was conducted using computational resources at the Maryland Advanced Research Computing Center (MARCC). The authors acknowledge the Texas Advanced Computing Center (TACC) at The University of Texas at Austin for providing HPC resources that have contributed to the research results reported within this paper \cite{10.1145/3311790.3396656}
(URL: \url{http://www.tacc.utexas.edu}).
This research made use of the following software: IPython~\cite{2007CSE.....9c..21P}, SciPy~\cite{2020NatMe..17..261V},  Matplotlib~\cite{2007CSE.....9...90H}, NumPy~\cite{2011CSE....13b..22V}, SymPy~\cite{Meurer:2017yhf}, \texttt{mpmath}~\cite{mpmath}, \texttt{scikit-learn}~\cite{2012arXiv1201.0490P}, \texttt{filltex}~\cite{2017JOSS....2..222G}.
\end{acknowledgments}

\appendix
\section{Timescales}

Let $\mathbf{L}$ be the intrinsic orbital angular momentum of a BBH and $\mathbf{S}_i = \chi_i m_i^2 \mathbf{n}_i$, $i=1,2$, the spins of the BBH's black holes, where $\chi_i$ are the unitless rotation parameters of the black holes, $0\leq\chi\leq 1$, and $\mathbf{n}_i$ are unit vectors indicating direction. 

\subsection{Effects of the spins of the BBH's black holes\label{subsec:BBHspin}}

Interaction among $\mathbf{S}_1$, $\mathbf{S}_2$ and $\mathbf{L}$ leads to precession of these vectors. For simplicity, consider an equal-mass binary $m_1=m_2=m$ and $\chi_1\sim\chi_2=\chi$. Then timescale~$T_{\rm S}$ for precession of the spins and timescale~$T_{\rm L}^{(1)}$ for the precession of~$\mathbf{L}$ as a result of the spin-orbit interaction are as follows (see, for example,~\cite{1994PhRvD..49.6274A,1995PhRvD..52..821K,2003PhRvD..67j4025B}):
\beqa
T_{\rm S} &\sim& \frac{a^{5/2}}{\eta (2m)^{3/2}} \nonumber \\
&\sim& 10\,\mbox{yr}\,\left(\frac{a}{0.01\,\mbox{AU}}\right)^{5/2}\left(\frac{m}{20\,M_\odot}\right)^{-3/2}\,, \\
T_{\rm L}^{(1)} &\gtrsim& \frac{ a^3}{\chi m^2} \nonumber \\
&\sim& 1\,\mbox{kyr}\,\left(\frac{a}{0.01\,\mbox{AU}}\right)^3 \left(\frac{m}{10\,M_\odot}\right)^{-2}\chi^{-1}\,.
\eeqa

\subsection{Effects of IMBH spin\label{subsec:IMBHspin}}

If an IMBH orbited by the BBH also has a spin $\mathbf{S}_{\rm IMBH}=\chi_{\rm IMBH}M_{\rm IMBH}^{2}\mathbf{n}$, it can interact with both~$\mathbf{L}$ (for the binaries under consideration $|\mathbf{L}|>>|\mathbf{S}_i|$) and the orbital angular momentum of the BBH around the IMBH. This contribution to the precession of~$\mathbf{L}$ happens on a timescale
\beqa
T_{\rm L}^{(2)} &\sim& \frac{R^{5/2}}{M_{\rm IMBH}^{3/2} } \sim 100\,\mbox{kyr}\,\left(\frac{R}{10\,\mbox{AU}}\right)^{5/2}\left(\frac{M_{\rm IMBH}}{10^3\,M_\odot}\right)^{-3/2}. \nonumber \\
&{}&
\eeqa
The precession timescale of the IMBH's spin is much longer,
\beqa
T_{\rm IMBH}&\sim& \frac{T_{\rm L}^{(2)}}{\eta} \sim 10\,\mbox{Myr}\,\left(\frac{R}{10\,\mbox{AU}}\right)^{5/2} \nonumber \\
&\times& \left(\frac{M_{\rm IMBH}}{10^3\,M_\odot}\right)^{-1/2}\left(\frac{m_1+m_2}{20\,M_\odot}\right)^{-1}\,.
\eeqa

Finally, $\mathbf{S}_{\rm IMBH}$ interacts with the angular momentum of the BBH's orbit around the IMBH with timescale
\beqa
T_{\rm LT} &\sim& \frac{R^3}{\chi M_{\rm IMBH}^2} \nonumber \\
&\sim & 100\,\mbox{Myr}\,\left(\frac{R}{10\,\mbox{AU}}\right)^3 \left(\frac{M_{\rm IMBH}}{10^3\,M_\odot}\right)^{-2}\chi^{-1}\,.
\eeqa

\subsection{Von Zeipel--Kozai--Lidov (ZKL) oscillations\label{subsec:ZKL}}

The ZKL timescale is~\cite{1910AN....183..345V,1962P&SS....9..719L,1962AJ.....67..591K,2019MEEP....7....1I}
\beqa
T_{\rm ZKL} &=& \frac{P^2}{P_{12}} = 10^4\,\mbox{yr}\,\left(\frac{R}{10\,\mbox{AU}}\right)^3\left(\frac{M_{\rm IMBH}}{10^3\,M_\odot}\right)^{-1} \nonumber \\
&\times& \left(\frac{a}{0.01\,\mbox{AU}}\right)^{-3/2}\left(\frac{m_1+m_2}{100\,M_\odot}\right)^{1/2}\,,
\eeqa
where $P_{12}$ refers to the proper orbital period of a BBH, and~$P$ to the orbital period around the IMBH.

Although the strong cubic dependence of~$T_{\rm ZKL}$ on~$R$ may bring this timescale down to $\sim 10$~yr for $R\sim 1$~AU, the ZKL contribution is only important if its timescale is below the period of the GR periapsis precession, which is $\sim a^{5/2}/(2m)^{3/2}\sim T_{\rm S}$~\cite{1997Natur.386..254H}. A more precise criterion for the ZKL effect to take over the GR precession reads~\cite{2002ApJ...578..775B}
\beqa
R &\lesssim& 1\,\mbox{AU}\,\left(\frac{a}{0.01\,\mbox{AU}}\right)^{4/3} \nonumber \\
&\times& \left(\frac{M_{\rm IMBH}}{10^3\,M_\odot}\right)^{1/3}\left(\frac{m_1+m_2}{20\,M_\odot}\right)^{-2/3}\,.
\eeqa

\subsection{Evaporation\label{subsec:evaporation}}

The interaction of a BBH with surrounding stars may lead to its disruption if the semimajor axis is on the order of the hard-binary limit $a_{\rm h}$, such that the BBH's binding energy is comparable to the kinetic energy of the stars: $Gm^2/a_{\rm h} \simeq m_\star\sigma_\star^2$, i.e.
\beqa
a_{\rm h} &\simeq& 100\,\mbox{AU}\,\left(\frac{m}{10\,M_\odot}\right)^2\left(\frac{m_\star}{M_\odot}\right)^{-1}\left(\frac{\sigma_\star}{10\,\mbox{km\;s}^{-1}}\right)^{-2}\,. \nonumber \\
&{}&
\eeqa
Thus, the BBHs we consider are obviously hard. This is also confirmed by their evaporation timescale~\cite{1987degc.book.....S,2008gady.book.....B}, which is comparable to or exceeds the Hubble time:
\beqa
T_{\rm ev} &=& \frac{\sqrt{3}\sigma_\star}{32\sqrt{\pi}G\rho_{\rm h}a\,\ln{\Lambda}}\frac{m_1+m_2}{m_\star}\nonumber \\
&\approx& 6\,\mbox{Gyr}\,\frac{\sigma_\star}{10\,\mbox{km\;s}^{-1}}\left(\frac{\rho}{10^5\,M_\odot\;\mbox{pc}^{-3}}\right)^{-1}\nonumber\\
&\times&\left(\frac{a}{0.1\,\mbox{AU}}\right)^{-1}\frac{m_1+m_2}{20\,M_\odot}\left(\frac{m_\star}{M_\odot}\right)^{-1}\,,
\eeqa
where we assumed $\ln{\Lambda}\approx 15$.

\bibliography{refs}

\begin{thebibliography}{144}%
\makeatletter
\providecommand \@ifxundefined [1]{%
 \@ifx{#1\undefined}
}%
\providecommand \@ifnum [1]{%
 \ifnum #1\expandafter \@firstoftwo
 \else \expandafter \@secondoftwo
 \fi
}%
\providecommand \@ifx [1]{%
 \ifx #1\expandafter \@firstoftwo
 \else \expandafter \@secondoftwo
 \fi
}%
\providecommand \natexlab [1]{#1}%
\providecommand \enquote  [1]{``#1''}%
\providecommand \bibnamefont  [1]{#1}%
\providecommand \bibfnamefont [1]{#1}%
\providecommand \citenamefont [1]{#1}%
\providecommand \href@noop [0]{\@secondoftwo}%
\providecommand \href [0]{\begingroup \@sanitize@url \@href}%
\providecommand \@href[1]{\@@startlink{#1}\@@href}%
\providecommand \@@href[1]{\endgroup#1\@@endlink}%
\providecommand \@sanitize@url [0]{\catcode `\\12\catcode `\$12\catcode
  `\&12\catcode `\#12\catcode `\^12\catcode `\_12\catcode `\%12\relax}%
\providecommand \@@startlink[1]{}%
\providecommand \@@endlink[0]{}%
\providecommand \url  [0]{\begingroup\@sanitize@url \@url }%
\providecommand \@url [1]{\endgroup\@href {#1}{\urlprefix }}%
\providecommand \urlprefix  [0]{URL }%
\providecommand \Eprint [0]{\href }%
\providecommand \doibase [0]{https://doi.org/}%
\providecommand \selectlanguage [0]{\@gobble}%
\providecommand \bibinfo  [0]{\@secondoftwo}%
\providecommand \bibfield  [0]{\@secondoftwo}%
\providecommand \translation [1]{[#1]}%
\providecommand \BibitemOpen [0]{}%
\providecommand \bibitemStop [0]{}%
\providecommand \bibitemNoStop [0]{.\EOS\space}%
\providecommand \EOS [0]{\spacefactor3000\relax}%
\providecommand \BibitemShut  [1]{\csname bibitem#1\endcsname}%
\let\auto@bib@innerbib\@empty
\bibitem [{\citenamefont {{Baldassare}}\ \emph {et~al.}(2018)\citenamefont
  {{Baldassare}}, \citenamefont {{Geha}},\ and\ \citenamefont
  {{Greene}}}]{2018ApJ...868..152B}%
  \BibitemOpen
  \bibfield  {author} {\bibinfo {author} {\bibfnamefont {V.~F.}\ \bibnamefont
  {{Baldassare}}}, \bibinfo {author} {\bibfnamefont {M.}~\bibnamefont
  {{Geha}}},\ and\ \bibinfo {author} {\bibfnamefont {J.}~\bibnamefont
  {{Greene}}},\ }\href {https://doi.org/10.3847/1538-4357/aae6cf} {\bibfield
  {journal} {\bibinfo  {journal} {\apj}\ }\textbf {\bibinfo {volume} {868}},\
  \bibinfo {eid} {152} (\bibinfo {year} {2018})},\ \Eprint
  {https://arxiv.org/abs/1808.09578} {arXiv:1808.09578 [astro-ph.GA]}
  \BibitemShut {NoStop}%
\bibitem [{\citenamefont {{Chilingarian}}\ \emph {et~al.}(2018)\citenamefont
  {{Chilingarian}}, \citenamefont {{Katkov}}, \citenamefont {{Zolotukhin}},
  \citenamefont {{Grishin}}, \citenamefont {{Beletsky}}, \citenamefont
  {{Boutsia}},\ and\ \citenamefont {{Osip}}}]{2018ApJ...863....1C}%
  \BibitemOpen
  \bibfield  {author} {\bibinfo {author} {\bibfnamefont {I.~V.}\ \bibnamefont
  {{Chilingarian}}}, \bibinfo {author} {\bibfnamefont {I.~Y.}\ \bibnamefont
  {{Katkov}}}, \bibinfo {author} {\bibfnamefont {I.~Y.}\ \bibnamefont
  {{Zolotukhin}}}, \bibinfo {author} {\bibfnamefont {K.~A.}\ \bibnamefont
  {{Grishin}}}, \bibinfo {author} {\bibfnamefont {Y.}~\bibnamefont
  {{Beletsky}}}, \bibinfo {author} {\bibfnamefont {K.}~\bibnamefont
  {{Boutsia}}},\ and\ \bibinfo {author} {\bibfnamefont {D.~J.}\ \bibnamefont
  {{Osip}}},\ }\href {https://doi.org/10.3847/1538-4357/aad184} {\bibfield
  {journal} {\bibinfo  {journal} {\apj}\ }\textbf {\bibinfo {volume} {863}},\
  \bibinfo {eid} {1} (\bibinfo {year} {2018})},\ \Eprint
  {https://arxiv.org/abs/1805.01467} {arXiv:1805.01467 [astro-ph.GA]}
  \BibitemShut {NoStop}%
\bibitem [{\citenamefont {{Lin}}\ \emph {et~al.}(2018)\citenamefont {{Lin}},
  \citenamefont {{Strader}}, \citenamefont {{Carrasco}}, \citenamefont
  {{Page}}, \citenamefont {{Romanowsky}}, \citenamefont {{Homan}},
  \citenamefont {{Irwin}}, \citenamefont {{Remillard}}, \citenamefont
  {{Godet}}, \citenamefont {{Webb}}, \citenamefont {{Baumgardt}}, \citenamefont
  {{Wijnands}}, \citenamefont {{Barret}}, \citenamefont {{Duc}}, \citenamefont
  {{Brodie}},\ and\ \citenamefont {{Gwyn}}}]{2018NatAs...2..656L}%
  \BibitemOpen
  \bibfield  {author} {\bibinfo {author} {\bibfnamefont {D.}~\bibnamefont
  {{Lin}}}, \bibinfo {author} {\bibfnamefont {J.}~\bibnamefont {{Strader}}},
  \bibinfo {author} {\bibfnamefont {E.~R.}\ \bibnamefont {{Carrasco}}},
  \bibinfo {author} {\bibfnamefont {D.}~\bibnamefont {{Page}}}, \bibinfo
  {author} {\bibfnamefont {A.~J.}\ \bibnamefont {{Romanowsky}}}, \bibinfo
  {author} {\bibfnamefont {J.}~\bibnamefont {{Homan}}}, \bibinfo {author}
  {\bibfnamefont {J.~A.}\ \bibnamefont {{Irwin}}}, \bibinfo {author}
  {\bibfnamefont {R.~A.}\ \bibnamefont {{Remillard}}}, \bibinfo {author}
  {\bibfnamefont {O.}~\bibnamefont {{Godet}}}, \bibinfo {author} {\bibfnamefont
  {N.~A.}\ \bibnamefont {{Webb}}}, \bibinfo {author} {\bibfnamefont
  {H.}~\bibnamefont {{Baumgardt}}}, \bibinfo {author} {\bibfnamefont
  {R.}~\bibnamefont {{Wijnands}}}, \bibinfo {author} {\bibfnamefont
  {D.}~\bibnamefont {{Barret}}}, \bibinfo {author} {\bibfnamefont {P.-A.}\
  \bibnamefont {{Duc}}}, \bibinfo {author} {\bibfnamefont {J.~P.}\ \bibnamefont
  {{Brodie}}},\ and\ \bibinfo {author} {\bibfnamefont {S.~D.~J.}\ \bibnamefont
  {{Gwyn}}},\ }\href {https://doi.org/10.1038/s41550-018-0493-1} {\bibfield
  {journal} {\bibinfo  {journal} {Nature Astronomy}\ }\textbf {\bibinfo
  {volume} {2}},\ \bibinfo {pages} {656} (\bibinfo {year} {2018})},\ \Eprint
  {https://arxiv.org/abs/1806.05692} {arXiv:1806.05692 [astro-ph.HE]}
  \BibitemShut {NoStop}%
\bibitem [{\citenamefont {Greene}\ \emph {et~al.}(2020)\citenamefont {Greene},
  \citenamefont {Strader},\ and\ \citenamefont {Ho}}]{2020ARA&A..58..257G}%
  \BibitemOpen
  \bibfield  {author} {\bibinfo {author} {\bibfnamefont {J.~E.}\ \bibnamefont
  {Greene}}, \bibinfo {author} {\bibfnamefont {J.}~\bibnamefont {Strader}},\
  and\ \bibinfo {author} {\bibfnamefont {L.~C.}\ \bibnamefont {Ho}},\ }\href
  {https://doi.org/10.1146/annurev-astro-032620-021835} {\bibfield  {journal}
  {\bibinfo  {journal} {Annual Review of Astronomy and Astrophysics}\ }\textbf
  {\bibinfo {volume} {58}},\ \bibinfo {pages} {257} (\bibinfo {year}
  {2020})}\BibitemShut {NoStop}%
\bibitem [{\citenamefont {{Rees}}(1978)}]{1978Obs....98..210R}%
  \BibitemOpen
  \bibfield  {author} {\bibinfo {author} {\bibfnamefont {M.~J.}\ \bibnamefont
  {{Rees}}},\ }\href@noop {} {\bibfield  {journal} {\bibinfo  {journal} {The
  Observatory}\ }\textbf {\bibinfo {volume} {98}},\ \bibinfo {pages} {210}
  (\bibinfo {year} {1978})}\BibitemShut {NoStop}%
\bibitem [{\citenamefont {{Loeb}}\ and\ \citenamefont
  {{Rasio}}(1994)}]{1994ApJ...432...52L}%
  \BibitemOpen
  \bibfield  {author} {\bibinfo {author} {\bibfnamefont {A.}~\bibnamefont
  {{Loeb}}}\ and\ \bibinfo {author} {\bibfnamefont {F.~A.}\ \bibnamefont
  {{Rasio}}},\ }\href {https://doi.org/10.1086/174548} {\bibfield  {journal}
  {\bibinfo  {journal} {\apj}\ }\textbf {\bibinfo {volume} {432}},\ \bibinfo
  {pages} {52} (\bibinfo {year} {1994})},\ \Eprint
  {https://arxiv.org/abs/astro-ph/9401026} {arXiv:astro-ph/9401026 [astro-ph]}
  \BibitemShut {NoStop}%
\bibitem [{\citenamefont {{Bromm}}\ and\ \citenamefont
  {{Loeb}}(2003)}]{2003ApJ...596...34B}%
  \BibitemOpen
  \bibfield  {author} {\bibinfo {author} {\bibfnamefont {V.}~\bibnamefont
  {{Bromm}}}\ and\ \bibinfo {author} {\bibfnamefont {A.}~\bibnamefont
  {{Loeb}}},\ }\href {https://doi.org/10.1086/377529} {\bibfield  {journal}
  {\bibinfo  {journal} {\apj}\ }\textbf {\bibinfo {volume} {596}},\ \bibinfo
  {pages} {34} (\bibinfo {year} {2003})},\ \Eprint
  {https://arxiv.org/abs/astro-ph/0212400} {arXiv:astro-ph/0212400 [astro-ph]}
  \BibitemShut {NoStop}%
\bibitem [{\citenamefont {{Latif}}\ \emph {et~al.}(2013)\citenamefont
  {{Latif}}, \citenamefont {{Schleicher}}, \citenamefont {{Schmidt}},\ and\
  \citenamefont {{Niemeyer}}}]{2013MNRAS.433.1607L}%
  \BibitemOpen
  \bibfield  {author} {\bibinfo {author} {\bibfnamefont {M.~A.}\ \bibnamefont
  {{Latif}}}, \bibinfo {author} {\bibfnamefont {D.~R.~G.}\ \bibnamefont
  {{Schleicher}}}, \bibinfo {author} {\bibfnamefont {W.}~\bibnamefont
  {{Schmidt}}},\ and\ \bibinfo {author} {\bibfnamefont {J.}~\bibnamefont
  {{Niemeyer}}},\ }\href {https://doi.org/10.1093/mnras/stt834} {\bibfield
  {journal} {\bibinfo  {journal} {\mnras}\ }\textbf {\bibinfo {volume} {433}},\
  \bibinfo {pages} {1607} (\bibinfo {year} {2013})},\ \Eprint
  {https://arxiv.org/abs/1304.0962} {arXiv:1304.0962 [astro-ph.CO]}
  \BibitemShut {NoStop}%
\bibitem [{\citenamefont {{Regan}}\ \emph {et~al.}(2017)\citenamefont
  {{Regan}}, \citenamefont {{Visbal}}, \citenamefont {{Wise}}, \citenamefont
  {{Haiman}}, \citenamefont {{Johansson}},\ and\ \citenamefont
  {{Bryan}}}]{2017NatAs...1E..75R}%
  \BibitemOpen
  \bibfield  {author} {\bibinfo {author} {\bibfnamefont {J.~A.}\ \bibnamefont
  {{Regan}}}, \bibinfo {author} {\bibfnamefont {E.}~\bibnamefont {{Visbal}}},
  \bibinfo {author} {\bibfnamefont {J.~H.}\ \bibnamefont {{Wise}}}, \bibinfo
  {author} {\bibfnamefont {Z.}~\bibnamefont {{Haiman}}}, \bibinfo {author}
  {\bibfnamefont {P.~H.}\ \bibnamefont {{Johansson}}},\ and\ \bibinfo {author}
  {\bibfnamefont {G.~L.}\ \bibnamefont {{Bryan}}},\ }\href
  {https://doi.org/10.1038/s41550-017-0075} {\bibfield  {journal} {\bibinfo
  {journal} {Nature Astronomy}\ }\textbf {\bibinfo {volume} {1}},\ \bibinfo
  {eid} {0075} (\bibinfo {year} {2017})},\ \Eprint
  {https://arxiv.org/abs/1703.03805} {arXiv:1703.03805 [astro-ph.GA]}
  \BibitemShut {NoStop}%
\bibitem [{\citenamefont {{Madau}}\ and\ \citenamefont
  {{Rees}}(2001)}]{2001ApJ...551L..27M}%
  \BibitemOpen
  \bibfield  {author} {\bibinfo {author} {\bibfnamefont {P.}~\bibnamefont
  {{Madau}}}\ and\ \bibinfo {author} {\bibfnamefont {M.~J.}\ \bibnamefont
  {{Rees}}},\ }\href {https://doi.org/10.1086/319848} {\bibfield  {journal}
  {\bibinfo  {journal} {\apjl}\ }\textbf {\bibinfo {volume} {551}},\ \bibinfo
  {pages} {L27} (\bibinfo {year} {2001})},\ \Eprint
  {https://arxiv.org/abs/astro-ph/0101223} {arXiv:astro-ph/0101223 [astro-ph]}
  \BibitemShut {NoStop}%
\bibitem [{\citenamefont {{Bromm}}\ \emph {et~al.}(2002)\citenamefont
  {{Bromm}}, \citenamefont {{Coppi}},\ and\ \citenamefont
  {{Larson}}}]{2002ApJ...564...23B}%
  \BibitemOpen
  \bibfield  {author} {\bibinfo {author} {\bibfnamefont {V.}~\bibnamefont
  {{Bromm}}}, \bibinfo {author} {\bibfnamefont {P.~S.}\ \bibnamefont
  {{Coppi}}},\ and\ \bibinfo {author} {\bibfnamefont {R.~B.}\ \bibnamefont
  {{Larson}}},\ }\href {https://doi.org/10.1086/323947} {\bibfield  {journal}
  {\bibinfo  {journal} {\apj}\ }\textbf {\bibinfo {volume} {564}},\ \bibinfo
  {pages} {23} (\bibinfo {year} {2002})},\ \Eprint
  {https://arxiv.org/abs/astro-ph/0102503} {arXiv:astro-ph/0102503 [astro-ph]}
  \BibitemShut {NoStop}%
\bibitem [{\citenamefont {{Bromm}}\ and\ \citenamefont
  {{Larson}}(2004)}]{2004ARA&A..42...79B}%
  \BibitemOpen
  \bibfield  {author} {\bibinfo {author} {\bibfnamefont {V.}~\bibnamefont
  {{Bromm}}}\ and\ \bibinfo {author} {\bibfnamefont {R.~B.}\ \bibnamefont
  {{Larson}}},\ }\href {https://doi.org/10.1146/annurev.astro.42.053102.134034}
  {\bibfield  {journal} {\bibinfo  {journal} {\araa}\ }\textbf {\bibinfo
  {volume} {42}},\ \bibinfo {pages} {79} (\bibinfo {year} {2004})},\ \Eprint
  {https://arxiv.org/abs/astro-ph/0311019} {arXiv:astro-ph/0311019 [astro-ph]}
  \BibitemShut {NoStop}%
\bibitem [{\citenamefont {{Fryer}}\ \emph {et~al.}(2001)\citenamefont
  {{Fryer}}, \citenamefont {{Woosley}},\ and\ \citenamefont
  {{Heger}}}]{2001ApJ...550..372F}%
  \BibitemOpen
  \bibfield  {author} {\bibinfo {author} {\bibfnamefont {C.~L.}\ \bibnamefont
  {{Fryer}}}, \bibinfo {author} {\bibfnamefont {S.~E.}\ \bibnamefont
  {{Woosley}}},\ and\ \bibinfo {author} {\bibfnamefont {A.}~\bibnamefont
  {{Heger}}},\ }\href {https://doi.org/10.1086/319719} {\bibfield  {journal}
  {\bibinfo  {journal} {\apj}\ }\textbf {\bibinfo {volume} {550}},\ \bibinfo
  {pages} {372} (\bibinfo {year} {2001})},\ \Eprint
  {https://arxiv.org/abs/astro-ph/0007176} {arXiv:astro-ph/0007176 [astro-ph]}
  \BibitemShut {NoStop}%
\bibitem [{\citenamefont {{Portegies Zwart}}\ and\ \citenamefont
  {{McMillan}}(2002)}]{2002ApJ...576..899P}%
  \BibitemOpen
  \bibfield  {author} {\bibinfo {author} {\bibfnamefont {S.~F.}\ \bibnamefont
  {{Portegies Zwart}}}\ and\ \bibinfo {author} {\bibfnamefont {S.~L.~W.}\
  \bibnamefont {{McMillan}}},\ }\href {https://doi.org/10.1086/341798}
  {\bibfield  {journal} {\bibinfo  {journal} {\apj}\ }\textbf {\bibinfo
  {volume} {576}},\ \bibinfo {pages} {899} (\bibinfo {year} {2002})},\ \Eprint
  {https://arxiv.org/abs/astro-ph/0201055} {arXiv:astro-ph/0201055 [astro-ph]}
  \BibitemShut {NoStop}%
\bibitem [{\citenamefont {{Mapelli}}(2016)}]{2016MNRAS.459.3432M}%
  \BibitemOpen
  \bibfield  {author} {\bibinfo {author} {\bibfnamefont {M.}~\bibnamefont
  {{Mapelli}}},\ }\href {https://doi.org/10.1093/mnras/stw869} {\bibfield
  {journal} {\bibinfo  {journal} {\mnras}\ }\textbf {\bibinfo {volume} {459}},\
  \bibinfo {pages} {3432} (\bibinfo {year} {2016})},\ \Eprint
  {https://arxiv.org/abs/1604.03559} {arXiv:1604.03559 [astro-ph.GA]}
  \BibitemShut {NoStop}%
\bibitem [{\citenamefont {{Di Carlo}}\ \emph {et~al.}(2019)\citenamefont {{Di
  Carlo}}, \citenamefont {{Giacobbo}}, \citenamefont {{Mapelli}}, \citenamefont
  {{Pasquato}}, \citenamefont {{Spera}}, \citenamefont {{Wang}},\ and\
  \citenamefont {{Haardt}}}]{2019MNRAS.487.2947D}%
  \BibitemOpen
  \bibfield  {author} {\bibinfo {author} {\bibfnamefont {U.~N.}\ \bibnamefont
  {{Di Carlo}}}, \bibinfo {author} {\bibfnamefont {N.}~\bibnamefont
  {{Giacobbo}}}, \bibinfo {author} {\bibfnamefont {M.}~\bibnamefont
  {{Mapelli}}}, \bibinfo {author} {\bibfnamefont {M.}~\bibnamefont
  {{Pasquato}}}, \bibinfo {author} {\bibfnamefont {M.}~\bibnamefont {{Spera}}},
  \bibinfo {author} {\bibfnamefont {L.}~\bibnamefont {{Wang}}},\ and\ \bibinfo
  {author} {\bibfnamefont {F.}~\bibnamefont {{Haardt}}},\ }\href
  {https://doi.org/10.1093/mnras/stz1453} {\bibfield  {journal} {\bibinfo
  {journal} {\mnras}\ }\textbf {\bibinfo {volume} {487}},\ \bibinfo {pages}
  {2947} (\bibinfo {year} {2019})},\ \Eprint {https://arxiv.org/abs/1901.00863}
  {arXiv:1901.00863 [astro-ph.HE]} \BibitemShut {NoStop}%
\bibitem [{\citenamefont {{Rodriguez}}\ \emph {et~al.}(2019)\citenamefont
  {{Rodriguez}}, \citenamefont {{Zevin}}, \citenamefont {{Amaro-Seoane}},
  \citenamefont {{Chatterjee}}, \citenamefont {{Kremer}}, \citenamefont
  {{Rasio}},\ and\ \citenamefont {{Ye}}}]{2019PhRvD.100d3027R}%
  \BibitemOpen
  \bibfield  {author} {\bibinfo {author} {\bibfnamefont {C.~L.}\ \bibnamefont
  {{Rodriguez}}}, \bibinfo {author} {\bibfnamefont {M.}~\bibnamefont
  {{Zevin}}}, \bibinfo {author} {\bibfnamefont {P.}~\bibnamefont
  {{Amaro-Seoane}}}, \bibinfo {author} {\bibfnamefont {S.}~\bibnamefont
  {{Chatterjee}}}, \bibinfo {author} {\bibfnamefont {K.}~\bibnamefont
  {{Kremer}}}, \bibinfo {author} {\bibfnamefont {F.~A.}\ \bibnamefont
  {{Rasio}}},\ and\ \bibinfo {author} {\bibfnamefont {C.~S.}\ \bibnamefont
  {{Ye}}},\ }\href {https://doi.org/10.1103/PhysRevD.100.043027} {\bibfield
  {journal} {\bibinfo  {journal} {\prd}\ }\textbf {\bibinfo {volume} {100}},\
  \bibinfo {eid} {043027} (\bibinfo {year} {2019})},\ \Eprint
  {https://arxiv.org/abs/1906.10260} {arXiv:1906.10260 [astro-ph.HE]}
  \BibitemShut {NoStop}%
\bibitem [{\citenamefont {{Rizzuto}}\ \emph {et~al.}(2021)\citenamefont
  {{Rizzuto}}, \citenamefont {{Naab}}, \citenamefont {{Spurzem}}, \citenamefont
  {{Giersz}}, \citenamefont {{Ostriker}}, \citenamefont {{Stone}},
  \citenamefont {{Wang}}, \citenamefont {{Berczik}},\ and\ \citenamefont
  {{Rampp}}}]{2021MNRAS.501.5257R}%
  \BibitemOpen
  \bibfield  {author} {\bibinfo {author} {\bibfnamefont {F.~P.}\ \bibnamefont
  {{Rizzuto}}}, \bibinfo {author} {\bibfnamefont {T.}~\bibnamefont {{Naab}}},
  \bibinfo {author} {\bibfnamefont {R.}~\bibnamefont {{Spurzem}}}, \bibinfo
  {author} {\bibfnamefont {M.}~\bibnamefont {{Giersz}}}, \bibinfo {author}
  {\bibfnamefont {J.~P.}\ \bibnamefont {{Ostriker}}}, \bibinfo {author}
  {\bibfnamefont {N.~C.}\ \bibnamefont {{Stone}}}, \bibinfo {author}
  {\bibfnamefont {L.}~\bibnamefont {{Wang}}}, \bibinfo {author} {\bibfnamefont
  {P.}~\bibnamefont {{Berczik}}},\ and\ \bibinfo {author} {\bibfnamefont
  {M.}~\bibnamefont {{Rampp}}},\ }\href
  {https://doi.org/10.1093/mnras/staa3634} {\bibfield  {journal} {\bibinfo
  {journal} {\mnras}\ }\textbf {\bibinfo {volume} {501}},\ \bibinfo {pages}
  {5257} (\bibinfo {year} {2021})},\ \Eprint {https://arxiv.org/abs/2008.09571}
  {arXiv:2008.09571 [astro-ph.GA]} \BibitemShut {NoStop}%
\bibitem [{\citenamefont {{Di Carlo}}\ \emph {et~al.}(2021)\citenamefont {{Di
  Carlo}}, \citenamefont {{Mapelli}}, \citenamefont {{Pasquato}}, \citenamefont
  {{Rastello}}, \citenamefont {{Ballone}}, \citenamefont {{Dall'Amico}},
  \citenamefont {{Giacobbo}}, \citenamefont {{Iorio}}, \citenamefont {{Spera}},
  \citenamefont {{Torniamenti}},\ and\ \citenamefont
  {{Haardt}}}]{2021MNRAS.507.5132D}%
  \BibitemOpen
  \bibfield  {author} {\bibinfo {author} {\bibfnamefont {U.~N.}\ \bibnamefont
  {{Di Carlo}}}, \bibinfo {author} {\bibfnamefont {M.}~\bibnamefont
  {{Mapelli}}}, \bibinfo {author} {\bibfnamefont {M.}~\bibnamefont
  {{Pasquato}}}, \bibinfo {author} {\bibfnamefont {S.}~\bibnamefont
  {{Rastello}}}, \bibinfo {author} {\bibfnamefont {A.}~\bibnamefont
  {{Ballone}}}, \bibinfo {author} {\bibfnamefont {M.}~\bibnamefont
  {{Dall'Amico}}}, \bibinfo {author} {\bibfnamefont {N.}~\bibnamefont
  {{Giacobbo}}}, \bibinfo {author} {\bibfnamefont {G.}~\bibnamefont {{Iorio}}},
  \bibinfo {author} {\bibfnamefont {M.}~\bibnamefont {{Spera}}}, \bibinfo
  {author} {\bibfnamefont {S.}~\bibnamefont {{Torniamenti}}},\ and\ \bibinfo
  {author} {\bibfnamefont {F.}~\bibnamefont {{Haardt}}},\ }\href
  {https://doi.org/10.1093/mnras/stab2390} {\bibfield  {journal} {\bibinfo
  {journal} {\mnras}\ }\textbf {\bibinfo {volume} {507}},\ \bibinfo {pages}
  {5132} (\bibinfo {year} {2021})},\ \Eprint {https://arxiv.org/abs/2105.01085}
  {arXiv:2105.01085 [astro-ph.GA]} \BibitemShut {NoStop}%
\bibitem [{\citenamefont {{G{\"u}rkan}}\ \emph {et~al.}(2004)\citenamefont
  {{G{\"u}rkan}}, \citenamefont {{Freitag}},\ and\ \citenamefont
  {{Rasio}}}]{2004ApJ...604..632G}%
  \BibitemOpen
  \bibfield  {author} {\bibinfo {author} {\bibfnamefont {M.~A.}\ \bibnamefont
  {{G{\"u}rkan}}}, \bibinfo {author} {\bibfnamefont {M.}~\bibnamefont
  {{Freitag}}},\ and\ \bibinfo {author} {\bibfnamefont {F.~A.}\ \bibnamefont
  {{Rasio}}},\ }\href {https://doi.org/10.1086/381968} {\bibfield  {journal}
  {\bibinfo  {journal} {\apj}\ }\textbf {\bibinfo {volume} {604}},\ \bibinfo
  {pages} {632} (\bibinfo {year} {2004})},\ \Eprint
  {https://arxiv.org/abs/astro-ph/0308449} {arXiv:astro-ph/0308449 [astro-ph]}
  \BibitemShut {NoStop}%
\bibitem [{\citenamefont {{Giersz}}\ \emph {et~al.}(2015)\citenamefont
  {{Giersz}}, \citenamefont {{Leigh}}, \citenamefont {{Hypki}}, \citenamefont
  {{L\"{u}tzgendorf}},\ and\ \citenamefont {{Askar}}}]{2015MNRAS.454.3150G}%
  \BibitemOpen
  \bibfield  {author} {\bibinfo {author} {\bibfnamefont {M.}~\bibnamefont
  {{Giersz}}}, \bibinfo {author} {\bibfnamefont {N.~W.}\ \bibnamefont
  {{Leigh}}}, \bibinfo {author} {\bibfnamefont {A.}~\bibnamefont {{Hypki}}},
  \bibinfo {author} {\bibfnamefont {N.}~\bibnamefont {{L\"{u}tzgendorf}}},\
  and\ \bibinfo {author} {\bibfnamefont {A.}~\bibnamefont {{Askar}}},\ }\href
  {https://doi.org/10.1093/mnras/stv2162} {\bibfield  {journal} {\bibinfo
  {journal} {\mnras}\ }\textbf {\bibinfo {volume} {454}},\ \bibinfo {pages}
  {3150} (\bibinfo {year} {2015})}\BibitemShut {NoStop}%
\bibitem [{\citenamefont {{Arca Sedda}}\ \emph {et~al.}(2019)\citenamefont
  {{Arca Sedda}}, \citenamefont {{Askar}},\ and\ \citenamefont
  {{Giersz}}}]{2019arXiv190500902A}%
  \BibitemOpen
  \bibfield  {author} {\bibinfo {author} {\bibfnamefont {M.}~\bibnamefont
  {{Arca Sedda}}}, \bibinfo {author} {\bibfnamefont {A.}~\bibnamefont
  {{Askar}}},\ and\ \bibinfo {author} {\bibfnamefont {M.}~\bibnamefont
  {{Giersz}}},\ }\href@noop {} {\bibfield  {journal} {\bibinfo  {journal}
  {arXiv e-prints}\ ,\ \bibinfo {eid} {arXiv:1905.00902}} (\bibinfo {year}
  {2019})},\ \Eprint {https://arxiv.org/abs/1905.00902} {arXiv:1905.00902
  [astro-ph.GA]} \BibitemShut {NoStop}%
\bibitem [{\citenamefont {{Gonz{\'a}lez}}\ \emph {et~al.}(2021)\citenamefont
  {{Gonz{\'a}lez}}, \citenamefont {{Kremer}}, \citenamefont {{Chatterjee}},
  \citenamefont {{Fragione}},\ and\ \citenamefont
  {et~al.}}]{2021ApJ...908L..29G}%
  \BibitemOpen
  \bibfield  {author} {\bibinfo {author} {\bibfnamefont {E.}~\bibnamefont
  {{Gonz{\'a}lez}}}, \bibinfo {author} {\bibfnamefont {K.}~\bibnamefont
  {{Kremer}}}, \bibinfo {author} {\bibfnamefont {S.}~\bibnamefont
  {{Chatterjee}}}, \bibinfo {author} {\bibfnamefont {G.}~\bibnamefont
  {{Fragione}}},\ and\ \bibinfo {author} {\bibnamefont {et~al.}},\ }\href
  {https://doi.org/10.3847/2041-8213/abdf5b} {\bibfield  {journal} {\bibinfo
  {journal} {\apjl}\ }\textbf {\bibinfo {volume} {908}},\ \bibinfo {eid} {L29}
  (\bibinfo {year} {2021})},\ \Eprint {https://arxiv.org/abs/2012.10497}
  {arXiv:2012.10497 [astro-ph.HE]} \BibitemShut {NoStop}%
\bibitem [{\citenamefont {{Maliszewski}}\ \emph {et~al.}(2021)\citenamefont
  {{Maliszewski}}, \citenamefont {{Giersz}}, \citenamefont
  {{Gondek-Rosi{\'n}ska}}, \citenamefont {{Askar}},\ and\ \citenamefont
  {{Hypki}}}]{2021arXiv211109223M}%
  \BibitemOpen
  \bibfield  {author} {\bibinfo {author} {\bibfnamefont {K.}~\bibnamefont
  {{Maliszewski}}}, \bibinfo {author} {\bibfnamefont {M.}~\bibnamefont
  {{Giersz}}}, \bibinfo {author} {\bibfnamefont {D.}~\bibnamefont
  {{Gondek-Rosi{\'n}ska}}}, \bibinfo {author} {\bibfnamefont {A.}~\bibnamefont
  {{Askar}}},\ and\ \bibinfo {author} {\bibfnamefont {A.}~\bibnamefont
  {{Hypki}}},\ }\href@noop {} {\bibfield  {journal} {\bibinfo  {journal} {arXiv
  e-prints}\ ,\ \bibinfo {eid} {arXiv:2111.09223}} (\bibinfo {year} {2021})},\
  \Eprint {https://arxiv.org/abs/2111.09223} {arXiv:2111.09223 [astro-ph.GA]}
  \BibitemShut {NoStop}%
\bibitem [{\citenamefont {{Miller}}\ and\ \citenamefont
  {{Hamilton}}(2002)}]{2002MNRAS.330..232C}%
  \BibitemOpen
  \bibfield  {author} {\bibinfo {author} {\bibfnamefont {M.~C.}\ \bibnamefont
  {{Miller}}}\ and\ \bibinfo {author} {\bibfnamefont {D.~P.}\ \bibnamefont
  {{Hamilton}}},\ }\href {https://doi.org/10.1046/j.1365-8711.2002.05112.x}
  {\bibfield  {journal} {\bibinfo  {journal} {\mnras}\ }\textbf {\bibinfo
  {volume} {330}},\ \bibinfo {pages} {232} (\bibinfo {year} {2002})},\ \Eprint
  {https://arxiv.org/abs/astro-ph/0106188} {arXiv:astro-ph/0106188 [astro-ph]}
  \BibitemShut {NoStop}%
\bibitem [{\citenamefont {{Miller}}\ and\ \citenamefont
  {{Colbert}}(2004)}]{2004IJMPD..13....1M}%
  \BibitemOpen
  \bibfield  {author} {\bibinfo {author} {\bibfnamefont {M.~C.}\ \bibnamefont
  {{Miller}}}\ and\ \bibinfo {author} {\bibfnamefont {E.~J.~M.}\ \bibnamefont
  {{Colbert}}},\ }\href {https://doi.org/10.1142/S0218271804004426} {\bibfield
  {journal} {\bibinfo  {journal} {International Journal of Modern Physics D}\
  }\textbf {\bibinfo {volume} {13}},\ \bibinfo {pages} {1} (\bibinfo {year}
  {2004})},\ \Eprint {https://arxiv.org/abs/astro-ph/0308402}
  {arXiv:astro-ph/0308402 [astro-ph]} \BibitemShut {NoStop}%
\bibitem [{\citenamefont {{Antonini}}\ \emph {et~al.}(2019)\citenamefont
  {{Antonini}}, \citenamefont {{Gieles}},\ and\ \citenamefont
  {{Gualandris}}}]{2019MNRAS.486.5008A}%
  \BibitemOpen
  \bibfield  {author} {\bibinfo {author} {\bibfnamefont {F.}~\bibnamefont
  {{Antonini}}}, \bibinfo {author} {\bibfnamefont {M.}~\bibnamefont
  {{Gieles}}},\ and\ \bibinfo {author} {\bibfnamefont {A.}~\bibnamefont
  {{Gualandris}}},\ }\href {https://doi.org/10.1093/mnras/stz1149} {\bibfield
  {journal} {\bibinfo  {journal} {\mnras}\ }\textbf {\bibinfo {volume} {486}},\
  \bibinfo {pages} {5008} (\bibinfo {year} {2019})},\ \Eprint
  {https://arxiv.org/abs/1811.03640} {arXiv:1811.03640 [astro-ph.HE]}
  \BibitemShut {NoStop}%
\bibitem [{\citenamefont {{Fragione}}\ and\ \citenamefont
  {{Silk}}(2020)}]{2020MNRAS.498.4591F}%
  \BibitemOpen
  \bibfield  {author} {\bibinfo {author} {\bibfnamefont {G.}~\bibnamefont
  {{Fragione}}}\ and\ \bibinfo {author} {\bibfnamefont {J.}~\bibnamefont
  {{Silk}}},\ }\href {https://doi.org/10.1093/mnras/staa2629} {\bibfield
  {journal} {\bibinfo  {journal} {\mnras}\ }\textbf {\bibinfo {volume} {498}},\
  \bibinfo {pages} {4591} (\bibinfo {year} {2020})},\ \Eprint
  {https://arxiv.org/abs/2006.01867} {arXiv:2006.01867 [astro-ph.GA]}
  \BibitemShut {NoStop}%
\bibitem [{\citenamefont {{Fragione}}\ \emph {et~al.}(2020)\citenamefont
  {{Fragione}}, \citenamefont {{Loeb}},\ and\ \citenamefont
  {{Rasio}}}]{2020ApJ...902L..26F}%
  \BibitemOpen
  \bibfield  {author} {\bibinfo {author} {\bibfnamefont {G.}~\bibnamefont
  {{Fragione}}}, \bibinfo {author} {\bibfnamefont {A.}~\bibnamefont {{Loeb}}},\
  and\ \bibinfo {author} {\bibfnamefont {F.~A.}\ \bibnamefont {{Rasio}}},\
  }\href {https://doi.org/10.3847/2041-8213/abbc0a} {\bibfield  {journal}
  {\bibinfo  {journal} {\apjl}\ }\textbf {\bibinfo {volume} {902}},\ \bibinfo
  {eid} {L26} (\bibinfo {year} {2020})},\ \Eprint
  {https://arxiv.org/abs/2009.05065} {arXiv:2009.05065 [astro-ph.GA]}
  \BibitemShut {NoStop}%
\bibitem [{\citenamefont {{Mapelli}}\ \emph {et~al.}(2021)\citenamefont
  {{Mapelli}}, \citenamefont {{Dall'Amico}}, \citenamefont {{Bouffanais}},
  \citenamefont {{Giacobbo}},\ and\ \citenamefont
  {et~al.}}]{2021MNRAS.505..339M}%
  \BibitemOpen
  \bibfield  {author} {\bibinfo {author} {\bibfnamefont {M.}~\bibnamefont
  {{Mapelli}}}, \bibinfo {author} {\bibfnamefont {M.}~\bibnamefont
  {{Dall'Amico}}}, \bibinfo {author} {\bibfnamefont {Y.}~\bibnamefont
  {{Bouffanais}}}, \bibinfo {author} {\bibfnamefont {N.}~\bibnamefont
  {{Giacobbo}}},\ and\ \bibinfo {author} {\bibnamefont {et~al.}},\ }\href
  {https://doi.org/10.1093/mnras/stab1334} {\bibfield  {journal} {\bibinfo
  {journal} {\mnras}\ }\textbf {\bibinfo {volume} {505}},\ \bibinfo {pages}
  {339} (\bibinfo {year} {2021})},\ \Eprint {https://arxiv.org/abs/2103.05016}
  {arXiv:2103.05016 [astro-ph.HE]} \BibitemShut {NoStop}%
\bibitem [{\citenamefont {{Greene}}\ and\ \citenamefont
  {{Ho}}(2007)}]{2007ApJ...670...92G}%
  \BibitemOpen
  \bibfield  {author} {\bibinfo {author} {\bibfnamefont {J.~E.}\ \bibnamefont
  {{Greene}}}\ and\ \bibinfo {author} {\bibfnamefont {L.~C.}\ \bibnamefont
  {{Ho}}},\ }\href {https://doi.org/10.1086/522082} {\bibfield  {journal}
  {\bibinfo  {journal} {\apj}\ }\textbf {\bibinfo {volume} {670}},\ \bibinfo
  {pages} {92} (\bibinfo {year} {2007})},\ \Eprint
  {https://arxiv.org/abs/0707.2617} {arXiv:0707.2617 [astro-ph]} \BibitemShut
  {NoStop}%
\bibitem [{\citenamefont {{Kaaret}}\ \emph {et~al.}(2017)\citenamefont
  {{Kaaret}}, \citenamefont {{Feng}},\ and\ \citenamefont
  {{Roberts}}}]{2017ARA&A..55..303K}%
  \BibitemOpen
  \bibfield  {author} {\bibinfo {author} {\bibfnamefont {P.}~\bibnamefont
  {{Kaaret}}}, \bibinfo {author} {\bibfnamefont {H.}~\bibnamefont {{Feng}}},\
  and\ \bibinfo {author} {\bibfnamefont {T.~P.}\ \bibnamefont {{Roberts}}},\
  }\href {https://doi.org/10.1146/annurev-astro-091916-055259} {\bibfield
  {journal} {\bibinfo  {journal} {\araa}\ }\textbf {\bibinfo {volume} {55}},\
  \bibinfo {pages} {303} (\bibinfo {year} {2017})},\ \Eprint
  {https://arxiv.org/abs/1703.10728} {arXiv:1703.10728 [astro-ph.HE]}
  \BibitemShut {NoStop}%
\bibitem [{\citenamefont {{Gualandris}}\ \emph {et~al.}(2010)\citenamefont
  {{Gualandris}}, \citenamefont {{Gillessen}},\ and\ \citenamefont
  {{2013degn.book.....M}}}]{2010MNRAS.409.1146G}%
  \BibitemOpen
  \bibfield  {author} {\bibinfo {author} {\bibfnamefont {A.}~\bibnamefont
  {{Gualandris}}}, \bibinfo {author} {\bibfnamefont {S.}~\bibnamefont
  {{Gillessen}}},\ and\ \bibinfo {author} {\bibfnamefont {D.}~\bibnamefont
  {{2013degn.book.....M}}},\ }\href
  {https://doi.org/10.1111/j.1365-2966.2010.17373.x} {\bibfield  {journal}
  {\bibinfo  {journal} {\mnras}\ }\textbf {\bibinfo {volume} {409}},\ \bibinfo
  {pages} {1146} (\bibinfo {year} {2010})},\ \Eprint
  {https://arxiv.org/abs/1006.3563} {arXiv:1006.3563 [astro-ph.GA]}
  \BibitemShut {NoStop}%
\bibitem [{\citenamefont {{Girma}}\ and\ \citenamefont
  {{Loeb}}(2019)}]{2019MNRAS.482.3669G}%
  \BibitemOpen
  \bibfield  {author} {\bibinfo {author} {\bibfnamefont {E.}~\bibnamefont
  {{Girma}}}\ and\ \bibinfo {author} {\bibfnamefont {A.}~\bibnamefont
  {{Loeb}}},\ }\href {https://doi.org/10.1093/mnras/sty2643} {\bibfield
  {journal} {\bibinfo  {journal} {\mnras}\ }\textbf {\bibinfo {volume} {482}},\
  \bibinfo {pages} {3669} (\bibinfo {year} {2019})},\ \Eprint
  {https://arxiv.org/abs/1807.02469} {arXiv:1807.02469 [astro-ph.GA]}
  \BibitemShut {NoStop}%
\bibitem [{\citenamefont {{Baumgardt}}\ \emph {et~al.}(2019)\citenamefont
  {{Baumgardt}}, \citenamefont {{He}}, \citenamefont {{Sweet}}, \citenamefont
  {{Drinkwater}}, \citenamefont {{Sollima}}, \citenamefont {{Hurley}},
  \citenamefont {{Usher}}, \citenamefont {{Kamann}}, \citenamefont
  {{Dalgleish}}, \citenamefont {{Dreizler}},\ and\ \citenamefont
  {{Husser}}}]{2019MNRAS.488.5340B}%
  \BibitemOpen
  \bibfield  {author} {\bibinfo {author} {\bibfnamefont {H.}~\bibnamefont
  {{Baumgardt}}}, \bibinfo {author} {\bibfnamefont {C.}~\bibnamefont {{He}}},
  \bibinfo {author} {\bibfnamefont {S.~M.}\ \bibnamefont {{Sweet}}}, \bibinfo
  {author} {\bibfnamefont {M.}~\bibnamefont {{Drinkwater}}}, \bibinfo {author}
  {\bibfnamefont {A.}~\bibnamefont {{Sollima}}}, \bibinfo {author}
  {\bibfnamefont {J.}~\bibnamefont {{Hurley}}}, \bibinfo {author}
  {\bibfnamefont {C.}~\bibnamefont {{Usher}}}, \bibinfo {author} {\bibfnamefont
  {S.}~\bibnamefont {{Kamann}}}, \bibinfo {author} {\bibfnamefont
  {H.}~\bibnamefont {{Dalgleish}}}, \bibinfo {author} {\bibfnamefont
  {S.}~\bibnamefont {{Dreizler}}},\ and\ \bibinfo {author} {\bibfnamefont
  {T.~O.}\ \bibnamefont {{Husser}}},\ }\href
  {https://doi.org/10.1093/mnras/stz2060} {\bibfield  {journal} {\bibinfo
  {journal} {\mnras}\ }\textbf {\bibinfo {volume} {488}},\ \bibinfo {pages}
  {5340} (\bibinfo {year} {2019})},\ \Eprint {https://arxiv.org/abs/1907.10845}
  {arXiv:1907.10845 [astro-ph.GA]} \BibitemShut {NoStop}%
\bibitem [{\citenamefont {{van der Marel}}\ and\ \citenamefont
  {{Anderson}}(2010)}]{2010ApJ...710.1063V}%
  \BibitemOpen
  \bibfield  {author} {\bibinfo {author} {\bibfnamefont {R.~P.}\ \bibnamefont
  {{van der Marel}}}\ and\ \bibinfo {author} {\bibfnamefont {J.}~\bibnamefont
  {{Anderson}}},\ }\href {https://doi.org/10.1088/0004-637X/710/2/1063}
  {\bibfield  {journal} {\bibinfo  {journal} {\apj}\ }\textbf {\bibinfo
  {volume} {710}},\ \bibinfo {pages} {1063} (\bibinfo {year} {2010})},\ \Eprint
  {https://arxiv.org/abs/0905.0638} {arXiv:0905.0638 [astro-ph.GA]}
  \BibitemShut {NoStop}%
\bibitem [{\citenamefont {{Noyola}}\ \emph {et~al.}(2010)\citenamefont
  {{Noyola}}, \citenamefont {{Gebhardt}}, \citenamefont {{Kissler-Patig}},
  \citenamefont {{L{\"u}tzgendorf}}, \citenamefont {{Jalali}}, \citenamefont
  {{de Zeeuw}},\ and\ \citenamefont {{Baumgardt}}}]{2010ApJ...719L..60N}%
  \BibitemOpen
  \bibfield  {author} {\bibinfo {author} {\bibfnamefont {E.}~\bibnamefont
  {{Noyola}}}, \bibinfo {author} {\bibfnamefont {K.}~\bibnamefont
  {{Gebhardt}}}, \bibinfo {author} {\bibfnamefont {M.}~\bibnamefont
  {{Kissler-Patig}}}, \bibinfo {author} {\bibfnamefont {N.}~\bibnamefont
  {{L{\"u}tzgendorf}}}, \bibinfo {author} {\bibfnamefont {B.}~\bibnamefont
  {{Jalali}}}, \bibinfo {author} {\bibfnamefont {P.~T.}\ \bibnamefont {{de
  Zeeuw}}},\ and\ \bibinfo {author} {\bibfnamefont {H.}~\bibnamefont
  {{Baumgardt}}},\ }\href {https://doi.org/10.1088/2041-8205/719/1/L60}
  {\bibfield  {journal} {\bibinfo  {journal} {\apjl}\ }\textbf {\bibinfo
  {volume} {719}},\ \bibinfo {pages} {L60} (\bibinfo {year} {2010})},\ \Eprint
  {https://arxiv.org/abs/1007.4559} {arXiv:1007.4559 [astro-ph.GA]}
  \BibitemShut {NoStop}%
\bibitem [{\citenamefont {{L{\"u}tzgendorf}}\ \emph {et~al.}(2011)\citenamefont
  {{L{\"u}tzgendorf}}, \citenamefont {{Kissler-Patig}}, \citenamefont
  {{Noyola}}, \citenamefont {{Jalali}}, \citenamefont {{de Zeeuw}},
  \citenamefont {{Gebhardt}},\ and\ \citenamefont
  {{Baumgardt}}}]{2011A&A...533A..36L}%
  \BibitemOpen
  \bibfield  {author} {\bibinfo {author} {\bibfnamefont {N.}~\bibnamefont
  {{L{\"u}tzgendorf}}}, \bibinfo {author} {\bibfnamefont {M.}~\bibnamefont
  {{Kissler-Patig}}}, \bibinfo {author} {\bibfnamefont {E.}~\bibnamefont
  {{Noyola}}}, \bibinfo {author} {\bibfnamefont {B.}~\bibnamefont {{Jalali}}},
  \bibinfo {author} {\bibfnamefont {P.~T.}\ \bibnamefont {{de Zeeuw}}},
  \bibinfo {author} {\bibfnamefont {K.}~\bibnamefont {{Gebhardt}}},\ and\
  \bibinfo {author} {\bibfnamefont {H.}~\bibnamefont {{Baumgardt}}},\ }\href
  {https://doi.org/10.1051/0004-6361/201116618} {\bibfield  {journal} {\bibinfo
   {journal} {\aap}\ }\textbf {\bibinfo {volume} {533}},\ \bibinfo {eid} {A36}
  (\bibinfo {year} {2011})},\ \Eprint {https://arxiv.org/abs/1107.4243}
  {arXiv:1107.4243 [astro-ph.GA]} \BibitemShut {NoStop}%
\bibitem [{\citenamefont {{Lanzoni}}\ \emph {et~al.}(2013)\citenamefont
  {{Lanzoni}}, \citenamefont {{Mucciarelli}}, \citenamefont {{Origlia}},
  \citenamefont {{Bellazzini}}, \citenamefont {{Ferraro}}, \citenamefont
  {{Valenti}}, \citenamefont {{Miocchi}}, \citenamefont {{Dalessandro}},
  \citenamefont {{Pallanca}},\ and\ \citenamefont
  {{Massari}}}]{2013ApJ...769..107L}%
  \BibitemOpen
  \bibfield  {author} {\bibinfo {author} {\bibfnamefont {B.}~\bibnamefont
  {{Lanzoni}}}, \bibinfo {author} {\bibfnamefont {A.}~\bibnamefont
  {{Mucciarelli}}}, \bibinfo {author} {\bibfnamefont {L.}~\bibnamefont
  {{Origlia}}}, \bibinfo {author} {\bibfnamefont {M.}~\bibnamefont
  {{Bellazzini}}}, \bibinfo {author} {\bibfnamefont {F.~R.}\ \bibnamefont
  {{Ferraro}}}, \bibinfo {author} {\bibfnamefont {E.}~\bibnamefont
  {{Valenti}}}, \bibinfo {author} {\bibfnamefont {P.}~\bibnamefont
  {{Miocchi}}}, \bibinfo {author} {\bibfnamefont {E.}~\bibnamefont
  {{Dalessandro}}}, \bibinfo {author} {\bibfnamefont {C.}~\bibnamefont
  {{Pallanca}}},\ and\ \bibinfo {author} {\bibfnamefont {D.}~\bibnamefont
  {{Massari}}},\ }\href {https://doi.org/10.1088/0004-637X/769/2/107}
  {\bibfield  {journal} {\bibinfo  {journal} {\apj}\ }\textbf {\bibinfo
  {volume} {769}},\ \bibinfo {eid} {107} (\bibinfo {year} {2013})},\ \Eprint
  {https://arxiv.org/abs/1304.2953} {arXiv:1304.2953 [astro-ph.SR]}
  \BibitemShut {NoStop}%
\bibitem [{\citenamefont {{L{\"u}tzgendorf}}\ \emph {et~al.}(2013)\citenamefont
  {{L{\"u}tzgendorf}}, \citenamefont {{Kissler-Patig}}, \citenamefont
  {{Gebhardt}}, \citenamefont {{Baumgardt}}, \citenamefont {{Noyola}},
  \citenamefont {{de Zeeuw}}, \citenamefont {{Neumayer}}, \citenamefont
  {{Jalali}},\ and\ \citenamefont {{Feldmeier}}}]{2013A&A...552A..49L}%
  \BibitemOpen
  \bibfield  {author} {\bibinfo {author} {\bibfnamefont {N.}~\bibnamefont
  {{L{\"u}tzgendorf}}}, \bibinfo {author} {\bibfnamefont {M.}~\bibnamefont
  {{Kissler-Patig}}}, \bibinfo {author} {\bibfnamefont {K.}~\bibnamefont
  {{Gebhardt}}}, \bibinfo {author} {\bibfnamefont {H.}~\bibnamefont
  {{Baumgardt}}}, \bibinfo {author} {\bibfnamefont {E.}~\bibnamefont
  {{Noyola}}}, \bibinfo {author} {\bibfnamefont {P.~T.}\ \bibnamefont {{de
  Zeeuw}}}, \bibinfo {author} {\bibfnamefont {N.}~\bibnamefont {{Neumayer}}},
  \bibinfo {author} {\bibfnamefont {B.}~\bibnamefont {{Jalali}}},\ and\
  \bibinfo {author} {\bibfnamefont {A.}~\bibnamefont {{Feldmeier}}},\ }\href
  {https://doi.org/10.1051/0004-6361/201220307} {\bibfield  {journal} {\bibinfo
   {journal} {\aap}\ }\textbf {\bibinfo {volume} {552}},\ \bibinfo {eid} {A49}
  (\bibinfo {year} {2013})},\ \Eprint {https://arxiv.org/abs/1212.3475}
  {arXiv:1212.3475 [astro-ph.GA]} \BibitemShut {NoStop}%
\bibitem [{\citenamefont {{Shen}}\ and\ \citenamefont
  {{Matzner}}(2014)}]{2014ApJ...784...87S}%
  \BibitemOpen
  \bibfield  {author} {\bibinfo {author} {\bibfnamefont {R.-F.}\ \bibnamefont
  {{Shen}}}\ and\ \bibinfo {author} {\bibfnamefont {C.~D.}\ \bibnamefont
  {{Matzner}}},\ }\href {https://doi.org/10.1088/0004-637X/784/2/87} {\bibfield
   {journal} {\bibinfo  {journal} {\apj}\ }\textbf {\bibinfo {volume} {784}},\
  \bibinfo {eid} {87} (\bibinfo {year} {2014})},\ \Eprint
  {https://arxiv.org/abs/1305.5570} {arXiv:1305.5570 [astro-ph.HE]}
  \BibitemShut {NoStop}%
\bibitem [{\citenamefont {{Lin}}\ \emph {et~al.}(2016)\citenamefont {{Lin}},
  \citenamefont {{Carrasco}}, \citenamefont {{Webb}}, \citenamefont {{Irwin}},
  \citenamefont {{Dupke}}, \citenamefont {{Romanowsky}}, \citenamefont
  {{Ramirez-Ruiz}}, \citenamefont {{Strader}}, \citenamefont {{Homan}},
  \citenamefont {{Barret}},\ and\ \citenamefont
  {{Godet}}}]{2016ApJ...821...25L}%
  \BibitemOpen
  \bibfield  {author} {\bibinfo {author} {\bibfnamefont {D.}~\bibnamefont
  {{Lin}}}, \bibinfo {author} {\bibfnamefont {E.~R.}\ \bibnamefont
  {{Carrasco}}}, \bibinfo {author} {\bibfnamefont {N.~A.}\ \bibnamefont
  {{Webb}}}, \bibinfo {author} {\bibfnamefont {J.~A.}\ \bibnamefont {{Irwin}}},
  \bibinfo {author} {\bibfnamefont {R.}~\bibnamefont {{Dupke}}}, \bibinfo
  {author} {\bibfnamefont {A.~J.}\ \bibnamefont {{Romanowsky}}}, \bibinfo
  {author} {\bibfnamefont {E.}~\bibnamefont {{Ramirez-Ruiz}}}, \bibinfo
  {author} {\bibfnamefont {J.}~\bibnamefont {{Strader}}}, \bibinfo {author}
  {\bibfnamefont {J.}~\bibnamefont {{Homan}}}, \bibinfo {author} {\bibfnamefont
  {D.}~\bibnamefont {{Barret}}},\ and\ \bibinfo {author} {\bibfnamefont
  {O.}~\bibnamefont {{Godet}}},\ }\href
  {https://doi.org/10.3847/0004-637X/821/1/25} {\bibfield  {journal} {\bibinfo
  {journal} {\apj}\ }\textbf {\bibinfo {volume} {821}},\ \bibinfo {eid} {25}
  (\bibinfo {year} {2016})},\ \Eprint {https://arxiv.org/abs/1603.00455}
  {arXiv:1603.00455 [astro-ph.HE]} \BibitemShut {NoStop}%
\bibitem [{\citenamefont {{Chen}}\ and\ \citenamefont
  {{Shen}}(2018)}]{2018ApJ...867...20C}%
  \BibitemOpen
  \bibfield  {author} {\bibinfo {author} {\bibfnamefont {J.-H.}\ \bibnamefont
  {{Chen}}}\ and\ \bibinfo {author} {\bibfnamefont {R.-F.}\ \bibnamefont
  {{Shen}}},\ }\href {https://doi.org/10.3847/1538-4357/aadfda} {\bibfield
  {journal} {\bibinfo  {journal} {\apj}\ }\textbf {\bibinfo {volume} {867}},\
  \bibinfo {eid} {20} (\bibinfo {year} {2018})},\ \Eprint
  {https://arxiv.org/abs/1806.08093} {arXiv:1806.08093 [astro-ph.HE]}
  \BibitemShut {NoStop}%
\bibitem [{\citenamefont {{Fragione}}\ \emph
  {et~al.}(2018{\natexlab{a}})\citenamefont {{Fragione}}, \citenamefont
  {{Leigh}}, \citenamefont {{Ginsburg}},\ and\ \citenamefont
  {{Kocsis}}}]{2018ApJ...867..119F}%
  \BibitemOpen
  \bibfield  {author} {\bibinfo {author} {\bibfnamefont {G.}~\bibnamefont
  {{Fragione}}}, \bibinfo {author} {\bibfnamefont {N.~W.~C.}\ \bibnamefont
  {{Leigh}}}, \bibinfo {author} {\bibfnamefont {I.}~\bibnamefont
  {{Ginsburg}}},\ and\ \bibinfo {author} {\bibfnamefont {B.}~\bibnamefont
  {{Kocsis}}},\ }\href {https://doi.org/10.3847/1538-4357/aae486} {\bibfield
  {journal} {\bibinfo  {journal} {\apj}\ }\textbf {\bibinfo {volume} {867}},\
  \bibinfo {eid} {119} (\bibinfo {year} {2018}{\natexlab{a}})},\ \Eprint
  {https://arxiv.org/abs/1806.08385} {arXiv:1806.08385 [astro-ph.GA]}
  \BibitemShut {NoStop}%
\bibitem [{\citenamefont {{Sakurai}}\ \emph {et~al.}(2019)\citenamefont
  {{Sakurai}}, \citenamefont {{Yoshida}},\ and\ \citenamefont
  {{Fujii}}}]{2019MNRAS.484.4665S}%
  \BibitemOpen
  \bibfield  {author} {\bibinfo {author} {\bibfnamefont {Y.}~\bibnamefont
  {{Sakurai}}}, \bibinfo {author} {\bibfnamefont {N.}~\bibnamefont
  {{Yoshida}}},\ and\ \bibinfo {author} {\bibfnamefont {M.~S.}\ \bibnamefont
  {{Fujii}}},\ }\href {https://doi.org/10.1093/mnras/stz315} {\bibfield
  {journal} {\bibinfo  {journal} {\mnras}\ }\textbf {\bibinfo {volume} {484}},\
  \bibinfo {pages} {4665} (\bibinfo {year} {2019})},\ \Eprint
  {https://arxiv.org/abs/1810.01985} {arXiv:1810.01985 [astro-ph.GA]}
  \BibitemShut {NoStop}%
\bibitem [{\citenamefont {{Wen}}\ \emph {et~al.}(2021)\citenamefont {{Wen}},
  \citenamefont {{Jonker}}, \citenamefont {{Stone}},\ and\ \citenamefont
  {{Zabludoff}}}]{2021ApJ...918...46W}%
  \BibitemOpen
  \bibfield  {author} {\bibinfo {author} {\bibfnamefont {S.}~\bibnamefont
  {{Wen}}}, \bibinfo {author} {\bibfnamefont {P.~G.}\ \bibnamefont {{Jonker}}},
  \bibinfo {author} {\bibfnamefont {N.~C.}\ \bibnamefont {{Stone}}},\ and\
  \bibinfo {author} {\bibfnamefont {A.~I.}\ \bibnamefont {{Zabludoff}}},\
  }\href {https://doi.org/10.3847/1538-4357/ac00b5} {\bibfield  {journal}
  {\bibinfo  {journal} {\apj}\ }\textbf {\bibinfo {volume} {918}},\ \bibinfo
  {eid} {46} (\bibinfo {year} {2021})},\ \Eprint
  {https://arxiv.org/abs/2104.01498} {arXiv:2104.01498 [astro-ph.HE]}
  \BibitemShut {NoStop}%
\bibitem [{\citenamefont {{Coughlin}}\ and\ \citenamefont
  {{Nixon}}(2015)}]{2015ApJ...808L..11C}%
  \BibitemOpen
  \bibfield  {author} {\bibinfo {author} {\bibfnamefont {E.~R.}\ \bibnamefont
  {{Coughlin}}}\ and\ \bibinfo {author} {\bibfnamefont {C.}~\bibnamefont
  {{Nixon}}},\ }\href {https://doi.org/10.1088/2041-8205/808/1/L11} {\bibfield
  {journal} {\bibinfo  {journal} {\apjl}\ }\textbf {\bibinfo {volume} {808}},\
  \bibinfo {eid} {L11} (\bibinfo {year} {2015})},\ \Eprint
  {https://arxiv.org/abs/1506.08194} {arXiv:1506.08194 [astro-ph.HE]}
  \BibitemShut {NoStop}%
\bibitem [{\citenamefont {{Bonnerot}}\ \emph {et~al.}(2021)\citenamefont
  {{Bonnerot}}, \citenamefont {{Lu}},\ and\ \citenamefont
  {{Hopkins}}}]{2021MNRAS.504.4885B}%
  \BibitemOpen
  \bibfield  {author} {\bibinfo {author} {\bibfnamefont {C.}~\bibnamefont
  {{Bonnerot}}}, \bibinfo {author} {\bibfnamefont {W.}~\bibnamefont {{Lu}}},\
  and\ \bibinfo {author} {\bibfnamefont {P.~F.}\ \bibnamefont {{Hopkins}}},\
  }\href {https://doi.org/10.1093/mnras/stab398} {\bibfield  {journal}
  {\bibinfo  {journal} {\mnras}\ }\textbf {\bibinfo {volume} {504}},\ \bibinfo
  {pages} {4885} (\bibinfo {year} {2021})},\ \Eprint
  {https://arxiv.org/abs/2012.12271} {arXiv:2012.12271 [astro-ph.HE]}
  \BibitemShut {NoStop}%
\bibitem [{\citenamefont {{Gezari}}(2021)}]{2021ARA&A..59...21G}%
  \BibitemOpen
  \bibfield  {author} {\bibinfo {author} {\bibfnamefont {S.}~\bibnamefont
  {{Gezari}}},\ }\bibfield  {journal} {\bibinfo  {journal} {\araa}\ }\textbf
  {\bibinfo {volume} {59}},\ \href
  {https://doi.org/10.1146/annurev-astro-111720-030029}
  {10.1146/annurev-astro-111720-030029} (\bibinfo {year} {2021}),\ \Eprint
  {https://arxiv.org/abs/2104.14580} {arXiv:2104.14580 [astro-ph.HE]}
  \BibitemShut {NoStop}%
\bibitem [{\citenamefont {{Andalman}}\ \emph {et~al.}(2021)\citenamefont
  {{Andalman}}, \citenamefont {{Liska}}, \citenamefont {{Tchekhovskoy}},
  \citenamefont {{Coughlin}},\ and\ \citenamefont
  {{Stone}}}]{2021MNRAS.tmp.3135A}%
  \BibitemOpen
  \bibfield  {author} {\bibinfo {author} {\bibfnamefont {Z.~L.}\ \bibnamefont
  {{Andalman}}}, \bibinfo {author} {\bibfnamefont {M.~T.~P.}\ \bibnamefont
  {{Liska}}}, \bibinfo {author} {\bibfnamefont {A.}~\bibnamefont
  {{Tchekhovskoy}}}, \bibinfo {author} {\bibfnamefont {E.~R.}\ \bibnamefont
  {{Coughlin}}},\ and\ \bibinfo {author} {\bibfnamefont {N.}~\bibnamefont
  {{Stone}}},\ }\bibfield  {journal} {\bibinfo  {journal} {\mnras}\ }\href
  {https://doi.org/10.1093/mnras/stab3444} {10.1093/mnras/stab3444} (\bibinfo
  {year} {2021}),\ \Eprint {https://arxiv.org/abs/2008.04922} {arXiv:2008.04922
  [astro-ph.HE]} \BibitemShut {NoStop}%
\bibitem [{\citenamefont {{Kocsis}}\ \emph {et~al.}(2012)\citenamefont
  {{Kocsis}}, \citenamefont {{Ray}},\ and\ \citenamefont {{Portegies
  Zwart}}}]{2012ApJ...752...67K}%
  \BibitemOpen
  \bibfield  {author} {\bibinfo {author} {\bibfnamefont {B.}~\bibnamefont
  {{Kocsis}}}, \bibinfo {author} {\bibfnamefont {A.}~\bibnamefont {{Ray}}},\
  and\ \bibinfo {author} {\bibfnamefont {S.}~\bibnamefont {{Portegies
  Zwart}}},\ }\href {https://doi.org/10.1088/0004-637X/752/1/67} {\bibfield
  {journal} {\bibinfo  {journal} {\apj}\ }\textbf {\bibinfo {volume} {752}},\
  \bibinfo {eid} {67} (\bibinfo {year} {2012})},\ \Eprint
  {https://arxiv.org/abs/1110.6172} {arXiv:1110.6172 [astro-ph.GA]}
  \BibitemShut {NoStop}%
\bibitem [{\citenamefont {{Amaro-Seoane}}\ \emph {et~al.}(2007)\citenamefont
  {{Amaro-Seoane}}, \citenamefont {{Gair}}, \citenamefont {{Freitag}},
  \citenamefont {{Miller}}, \citenamefont {{Mandel}}, \citenamefont
  {{Cutler}},\ and\ \citenamefont {{Babak}}}]{2007CQGra..24R.113A}%
  \BibitemOpen
  \bibfield  {author} {\bibinfo {author} {\bibfnamefont {P.}~\bibnamefont
  {{Amaro-Seoane}}}, \bibinfo {author} {\bibfnamefont {J.~R.}\ \bibnamefont
  {{Gair}}}, \bibinfo {author} {\bibfnamefont {M.}~\bibnamefont {{Freitag}}},
  \bibinfo {author} {\bibfnamefont {M.~C.}\ \bibnamefont {{Miller}}}, \bibinfo
  {author} {\bibfnamefont {I.}~\bibnamefont {{Mandel}}}, \bibinfo {author}
  {\bibfnamefont {C.~J.}\ \bibnamefont {{Cutler}}},\ and\ \bibinfo {author}
  {\bibfnamefont {S.}~\bibnamefont {{Babak}}},\ }\href
  {https://doi.org/10.1088/0264-9381/24/17/R01} {\bibfield  {journal} {\bibinfo
   {journal} {Classical and Quantum Gravity}\ }\textbf {\bibinfo {volume}
  {24}},\ \bibinfo {pages} {R113} (\bibinfo {year} {2007})},\ \Eprint
  {https://arxiv.org/abs/astro-ph/0703495} {arXiv:astro-ph/0703495 [astro-ph]}
  \BibitemShut {NoStop}%
\bibitem [{\citenamefont {{Cutler}}\ \emph {et~al.}(2019)\citenamefont
  {{Cutler}}, \citenamefont {{Berti}}, \citenamefont {{Holley-Bockelmann}},
  \citenamefont {{Jani}}, \citenamefont {{Kovetz}}, \citenamefont {{Larson}},
  \citenamefont {{Littenberg}}, \citenamefont {{McWilliams}}, \citenamefont
  {{Mueller}}, \citenamefont {{Randall}}, \citenamefont {{Schnittman}},
  \citenamefont {{Shoemaker}}, \citenamefont {{Vallisneri}}, \citenamefont
  {{Vitale}},\ and\ \citenamefont {{Wong}}}]{2019BAAS...51c.109C}%
  \BibitemOpen
  \bibfield  {author} {\bibinfo {author} {\bibfnamefont {C.}~\bibnamefont
  {{Cutler}}}, \bibinfo {author} {\bibfnamefont {E.}~\bibnamefont {{Berti}}},
  \bibinfo {author} {\bibfnamefont {K.}~\bibnamefont {{Holley-Bockelmann}}},
  \bibinfo {author} {\bibfnamefont {K.}~\bibnamefont {{Jani}}}, \bibinfo
  {author} {\bibfnamefont {E.~D.}\ \bibnamefont {{Kovetz}}}, \bibinfo {author}
  {\bibfnamefont {S.~L.}\ \bibnamefont {{Larson}}}, \bibinfo {author}
  {\bibfnamefont {T.}~\bibnamefont {{Littenberg}}}, \bibinfo {author}
  {\bibfnamefont {S.~T.}\ \bibnamefont {{McWilliams}}}, \bibinfo {author}
  {\bibfnamefont {G.}~\bibnamefont {{Mueller}}}, \bibinfo {author}
  {\bibfnamefont {L.}~\bibnamefont {{Randall}}}, \bibinfo {author}
  {\bibfnamefont {J.~D.}\ \bibnamefont {{Schnittman}}}, \bibinfo {author}
  {\bibfnamefont {D.~H.}\ \bibnamefont {{Shoemaker}}}, \bibinfo {author}
  {\bibfnamefont {M.}~\bibnamefont {{Vallisneri}}}, \bibinfo {author}
  {\bibfnamefont {S.}~\bibnamefont {{Vitale}}},\ and\ \bibinfo {author}
  {\bibfnamefont {K.~W.~K.}\ \bibnamefont {{Wong}}},\ }\href@noop {} {\bibfield
   {journal} {\bibinfo  {journal} {\baas}\ }\textbf {\bibinfo {volume} {51}},\
  \bibinfo {eid} {109} (\bibinfo {year} {2019})},\ \Eprint
  {https://arxiv.org/abs/1903.04069} {arXiv:1903.04069 [astro-ph.HE]}
  \BibitemShut {NoStop}%
\bibitem [{\citenamefont {{Arca Sedda}}\ \emph {et~al.}(2020)\citenamefont
  {{Arca Sedda}}, \citenamefont {{Berry}}, \citenamefont {{Jani}},
  \citenamefont {{Amaro-Seoane}}, \citenamefont {{Auclair}}, \citenamefont
  {{Baird}}, \citenamefont {{Baker}}, \citenamefont {{Berti}}, \citenamefont
  {{Breivik}}, \citenamefont {{Burrows}}, \citenamefont {{Caprini}},
  \citenamefont {{Chen}}, \citenamefont {{Doneva}}, \citenamefont {{Ezquiaga}},
  \citenamefont {{Saavik Ford}}, \citenamefont {{Katz}}, \citenamefont
  {{Kolkowitz}}, \citenamefont {{McKernan}}, \citenamefont {{Mueller}},
  \citenamefont {{Nardini}}, \citenamefont {{Pikovski}}, \citenamefont
  {{Rajendran}}, \citenamefont {{Sesana}}, \citenamefont {{Shao}},
  \citenamefont {{Tamanini}}, \citenamefont {{Vartanyan}}, \citenamefont
  {{Warburton}}, \citenamefont {{Witek}}, \citenamefont {{Wong}},\ and\
  \citenamefont {{Zevin}}}]{2020CQGra..37u5011A}%
  \BibitemOpen
  \bibfield  {author} {\bibinfo {author} {\bibfnamefont {M.}~\bibnamefont
  {{Arca Sedda}}}, \bibinfo {author} {\bibfnamefont {C.~P.~L.}\ \bibnamefont
  {{Berry}}}, \bibinfo {author} {\bibfnamefont {K.}~\bibnamefont {{Jani}}},
  \bibinfo {author} {\bibfnamefont {P.}~\bibnamefont {{Amaro-Seoane}}},
  \bibinfo {author} {\bibfnamefont {P.}~\bibnamefont {{Auclair}}}, \bibinfo
  {author} {\bibfnamefont {J.}~\bibnamefont {{Baird}}}, \bibinfo {author}
  {\bibfnamefont {T.}~\bibnamefont {{Baker}}}, \bibinfo {author} {\bibfnamefont
  {E.}~\bibnamefont {{Berti}}}, \bibinfo {author} {\bibfnamefont
  {K.}~\bibnamefont {{Breivik}}}, \bibinfo {author} {\bibfnamefont
  {A.}~\bibnamefont {{Burrows}}}, \bibinfo {author} {\bibfnamefont
  {C.}~\bibnamefont {{Caprini}}}, \bibinfo {author} {\bibfnamefont
  {X.}~\bibnamefont {{Chen}}}, \bibinfo {author} {\bibfnamefont
  {D.}~\bibnamefont {{Doneva}}}, \bibinfo {author} {\bibfnamefont {J.~M.}\
  \bibnamefont {{Ezquiaga}}}, \bibinfo {author} {\bibfnamefont {K.~E.}\
  \bibnamefont {{Saavik Ford}}}, \bibinfo {author} {\bibfnamefont {M.~L.}\
  \bibnamefont {{Katz}}}, \bibinfo {author} {\bibfnamefont {S.}~\bibnamefont
  {{Kolkowitz}}}, \bibinfo {author} {\bibfnamefont {B.}~\bibnamefont
  {{McKernan}}}, \bibinfo {author} {\bibfnamefont {G.}~\bibnamefont
  {{Mueller}}}, \bibinfo {author} {\bibfnamefont {G.}~\bibnamefont
  {{Nardini}}}, \bibinfo {author} {\bibfnamefont {I.}~\bibnamefont
  {{Pikovski}}}, \bibinfo {author} {\bibfnamefont {S.}~\bibnamefont
  {{Rajendran}}}, \bibinfo {author} {\bibfnamefont {A.}~\bibnamefont
  {{Sesana}}}, \bibinfo {author} {\bibfnamefont {L.}~\bibnamefont {{Shao}}},
  \bibinfo {author} {\bibfnamefont {N.}~\bibnamefont {{Tamanini}}}, \bibinfo
  {author} {\bibfnamefont {D.}~\bibnamefont {{Vartanyan}}}, \bibinfo {author}
  {\bibfnamefont {N.}~\bibnamefont {{Warburton}}}, \bibinfo {author}
  {\bibfnamefont {H.}~\bibnamefont {{Witek}}}, \bibinfo {author} {\bibfnamefont
  {K.}~\bibnamefont {{Wong}}},\ and\ \bibinfo {author} {\bibfnamefont
  {M.}~\bibnamefont {{Zevin}}},\ }\href
  {https://doi.org/10.1088/1361-6382/abb5c1} {\bibfield  {journal} {\bibinfo
  {journal} {Classical and Quantum Gravity}\ }\textbf {\bibinfo {volume}
  {37}},\ \bibinfo {eid} {215011} (\bibinfo {year} {2020})},\ \Eprint
  {https://arxiv.org/abs/1908.11375} {arXiv:1908.11375 [gr-qc]} \BibitemShut
  {NoStop}%
\bibitem [{\citenamefont {{Mandel}}\ \emph {et~al.}(2008)\citenamefont
  {{Mandel}}, \citenamefont {{Brown}}, \citenamefont {{Gair}},\ and\
  \citenamefont {{Miller}}}]{2008ApJ...681.1431M}%
  \BibitemOpen
  \bibfield  {author} {\bibinfo {author} {\bibfnamefont {I.}~\bibnamefont
  {{Mandel}}}, \bibinfo {author} {\bibfnamefont {D.~A.}\ \bibnamefont
  {{Brown}}}, \bibinfo {author} {\bibfnamefont {J.~R.}\ \bibnamefont
  {{Gair}}},\ and\ \bibinfo {author} {\bibfnamefont {M.~C.}\ \bibnamefont
  {{Miller}}},\ }\href {https://doi.org/10.1086/588246} {\bibfield  {journal}
  {\bibinfo  {journal} {\apj}\ }\textbf {\bibinfo {volume} {681}},\ \bibinfo
  {pages} {1431} (\bibinfo {year} {2008})},\ \Eprint
  {https://arxiv.org/abs/0705.0285} {arXiv:0705.0285 [astro-ph]} \BibitemShut
  {NoStop}%
\bibitem [{\citenamefont {{Leigh}}\ \emph {et~al.}(2014)\citenamefont
  {{Leigh}}, \citenamefont {{L{\"u}tzgendorf}}, \citenamefont {{Geller}},
  \citenamefont {{Maccarone}}, \citenamefont {{Heinke}},\ and\ \citenamefont
  {{Sesana}}}]{2014MNRAS.444...29L}%
  \BibitemOpen
  \bibfield  {author} {\bibinfo {author} {\bibfnamefont {N.~W.~C.}\
  \bibnamefont {{Leigh}}}, \bibinfo {author} {\bibfnamefont {N.}~\bibnamefont
  {{L{\"u}tzgendorf}}}, \bibinfo {author} {\bibfnamefont {A.~M.}\ \bibnamefont
  {{Geller}}}, \bibinfo {author} {\bibfnamefont {T.~J.}\ \bibnamefont
  {{Maccarone}}}, \bibinfo {author} {\bibfnamefont {C.}~\bibnamefont
  {{Heinke}}},\ and\ \bibinfo {author} {\bibfnamefont {A.}~\bibnamefont
  {{Sesana}}},\ }\href {https://doi.org/10.1093/mnras/stu1437} {\bibfield
  {journal} {\bibinfo  {journal} {\mnras}\ }\textbf {\bibinfo {volume} {444}},\
  \bibinfo {pages} {29} (\bibinfo {year} {2014})},\ \Eprint
  {https://arxiv.org/abs/1407.4459} {arXiv:1407.4459 [astro-ph.SR]}
  \BibitemShut {NoStop}%
\bibitem [{\citenamefont {{Haster}}\ \emph {et~al.}(2016)\citenamefont
  {{Haster}}, \citenamefont {{Antonini}}, \citenamefont {{Kalogera}},\ and\
  \citenamefont {{Mandel}}}]{2016ApJ...832..192H}%
  \BibitemOpen
  \bibfield  {author} {\bibinfo {author} {\bibfnamefont {C.-J.}\ \bibnamefont
  {{Haster}}}, \bibinfo {author} {\bibfnamefont {F.}~\bibnamefont
  {{Antonini}}}, \bibinfo {author} {\bibfnamefont {V.}~\bibnamefont
  {{Kalogera}}},\ and\ \bibinfo {author} {\bibfnamefont {I.}~\bibnamefont
  {{Mandel}}},\ }\href {https://doi.org/10.3847/0004-637X/832/2/192} {\bibfield
   {journal} {\bibinfo  {journal} {\apj}\ }\textbf {\bibinfo {volume} {832}},\
  \bibinfo {eid} {192} (\bibinfo {year} {2016})},\ \Eprint
  {https://arxiv.org/abs/1606.07097} {arXiv:1606.07097 [astro-ph.HE]}
  \BibitemShut {NoStop}%
\bibitem [{\citenamefont {{Fragione}}\ \emph
  {et~al.}(2018{\natexlab{b}})\citenamefont {{Fragione}}, \citenamefont
  {{Ginsburg}},\ and\ \citenamefont {{Kocsis}}}]{2018ApJ...856...92F}%
  \BibitemOpen
  \bibfield  {author} {\bibinfo {author} {\bibfnamefont {G.}~\bibnamefont
  {{Fragione}}}, \bibinfo {author} {\bibfnamefont {I.}~\bibnamefont
  {{Ginsburg}}},\ and\ \bibinfo {author} {\bibfnamefont {B.}~\bibnamefont
  {{Kocsis}}},\ }\href {https://doi.org/10.3847/1538-4357/aab368} {\bibfield
  {journal} {\bibinfo  {journal} {\apj}\ }\textbf {\bibinfo {volume} {856}},\
  \bibinfo {pages} {92} (\bibinfo {year} {2018}{\natexlab{b}})}\BibitemShut
  {NoStop}%
\bibitem [{\citenamefont {{Jani}}\ \emph {et~al.}(2020)\citenamefont {{Jani}},
  \citenamefont {{Shoemaker}},\ and\ \citenamefont
  {{Cutler}}}]{2020NatAs...4..260J}%
  \BibitemOpen
  \bibfield  {author} {\bibinfo {author} {\bibfnamefont {K.}~\bibnamefont
  {{Jani}}}, \bibinfo {author} {\bibfnamefont {D.}~\bibnamefont
  {{Shoemaker}}},\ and\ \bibinfo {author} {\bibfnamefont {C.}~\bibnamefont
  {{Cutler}}},\ }\href {https://doi.org/10.1038/s41550-019-0932-7} {\bibfield
  {journal} {\bibinfo  {journal} {Nature Astronomy}\ }\textbf {\bibinfo
  {volume} {4}},\ \bibinfo {pages} {260} (\bibinfo {year} {2020})},\ \Eprint
  {https://arxiv.org/abs/1908.04985} {arXiv:1908.04985 [gr-qc]} \BibitemShut
  {NoStop}%
\bibitem [{\citenamefont {Rasskazov}\ \emph {et~al.}(2020)\citenamefont
  {Rasskazov}, \citenamefont {Fragione},\ and\ \citenamefont
  {Kocsis}}]{2020ApJ...899..149R}%
  \BibitemOpen
  \bibfield  {author} {\bibinfo {author} {\bibfnamefont {A.}~\bibnamefont
  {Rasskazov}}, \bibinfo {author} {\bibfnamefont {G.}~\bibnamefont
  {Fragione}},\ and\ \bibinfo {author} {\bibfnamefont {B.}~\bibnamefont
  {Kocsis}},\ }\href {https://doi.org/10.3847/1538-4357/aba2f4} {\bibfield
  {journal} {\bibinfo  {journal} {The Astrophysical Journal}\ }\textbf
  {\bibinfo {volume} {899}},\ \bibinfo {pages} {149} (\bibinfo {year}
  {2020})}\BibitemShut {NoStop}%
\bibitem [{\citenamefont {{Arca Sedda}}\ \emph {et~al.}(2021)\citenamefont
  {{Arca Sedda}}, \citenamefont {{Amaro Seoane}},\ and\ \citenamefont
  {{Chen}}}]{2021A&A...652A..54A}%
  \BibitemOpen
  \bibfield  {author} {\bibinfo {author} {\bibfnamefont {M.}~\bibnamefont
  {{Arca Sedda}}}, \bibinfo {author} {\bibfnamefont {P.}~\bibnamefont {{Amaro
  Seoane}}},\ and\ \bibinfo {author} {\bibfnamefont {X.}~\bibnamefont
  {{Chen}}},\ }\href {https://doi.org/10.1051/0004-6361/202037785} {\bibfield
  {journal} {\bibinfo  {journal} {\aap}\ }\textbf {\bibinfo {volume} {652}},\
  \bibinfo {eid} {A54} (\bibinfo {year} {2021})},\ \Eprint
  {https://arxiv.org/abs/2007.13746} {arXiv:2007.13746 [astro-ph.GA]}
  \BibitemShut {NoStop}%
\bibitem [{\citenamefont {Abbott}\ \emph {et~al.}(2020)\citenamefont {Abbott}
  \emph {et~al.}}]{2020PhRvL.125j1102A}%
  \BibitemOpen
  \bibfield  {author} {\bibinfo {author} {\bibfnamefont {R.}~\bibnamefont
  {Abbott}} \emph {et~al.},\ }\bibfield  {journal} {\bibinfo  {journal}
  {Physical Review Letters}\ }\textbf {\bibinfo {volume} {125}},\ \href
  {https://doi.org/10.1103/physrevlett.125.101102}
  {10.1103/physrevlett.125.101102} (\bibinfo {year} {2020})\BibitemShut
  {NoStop}%
\bibitem [{\citenamefont {{Abbott}}\ \emph {et~al.}(2021)\citenamefont
  {{Abbott}}, \citenamefont {{Abbott}}, \citenamefont {{Abraham}},
  \citenamefont {{Acernese}}, \citenamefont {{Ackley}}, \citenamefont
  {{Adams}}, \citenamefont {{Adams}}, \citenamefont {{Adhikari}}, \citenamefont
  {{Adya}}, \citenamefont {{Affeldt}},\ and\ \citenamefont
  {et~al.}}]{2021PhRvX..11b1053A}%
  \BibitemOpen
  \bibfield  {author} {\bibinfo {author} {\bibfnamefont {R.}~\bibnamefont
  {{Abbott}}}, \bibinfo {author} {\bibfnamefont {T.~D.}\ \bibnamefont
  {{Abbott}}}, \bibinfo {author} {\bibfnamefont {S.}~\bibnamefont {{Abraham}}},
  \bibinfo {author} {\bibfnamefont {F.}~\bibnamefont {{Acernese}}}, \bibinfo
  {author} {\bibfnamefont {K.}~\bibnamefont {{Ackley}}}, \bibinfo {author}
  {\bibfnamefont {A.}~\bibnamefont {{Adams}}}, \bibinfo {author} {\bibfnamefont
  {C.}~\bibnamefont {{Adams}}}, \bibinfo {author} {\bibfnamefont {R.~X.}\
  \bibnamefont {{Adhikari}}}, \bibinfo {author} {\bibfnamefont {V.~B.}\
  \bibnamefont {{Adya}}}, \bibinfo {author} {\bibfnamefont {C.}~\bibnamefont
  {{Affeldt}}},\ and\ \bibinfo {author} {\bibnamefont {et~al.}},\ }\href
  {https://doi.org/10.1103/PhysRevX.11.021053} {\bibfield  {journal} {\bibinfo
  {journal} {Physical Review X}\ }\textbf {\bibinfo {volume} {11}},\ \bibinfo
  {eid} {021053} (\bibinfo {year} {2021})},\ \Eprint
  {https://arxiv.org/abs/2010.14527} {arXiv:2010.14527 [gr-qc]} \BibitemShut
  {NoStop}%
\bibitem [{\citenamefont {{The LIGO Scientific Collaboration}}\ \emph
  {et~al.}(2021)\citenamefont {{The LIGO Scientific Collaboration}},
  \citenamefont {{the Virgo Collaboration}}, \citenamefont {{the KAGRA
  Collaboration}}, \citenamefont {{Abbott}}, \citenamefont {{Abbott}},
  \citenamefont {{Acernese}}, \citenamefont {{Ackley}}, \citenamefont
  {{Adams}}, \citenamefont {{Adhikari}}, \citenamefont {{Adhikari}},\ and\
  \citenamefont {et~al.}}]{2021arXiv211103606T}%
  \BibitemOpen
  \bibfield  {author} {\bibinfo {author} {\bibnamefont {{The LIGO Scientific
  Collaboration}}}, \bibinfo {author} {\bibnamefont {{the Virgo
  Collaboration}}}, \bibinfo {author} {\bibnamefont {{the KAGRA
  Collaboration}}}, \bibinfo {author} {\bibfnamefont {R.}~\bibnamefont
  {{Abbott}}}, \bibinfo {author} {\bibfnamefont {T.~D.}\ \bibnamefont
  {{Abbott}}}, \bibinfo {author} {\bibfnamefont {F.}~\bibnamefont
  {{Acernese}}}, \bibinfo {author} {\bibfnamefont {K.}~\bibnamefont
  {{Ackley}}}, \bibinfo {author} {\bibfnamefont {C.}~\bibnamefont {{Adams}}},
  \bibinfo {author} {\bibfnamefont {N.}~\bibnamefont {{Adhikari}}}, \bibinfo
  {author} {\bibfnamefont {R.~X.}\ \bibnamefont {{Adhikari}}},\ and\ \bibinfo
  {author} {\bibnamefont {et~al.}},\ }\href@noop {} {\bibfield  {journal}
  {\bibinfo  {journal} {arXiv e-prints}\ ,\ \bibinfo {eid} {arXiv:2111.03606}}
  (\bibinfo {year} {2021})},\ \Eprint {https://arxiv.org/abs/2111.03606}
  {arXiv:2111.03606 [gr-qc]} \BibitemShut {NoStop}%
\bibitem [{\citenamefont {Wong}\ \emph {et~al.}(2019)\citenamefont {Wong},
  \citenamefont {Baibhav},\ and\ \citenamefont {Berti}}]{2019MNRAS.488.5665W}%
  \BibitemOpen
  \bibfield  {author} {\bibinfo {author} {\bibfnamefont {K.~W.~K.}\
  \bibnamefont {Wong}}, \bibinfo {author} {\bibfnamefont {V.}~\bibnamefont
  {Baibhav}},\ and\ \bibinfo {author} {\bibfnamefont {E.}~\bibnamefont
  {Berti}},\ }\href {https://doi.org/10.1093/mnras/stz2077} {\bibfield
  {journal} {\bibinfo  {journal} {Monthly Notices of the Royal Astronomical
  Society}\ }\textbf {\bibinfo {volume} {488}},\ \bibinfo {pages} {5665}
  (\bibinfo {year} {2019})}\BibitemShut {NoStop}%
\bibitem [{\citenamefont {{Inayoshi}}\ \emph {et~al.}(2017)\citenamefont
  {{Inayoshi}}, \citenamefont {{Tamanini}}, \citenamefont {{Caprini}},\ and\
  \citenamefont {{Haiman}}}]{2017PhRvD..96f3014I}%
  \BibitemOpen
  \bibfield  {author} {\bibinfo {author} {\bibfnamefont {K.}~\bibnamefont
  {{Inayoshi}}}, \bibinfo {author} {\bibfnamefont {N.}~\bibnamefont
  {{Tamanini}}}, \bibinfo {author} {\bibfnamefont {C.}~\bibnamefont
  {{Caprini}}},\ and\ \bibinfo {author} {\bibfnamefont {Z.}~\bibnamefont
  {{Haiman}}},\ }\href {https://doi.org/10.1103/PhysRevD.96.063014} {\bibfield
  {journal} {\bibinfo  {journal} {\prd}\ }\textbf {\bibinfo {volume} {96}},\
  \bibinfo {eid} {063014} (\bibinfo {year} {2017})},\ \Eprint
  {https://arxiv.org/abs/1702.06529} {arXiv:1702.06529 [astro-ph.HE]}
  \BibitemShut {NoStop}%
\bibitem [{\citenamefont {{Bonvin}}\ \emph {et~al.}(2017)\citenamefont
  {{Bonvin}}, \citenamefont {{Caprini}}, \citenamefont {{Sturani}},\ and\
  \citenamefont {{Tamanini}}}]{2017PhRvD..95d4029B}%
  \BibitemOpen
  \bibfield  {author} {\bibinfo {author} {\bibfnamefont {C.}~\bibnamefont
  {{Bonvin}}}, \bibinfo {author} {\bibfnamefont {C.}~\bibnamefont {{Caprini}}},
  \bibinfo {author} {\bibfnamefont {R.}~\bibnamefont {{Sturani}}},\ and\
  \bibinfo {author} {\bibfnamefont {N.}~\bibnamefont {{Tamanini}}},\ }\href
  {https://doi.org/10.1103/PhysRevD.95.044029} {\bibfield  {journal} {\bibinfo
  {journal} {\prd}\ }\textbf {\bibinfo {volume} {95}},\ \bibinfo {eid} {044029}
  (\bibinfo {year} {2017})},\ \Eprint {https://arxiv.org/abs/1609.08093}
  {arXiv:1609.08093 [astro-ph.CO]} \BibitemShut {NoStop}%
\bibitem [{\citenamefont {Tamanini}\ \emph {et~al.}(2020)\citenamefont
  {Tamanini}, \citenamefont {Klein}, \citenamefont {Bonvin}, \citenamefont
  {Barausse},\ and\ \citenamefont {Caprini}}]{2020PhRvD.101f3002T}%
  \BibitemOpen
  \bibfield  {author} {\bibinfo {author} {\bibfnamefont {N.}~\bibnamefont
  {Tamanini}}, \bibinfo {author} {\bibfnamefont {A.}~\bibnamefont {Klein}},
  \bibinfo {author} {\bibfnamefont {C.}~\bibnamefont {Bonvin}}, \bibinfo
  {author} {\bibfnamefont {E.}~\bibnamefont {Barausse}},\ and\ \bibinfo
  {author} {\bibfnamefont {C.}~\bibnamefont {Caprini}},\ }\bibfield  {journal}
  {\bibinfo  {journal} {Physical Review D}\ }\textbf {\bibinfo {volume}
  {101}},\ \href {https://doi.org/10.1103/physrevd.101.063002}
  {10.1103/physrevd.101.063002} (\bibinfo {year} {2020})\BibitemShut {NoStop}%
\bibitem [{\citenamefont {{Xuan}}\ \emph {et~al.}(2021)\citenamefont {{Xuan}},
  \citenamefont {{Peng}},\ and\ \citenamefont {{Chen}}}]{2021MNRAS.502.4199X}%
  \BibitemOpen
  \bibfield  {author} {\bibinfo {author} {\bibfnamefont {Z.}~\bibnamefont
  {{Xuan}}}, \bibinfo {author} {\bibfnamefont {P.}~\bibnamefont {{Peng}}},\
  and\ \bibinfo {author} {\bibfnamefont {X.}~\bibnamefont {{Chen}}},\ }\href
  {https://doi.org/10.1093/mnras/stab331} {\bibfield  {journal} {\bibinfo
  {journal} {\mnras}\ }\textbf {\bibinfo {volume} {502}},\ \bibinfo {pages}
  {4199} (\bibinfo {year} {2021})},\ \Eprint {https://arxiv.org/abs/2012.00049}
  {arXiv:2012.00049 [astro-ph.HE]} \BibitemShut {NoStop}%
\bibitem [{\citenamefont {{Yunes}}\ \emph {et~al.}(2011)\citenamefont
  {{Yunes}}, \citenamefont {{Miller}},\ and\ \citenamefont
  {{Thornburg}}}]{2011PhRvD..83d4030Y}%
  \BibitemOpen
  \bibfield  {author} {\bibinfo {author} {\bibfnamefont {N.}~\bibnamefont
  {{Yunes}}}, \bibinfo {author} {\bibfnamefont {M.~C.}\ \bibnamefont
  {{Miller}}},\ and\ \bibinfo {author} {\bibfnamefont {J.}~\bibnamefont
  {{Thornburg}}},\ }\href {https://doi.org/10.1103/PhysRevD.83.044030}
  {\bibfield  {journal} {\bibinfo  {journal} {\prd}\ }\textbf {\bibinfo
  {volume} {83}},\ \bibinfo {eid} {044030} (\bibinfo {year} {2011})},\ \Eprint
  {https://arxiv.org/abs/1010.1721} {arXiv:1010.1721 [astro-ph.GA]}
  \BibitemShut {NoStop}%
\bibitem [{\citenamefont {{Meiron}}\ \emph {et~al.}(2017)\citenamefont
  {{Meiron}}, \citenamefont {{Kocsis}},\ and\ \citenamefont
  {{Loeb}}}]{2017ApJ...834..200M}%
  \BibitemOpen
  \bibfield  {author} {\bibinfo {author} {\bibfnamefont {Y.}~\bibnamefont
  {{Meiron}}}, \bibinfo {author} {\bibfnamefont {B.}~\bibnamefont {{Kocsis}}},\
  and\ \bibinfo {author} {\bibfnamefont {A.}~\bibnamefont {{Loeb}}},\ }\href
  {https://doi.org/10.3847/1538-4357/834/2/200} {\bibfield  {journal} {\bibinfo
   {journal} {\apj}\ }\textbf {\bibinfo {volume} {834}},\ \bibinfo {eid} {200}
  (\bibinfo {year} {2017})},\ \Eprint {https://arxiv.org/abs/1604.02148}
  {arXiv:1604.02148 [astro-ph.HE]} \BibitemShut {NoStop}%
\bibitem [{\citenamefont {Cutler}(1998)}]{1998PhRvD..57.7089C}%
  \BibitemOpen
  \bibfield  {author} {\bibinfo {author} {\bibfnamefont {C.}~\bibnamefont
  {Cutler}},\ }\href {https://doi.org/10.1103/physrevd.57.7089} {\bibfield
  {journal} {\bibinfo  {journal} {Physical Review D}\ }\textbf {\bibinfo
  {volume} {57}},\ \bibinfo {pages} {7089} (\bibinfo {year}
  {1998})}\BibitemShut {NoStop}%
\bibitem [{\citenamefont {Berti}\ \emph {et~al.}(2005)\citenamefont {Berti},
  \citenamefont {Buonanno},\ and\ \citenamefont {Will}}]{2005PhRvD..71h4025B}%
  \BibitemOpen
  \bibfield  {author} {\bibinfo {author} {\bibfnamefont {E.}~\bibnamefont
  {Berti}}, \bibinfo {author} {\bibfnamefont {A.}~\bibnamefont {Buonanno}},\
  and\ \bibinfo {author} {\bibfnamefont {C.~M.}\ \bibnamefont {Will}},\
  }\bibfield  {journal} {\bibinfo  {journal} {Physical Review D}\ }\textbf
  {\bibinfo {volume} {71}},\ \href {https://doi.org/10.1103/physrevd.71.084025}
  {10.1103/physrevd.71.084025} (\bibinfo {year} {2005})\BibitemShut {NoStop}%
\bibitem [{\citenamefont {{Stavridis}}\ and\ \citenamefont
  {{Will}}(2009)}]{2009PhRvD..80d4002S}%
  \BibitemOpen
  \bibfield  {author} {\bibinfo {author} {\bibfnamefont {A.}~\bibnamefont
  {{Stavridis}}}\ and\ \bibinfo {author} {\bibfnamefont {C.~M.}\ \bibnamefont
  {{Will}}},\ }\href {https://doi.org/10.1103/PhysRevD.80.044002} {\bibfield
  {journal} {\bibinfo  {journal} {\prd}\ }\textbf {\bibinfo {volume} {80}},\
  \bibinfo {eid} {044002} (\bibinfo {year} {2009})},\ \Eprint
  {https://arxiv.org/abs/0906.3602} {arXiv:0906.3602 [gr-qc]} \BibitemShut
  {NoStop}%
\bibitem [{\citenamefont {Vallisneri}(2008)}]{2008PhRvD..77d2001V}%
  \BibitemOpen
  \bibfield  {author} {\bibinfo {author} {\bibfnamefont {M.}~\bibnamefont
  {Vallisneri}},\ }\bibfield  {journal} {\bibinfo  {journal} {Physical Review
  D}\ }\textbf {\bibinfo {volume} {77}},\ \href
  {https://doi.org/10.1103/physrevd.77.042001} {10.1103/physrevd.77.042001}
  (\bibinfo {year} {2008})\BibitemShut {NoStop}%
\bibitem [{\citenamefont {{Babak}}\ \emph {et~al.}(2021)\citenamefont
  {{Babak}}, \citenamefont {{Hewitson}},\ and\ \citenamefont
  {{Petiteau}}}]{2021arXiv210801167B}%
  \BibitemOpen
  \bibfield  {author} {\bibinfo {author} {\bibfnamefont {S.}~\bibnamefont
  {{Babak}}}, \bibinfo {author} {\bibfnamefont {M.}~\bibnamefont
  {{Hewitson}}},\ and\ \bibinfo {author} {\bibfnamefont {A.}~\bibnamefont
  {{Petiteau}}},\ }\href@noop {} {\bibfield  {journal} {\bibinfo  {journal}
  {arXiv e-prints}\ ,\ \bibinfo {eid} {arXiv:2108.01167}} (\bibinfo {year}
  {2021})},\ \Eprint {https://arxiv.org/abs/2108.01167} {arXiv:2108.01167
  [astro-ph.IM]} \BibitemShut {NoStop}%
\bibitem [{\citenamefont {{Amaro-Seoane}}\ \emph {et~al.}(2017)\citenamefont
  {{Amaro-Seoane}} \emph {et~al.}}]{2017arXiv170200786A}%
  \BibitemOpen
  \bibfield  {author} {\bibinfo {author} {\bibfnamefont {P.}~\bibnamefont
  {{Amaro-Seoane}}} \emph {et~al.},\ }\href@noop {} {\bibfield  {journal}
  {\bibinfo  {journal} {arXiv e-prints}\ ,\ \bibinfo {eid} {arXiv:1702.00786}}
  (\bibinfo {year} {2017})},\ \Eprint {https://arxiv.org/abs/1702.00786}
  {arXiv:1702.00786 [astro-ph.IM]} \BibitemShut {NoStop}%
\bibitem [{\citenamefont {{Amaro-Seoane}}\ \emph {et~al.}(2021)\citenamefont
  {{Amaro-Seoane}}, \citenamefont {{Arca Sedda}}, \citenamefont {{Babak}},
  \citenamefont {{Berry}}, \citenamefont {{Berti}}, \citenamefont {{Bertone}},
  \citenamefont {{Blas}}, \citenamefont {{Bogdanovi{\'c}}}, \citenamefont
  {{Bonetti}}, \citenamefont {{Breivik}}, \citenamefont {{Brito}},
  \citenamefont {{Caldwell}}, \citenamefont {{Capelo}}, \citenamefont
  {{Caprini}}, \citenamefont {{Cardoso}}, \citenamefont {{Carson}},
  \citenamefont {{Chen}}, \citenamefont {{Chua}}, \citenamefont {{Dvorkin}},
  \citenamefont {{Haiman}}, \citenamefont {{Heisenberg}}, \citenamefont
  {{Isi}}, \citenamefont {{Karnesis}}, \citenamefont {{Kavanagh}},
  \citenamefont {{Littenberg}}, \citenamefont {{Mangiagli}}, \citenamefont
  {{Marcoccia}}, \citenamefont {{Maselli}}, \citenamefont {{Nardini}},
  \citenamefont {{Pani}}, \citenamefont {{Peloso}}, \citenamefont {{Pieroni}},
  \citenamefont {{Ricciardone}}, \citenamefont {{Sesana}}, \citenamefont
  {{Tamanini}}, \citenamefont {{Toubiana}}, \citenamefont {{Valiante}},
  \citenamefont {{Vretinaris}}, \citenamefont {{Weir}}, \citenamefont
  {{Yagi}},\ and\ \citenamefont {{Zimmerman}}}]{2021arXiv210709665S}%
  \BibitemOpen
  \bibfield  {author} {\bibinfo {author} {\bibfnamefont {P.}~\bibnamefont
  {{Amaro-Seoane}}}, \bibinfo {author} {\bibfnamefont {M.}~\bibnamefont {{Arca
  Sedda}}}, \bibinfo {author} {\bibfnamefont {S.}~\bibnamefont {{Babak}}},
  \bibinfo {author} {\bibfnamefont {C.~P.~L.}\ \bibnamefont {{Berry}}},
  \bibinfo {author} {\bibfnamefont {E.}~\bibnamefont {{Berti}}}, \bibinfo
  {author} {\bibfnamefont {G.}~\bibnamefont {{Bertone}}}, \bibinfo {author}
  {\bibfnamefont {D.}~\bibnamefont {{Blas}}}, \bibinfo {author} {\bibfnamefont
  {T.}~\bibnamefont {{Bogdanovi{\'c}}}}, \bibinfo {author} {\bibfnamefont
  {M.}~\bibnamefont {{Bonetti}}}, \bibinfo {author} {\bibfnamefont
  {K.}~\bibnamefont {{Breivik}}}, \bibinfo {author} {\bibfnamefont
  {R.}~\bibnamefont {{Brito}}}, \bibinfo {author} {\bibfnamefont
  {R.}~\bibnamefont {{Caldwell}}}, \bibinfo {author} {\bibfnamefont {P.~R.}\
  \bibnamefont {{Capelo}}}, \bibinfo {author} {\bibfnamefont {C.}~\bibnamefont
  {{Caprini}}}, \bibinfo {author} {\bibfnamefont {V.}~\bibnamefont
  {{Cardoso}}}, \bibinfo {author} {\bibfnamefont {Z.}~\bibnamefont {{Carson}}},
  \bibinfo {author} {\bibfnamefont {H.-Y.}\ \bibnamefont {{Chen}}}, \bibinfo
  {author} {\bibfnamefont {A.~J.~K.}\ \bibnamefont {{Chua}}}, \bibinfo {author}
  {\bibfnamefont {I.}~\bibnamefont {{Dvorkin}}}, \bibinfo {author}
  {\bibfnamefont {Z.}~\bibnamefont {{Haiman}}}, \bibinfo {author}
  {\bibfnamefont {L.}~\bibnamefont {{Heisenberg}}}, \bibinfo {author}
  {\bibfnamefont {M.}~\bibnamefont {{Isi}}}, \bibinfo {author} {\bibfnamefont
  {N.}~\bibnamefont {{Karnesis}}}, \bibinfo {author} {\bibfnamefont {B.~J.}\
  \bibnamefont {{Kavanagh}}}, \bibinfo {author} {\bibfnamefont {T.~B.}\
  \bibnamefont {{Littenberg}}}, \bibinfo {author} {\bibfnamefont
  {A.}~\bibnamefont {{Mangiagli}}}, \bibinfo {author} {\bibfnamefont
  {P.}~\bibnamefont {{Marcoccia}}}, \bibinfo {author} {\bibfnamefont
  {A.}~\bibnamefont {{Maselli}}}, \bibinfo {author} {\bibfnamefont
  {G.}~\bibnamefont {{Nardini}}}, \bibinfo {author} {\bibfnamefont
  {P.}~\bibnamefont {{Pani}}}, \bibinfo {author} {\bibfnamefont
  {M.}~\bibnamefont {{Peloso}}}, \bibinfo {author} {\bibfnamefont
  {M.}~\bibnamefont {{Pieroni}}}, \bibinfo {author} {\bibfnamefont
  {A.}~\bibnamefont {{Ricciardone}}}, \bibinfo {author} {\bibfnamefont
  {A.}~\bibnamefont {{Sesana}}}, \bibinfo {author} {\bibfnamefont
  {N.}~\bibnamefont {{Tamanini}}}, \bibinfo {author} {\bibfnamefont
  {A.}~\bibnamefont {{Toubiana}}}, \bibinfo {author} {\bibfnamefont
  {R.}~\bibnamefont {{Valiante}}}, \bibinfo {author} {\bibfnamefont
  {S.}~\bibnamefont {{Vretinaris}}}, \bibinfo {author} {\bibfnamefont
  {D.}~\bibnamefont {{Weir}}}, \bibinfo {author} {\bibfnamefont
  {K.}~\bibnamefont {{Yagi}}},\ and\ \bibinfo {author} {\bibfnamefont
  {A.}~\bibnamefont {{Zimmerman}}},\ }\href@noop {} {\bibfield  {journal}
  {\bibinfo  {journal} {arXiv e-prints}\ ,\ \bibinfo {eid} {arXiv:2107.09665}}
  (\bibinfo {year} {2021})},\ \Eprint {https://arxiv.org/abs/2107.09665}
  {arXiv:2107.09665 [astro-ph.IM]} \BibitemShut {NoStop}%
\bibitem [{\citenamefont {Johansson}\ \emph {et~al.}(2013)\citenamefont
  {Johansson} \emph {et~al.}}]{mpmath}%
  \BibitemOpen
  \bibfield  {author} {\bibinfo {author} {\bibfnamefont {F.}~\bibnamefont
  {Johansson}} \emph {et~al.},\ }\href@noop {} {\bibinfo {title} {mpmath: a
  {P}ython library for arbitrary-precision floating-point arithmetic (version
  1.1.0)}} (\bibinfo {year} {2013}),\ \bibinfo {note} {{\tt
  http://mpmath.org/}}\BibitemShut {NoStop}%
\bibitem [{\citenamefont {{Harris}}(1996)}]{1996AJ....112.1487H}%
  \BibitemOpen
  \bibfield  {author} {\bibinfo {author} {\bibfnamefont {W.~E.}\ \bibnamefont
  {{Harris}}},\ }\href {https://doi.org/10.1086/118116} {\bibfield  {journal}
  {\bibinfo  {journal} {\aj}\ }\textbf {\bibinfo {volume} {112}},\ \bibinfo
  {pages} {1487} (\bibinfo {year} {1996})}\BibitemShut {NoStop}%
\bibitem [{\citenamefont {{Harris}}\ \emph {et~al.}(2017)\citenamefont
  {{Harris}}, \citenamefont {{Blakeslee}},\ and\ \citenamefont
  {{Harris}}}]{2017ApJ...836...67H}%
  \BibitemOpen
  \bibfield  {author} {\bibinfo {author} {\bibfnamefont {W.~E.}\ \bibnamefont
  {{Harris}}}, \bibinfo {author} {\bibfnamefont {J.~P.}\ \bibnamefont
  {{Blakeslee}}},\ and\ \bibinfo {author} {\bibfnamefont {G.~L.~H.}\
  \bibnamefont {{Harris}}},\ }\href
  {https://doi.org/10.3847/1538-4357/836/1/67} {\bibfield  {journal} {\bibinfo
  {journal} {\apj}\ }\textbf {\bibinfo {volume} {836}},\ \bibinfo {eid} {67}
  (\bibinfo {year} {2017})},\ \Eprint {https://arxiv.org/abs/1701.04845}
  {arXiv:1701.04845 [astro-ph.GA]} \BibitemShut {NoStop}%
\bibitem [{\citenamefont {McLaughlin}(2000)}]{2000ApJ...539..618M}%
  \BibitemOpen
  \bibfield  {author} {\bibinfo {author} {\bibfnamefont {D.~E.}\ \bibnamefont
  {McLaughlin}},\ }\href {https://doi.org/10.1086/309247} {\bibfield  {journal}
  {\bibinfo  {journal} {The Astrophysical Journal}\ }\textbf {\bibinfo {volume}
  {539}},\ \bibinfo {pages} {618} (\bibinfo {year} {2000})}\BibitemShut
  {NoStop}%
\bibitem [{\citenamefont {Rejkuba}\ \emph {et~al.}(2007)\citenamefont
  {Rejkuba}, \citenamefont {Dubath}, \citenamefont {Minniti},\ and\
  \citenamefont {Meylan}}]{2007A&A...469..147R}%
  \BibitemOpen
  \bibfield  {author} {\bibinfo {author} {\bibfnamefont {M.}~\bibnamefont
  {Rejkuba}}, \bibinfo {author} {\bibfnamefont {P.}~\bibnamefont {Dubath}},
  \bibinfo {author} {\bibfnamefont {D.}~\bibnamefont {Minniti}},\ and\ \bibinfo
  {author} {\bibfnamefont {G.}~\bibnamefont {Meylan}},\ }\href
  {https://doi.org/10.1051/0004-6361:20066493} {\bibfield  {journal} {\bibinfo
  {journal} {Astronomy {\&} Astrophysics}\ }\textbf {\bibinfo {volume} {469}},\
  \bibinfo {pages} {147} (\bibinfo {year} {2007})}\BibitemShut {NoStop}%
\bibitem [{\citenamefont {Baumgardt}(2016)}]{2017MNRAS.464.2174B}%
  \BibitemOpen
  \bibfield  {author} {\bibinfo {author} {\bibfnamefont {H.}~\bibnamefont
  {Baumgardt}},\ }\href {https://doi.org/10.1093/mnras/stw2488} {\bibfield
  {journal} {\bibinfo  {journal} {Monthly Notices of the Royal Astronomical
  Society}\ }\textbf {\bibinfo {volume} {464}},\ \bibinfo {pages} {2174}
  (\bibinfo {year} {2016})}\BibitemShut {NoStop}%
\bibitem [{\citenamefont {Kroupa}(2001)}]{2001MNRAS.322..231K}%
  \BibitemOpen
  \bibfield  {author} {\bibinfo {author} {\bibfnamefont {P.}~\bibnamefont
  {Kroupa}},\ }\href {https://doi.org/10.1046/j.1365-8711.2001.04022.x}
  {\bibfield  {journal} {\bibinfo  {journal} {Monthly Notices of the Royal
  Astronomical Society}\ }\textbf {\bibinfo {volume} {322}},\ \bibinfo {pages}
  {231} (\bibinfo {year} {2001})}\BibitemShut {NoStop}%
\bibitem [{\citenamefont {{Hurley}}\ \emph {et~al.}(2000)\citenamefont
  {{Hurley}}, \citenamefont {{Pols}},\ and\ \citenamefont
  {{Tout}}}]{2000MNRAS.315..543H}%
  \BibitemOpen
  \bibfield  {author} {\bibinfo {author} {\bibfnamefont {J.~R.}\ \bibnamefont
  {{Hurley}}}, \bibinfo {author} {\bibfnamefont {O.~R.}\ \bibnamefont
  {{Pols}}},\ and\ \bibinfo {author} {\bibfnamefont {C.~A.}\ \bibnamefont
  {{Tout}}},\ }\href {https://doi.org/10.1046/j.1365-8711.2000.03426.x}
  {\bibfield  {journal} {\bibinfo  {journal} {\mnras}\ }\textbf {\bibinfo
  {volume} {315}},\ \bibinfo {pages} {543} (\bibinfo {year} {2000})},\ \Eprint
  {https://arxiv.org/abs/astro-ph/0001295} {arXiv:astro-ph/0001295 [astro-ph]}
  \BibitemShut {NoStop}%
\bibitem [{\citenamefont {{Banerjee}}\ \emph {et~al.}(2020)\citenamefont
  {{Banerjee}}, \citenamefont {{Belczynski}}, \citenamefont {{Fryer}},
  \citenamefont {{Berczik}}, \citenamefont {{Hurley}}, \citenamefont
  {{Spurzem}},\ and\ \citenamefont {{Wang}}}]{2020A&A...639A..41B}%
  \BibitemOpen
  \bibfield  {author} {\bibinfo {author} {\bibfnamefont {S.}~\bibnamefont
  {{Banerjee}}}, \bibinfo {author} {\bibfnamefont {K.}~\bibnamefont
  {{Belczynski}}}, \bibinfo {author} {\bibfnamefont {C.~L.}\ \bibnamefont
  {{Fryer}}}, \bibinfo {author} {\bibfnamefont {P.}~\bibnamefont {{Berczik}}},
  \bibinfo {author} {\bibfnamefont {J.~R.}\ \bibnamefont {{Hurley}}}, \bibinfo
  {author} {\bibfnamefont {R.}~\bibnamefont {{Spurzem}}},\ and\ \bibinfo
  {author} {\bibfnamefont {L.}~\bibnamefont {{Wang}}},\ }\href
  {https://doi.org/10.1051/0004-6361/201935332} {\bibfield  {journal} {\bibinfo
   {journal} {\aap}\ }\textbf {\bibinfo {volume} {639}},\ \bibinfo {eid} {A41}
  (\bibinfo {year} {2020})},\ \Eprint {https://arxiv.org/abs/1902.07718}
  {arXiv:1902.07718 [astro-ph.SR]} \BibitemShut {NoStop}%
\bibitem [{\citenamefont {Anders}\ and\ \citenamefont
  {Grevesse}(1989)}]{1989GeCoA..53..197A}%
  \BibitemOpen
  \bibfield  {author} {\bibinfo {author} {\bibfnamefont {E.}~\bibnamefont
  {Anders}}\ and\ \bibinfo {author} {\bibfnamefont {N.}~\bibnamefont
  {Grevesse}},\ }\href {https://doi.org/10.1016/0016-7037(89)90286-x}
  {\bibfield  {journal} {\bibinfo  {journal} {Geochimica et Cosmochimica Acta}\
  }\textbf {\bibinfo {volume} {53}},\ \bibinfo {pages} {197} (\bibinfo {year}
  {1989})}\BibitemShut {NoStop}%
\bibitem [{\citenamefont {{Gnedin}}\ \emph {et~al.}(2014)\citenamefont
  {{Gnedin}}, \citenamefont {{Ostriker}},\ and\ \citenamefont
  {{Tremaine}}}]{2014ApJ...785...71G}%
  \BibitemOpen
  \bibfield  {author} {\bibinfo {author} {\bibfnamefont {O.~Y.}\ \bibnamefont
  {{Gnedin}}}, \bibinfo {author} {\bibfnamefont {J.~P.}\ \bibnamefont
  {{Ostriker}}},\ and\ \bibinfo {author} {\bibfnamefont {S.}~\bibnamefont
  {{Tremaine}}},\ }\href {https://doi.org/10.1088/0004-637X/785/1/71}
  {\bibfield  {journal} {\bibinfo  {journal} {\apj}\ }\textbf {\bibinfo
  {volume} {785}},\ \bibinfo {eid} {71} (\bibinfo {year} {2014})},\ \Eprint
  {https://arxiv.org/abs/1308.0021} {arXiv:1308.0021 [astro-ph.CO]}
  \BibitemShut {NoStop}%
\bibitem [{\citenamefont {{Mardling}}\ and\ \citenamefont
  {{Aarseth}}(2001)}]{2001MNRAS.321..398M}%
  \BibitemOpen
  \bibfield  {author} {\bibinfo {author} {\bibfnamefont {R.~A.}\ \bibnamefont
  {{Mardling}}}\ and\ \bibinfo {author} {\bibfnamefont {S.~J.}\ \bibnamefont
  {{Aarseth}}},\ }\href {https://doi.org/10.1046/j.1365-8711.2001.03974.x}
  {\bibfield  {journal} {\bibinfo  {journal} {\mnras}\ }\textbf {\bibinfo
  {volume} {321}},\ \bibinfo {pages} {398} (\bibinfo {year}
  {2001})}\BibitemShut {NoStop}%
\bibitem [{\citenamefont {{Antonini}}\ and\ \citenamefont
  {{Perets}}(2012)}]{2012ApJ...757...27A}%
  \BibitemOpen
  \bibfield  {author} {\bibinfo {author} {\bibfnamefont {F.}~\bibnamefont
  {{Antonini}}}\ and\ \bibinfo {author} {\bibfnamefont {H.~B.}\ \bibnamefont
  {{Perets}}},\ }\href {https://doi.org/10.1088/0004-637X/757/1/27} {\bibfield
  {journal} {\bibinfo  {journal} {\apj}\ }\textbf {\bibinfo {volume} {757}},\
  \bibinfo {eid} {27} (\bibinfo {year} {2012})},\ \Eprint
  {https://arxiv.org/abs/1203.2938} {arXiv:1203.2938 [astro-ph.GA]}
  \BibitemShut {NoStop}%
\bibitem [{\citenamefont {{Fragione}}\ \emph {et~al.}(2019)\citenamefont
  {{Fragione}}, \citenamefont {{Grishin}}, \citenamefont {{Leigh}},
  \citenamefont {{Perets}},\ and\ \citenamefont
  {et~al.}}]{2019MNRAS.488...47F}%
  \BibitemOpen
  \bibfield  {author} {\bibinfo {author} {\bibfnamefont {G.}~\bibnamefont
  {{Fragione}}}, \bibinfo {author} {\bibfnamefont {E.}~\bibnamefont
  {{Grishin}}}, \bibinfo {author} {\bibfnamefont {N.~W.~C.}\ \bibnamefont
  {{Leigh}}}, \bibinfo {author} {\bibfnamefont {H.~B.}\ \bibnamefont
  {{Perets}}},\ and\ \bibinfo {author} {\bibnamefont {et~al.}},\ }\href
  {https://doi.org/10.1093/mnras/stz1651} {\bibfield  {journal} {\bibinfo
  {journal} {\mnras}\ }\textbf {\bibinfo {volume} {488}},\ \bibinfo {pages}
  {47} (\bibinfo {year} {2019})},\ \Eprint {https://arxiv.org/abs/1811.10627}
  {arXiv:1811.10627 [astro-ph.GA]} \BibitemShut {NoStop}%
\bibitem [{\citenamefont {Morscher}\ \emph {et~al.}(2015)\citenamefont
  {Morscher}, \citenamefont {Pattabiraman}, \citenamefont {Rodriguez},
  \citenamefont {Rasio},\ and\ \citenamefont {Umbreit}}]{2015ApJ...800....9M}%
  \BibitemOpen
  \bibfield  {author} {\bibinfo {author} {\bibfnamefont {M.}~\bibnamefont
  {Morscher}}, \bibinfo {author} {\bibfnamefont {B.}~\bibnamefont
  {Pattabiraman}}, \bibinfo {author} {\bibfnamefont {C.}~\bibnamefont
  {Rodriguez}}, \bibinfo {author} {\bibfnamefont {F.~A.}\ \bibnamefont
  {Rasio}},\ and\ \bibinfo {author} {\bibfnamefont {S.}~\bibnamefont
  {Umbreit}},\ }\href {https://doi.org/10.1088/0004-637x/800/1/9} {\bibfield
  {journal} {\bibinfo  {journal} {The Astrophysical Journal}\ }\textbf
  {\bibinfo {volume} {800}},\ \bibinfo {pages} {9} (\bibinfo {year}
  {2015})}\BibitemShut {NoStop}%
\bibitem [{\citenamefont {Weatherford}\ \emph {et~al.}(2020)\citenamefont
  {Weatherford}, \citenamefont {Chatterjee}, \citenamefont {Kremer},\ and\
  \citenamefont {Rasio}}]{2020ApJ...898..162W}%
  \BibitemOpen
  \bibfield  {author} {\bibinfo {author} {\bibfnamefont {N.~C.}\ \bibnamefont
  {Weatherford}}, \bibinfo {author} {\bibfnamefont {S.}~\bibnamefont
  {Chatterjee}}, \bibinfo {author} {\bibfnamefont {K.}~\bibnamefont {Kremer}},\
  and\ \bibinfo {author} {\bibfnamefont {F.~A.}\ \bibnamefont {Rasio}},\ }\href
  {https://doi.org/10.3847/1538-4357/ab9f98} {\bibfield  {journal} {\bibinfo
  {journal} {The Astrophysical Journal}\ }\textbf {\bibinfo {volume} {898}},\
  \bibinfo {pages} {162} (\bibinfo {year} {2020})}\BibitemShut {NoStop}%
\bibitem [{\citenamefont {{Weatherford}}\ \emph {et~al.}(2021)\citenamefont
  {{Weatherford}}, \citenamefont {{Fragione}}, \citenamefont {{Kremer}},
  \citenamefont {{Chatterjee}},\ and\ \citenamefont
  {et~al.}}]{2021ApJ...907L..25W}%
  \BibitemOpen
  \bibfield  {author} {\bibinfo {author} {\bibfnamefont {N.~C.}\ \bibnamefont
  {{Weatherford}}}, \bibinfo {author} {\bibfnamefont {G.}~\bibnamefont
  {{Fragione}}}, \bibinfo {author} {\bibfnamefont {K.}~\bibnamefont
  {{Kremer}}}, \bibinfo {author} {\bibfnamefont {S.}~\bibnamefont
  {{Chatterjee}}},\ and\ \bibinfo {author} {\bibnamefont {et~al.}},\ }\href
  {https://doi.org/10.3847/2041-8213/abd79c} {\bibfield  {journal} {\bibinfo
  {journal} {\apjl}\ }\textbf {\bibinfo {volume} {907}},\ \bibinfo {eid} {L25}
  (\bibinfo {year} {2021})},\ \Eprint {https://arxiv.org/abs/2101.02217}
  {arXiv:2101.02217 [astro-ph.GA]} \BibitemShut {NoStop}%
\bibitem [{\citenamefont {{Kremer}}\ \emph
  {et~al.}(2020{\natexlab{a}})\citenamefont {{Kremer}}, \citenamefont
  {{Spera}}, \citenamefont {{Becker}}, \citenamefont {{Chatterjee}},\ and\
  \citenamefont {et~al.}}]{2020ApJ...903...45K}%
  \BibitemOpen
  \bibfield  {author} {\bibinfo {author} {\bibfnamefont {K.}~\bibnamefont
  {{Kremer}}}, \bibinfo {author} {\bibfnamefont {M.}~\bibnamefont {{Spera}}},
  \bibinfo {author} {\bibfnamefont {D.}~\bibnamefont {{Becker}}}, \bibinfo
  {author} {\bibfnamefont {S.}~\bibnamefont {{Chatterjee}}},\ and\ \bibinfo
  {author} {\bibnamefont {et~al.}},\ }\href
  {https://doi.org/10.3847/1538-4357/abb945} {\bibfield  {journal} {\bibinfo
  {journal} {\apj}\ }\textbf {\bibinfo {volume} {903}},\ \bibinfo {eid} {45}
  (\bibinfo {year} {2020}{\natexlab{a}})},\ \Eprint
  {https://arxiv.org/abs/2006.10771} {arXiv:2006.10771 [astro-ph.HE]}
  \BibitemShut {NoStop}%
\bibitem [{\citenamefont {{Baumgardt}}\ \emph {et~al.}(2004)\citenamefont
  {{Baumgardt}}, \citenamefont {{Makino}},\ and\ \citenamefont
  {{Ebisuzaki}}}]{2004ApJ...613.1143B}%
  \BibitemOpen
  \bibfield  {author} {\bibinfo {author} {\bibfnamefont {H.}~\bibnamefont
  {{Baumgardt}}}, \bibinfo {author} {\bibfnamefont {J.}~\bibnamefont
  {{Makino}}},\ and\ \bibinfo {author} {\bibfnamefont {T.}~\bibnamefont
  {{Ebisuzaki}}},\ }\href {https://doi.org/10.1086/423299} {\bibfield
  {journal} {\bibinfo  {journal} {\apj}\ }\textbf {\bibinfo {volume} {613}},\
  \bibinfo {pages} {1143} (\bibinfo {year} {2004})},\ \Eprint
  {https://arxiv.org/abs/astro-ph/0406231} {arXiv:astro-ph/0406231 [astro-ph]}
  \BibitemShut {NoStop}%
\bibitem [{\citenamefont {{MacLeod}}\ \emph {et~al.}(2016)\citenamefont
  {{MacLeod}}, \citenamefont {{Trenti}},\ and\ \citenamefont
  {{Ramirez-Ruiz}}}]{2016ApJ...819...70M}%
  \BibitemOpen
  \bibfield  {author} {\bibinfo {author} {\bibfnamefont {M.}~\bibnamefont
  {{MacLeod}}}, \bibinfo {author} {\bibfnamefont {M.}~\bibnamefont
  {{Trenti}}},\ and\ \bibinfo {author} {\bibfnamefont {E.}~\bibnamefont
  {{Ramirez-Ruiz}}},\ }\href {https://doi.org/10.3847/0004-637X/819/1/70}
  {\bibfield  {journal} {\bibinfo  {journal} {\apj}\ }\textbf {\bibinfo
  {volume} {819}},\ \bibinfo {eid} {70} (\bibinfo {year} {2016})},\ \Eprint
  {https://arxiv.org/abs/1508.07000} {arXiv:1508.07000 [astro-ph.HE]}
  \BibitemShut {NoStop}%
\bibitem [{\citenamefont {{Kremer}}\ \emph {et~al.}(2018)\citenamefont
  {{Kremer}}, \citenamefont {{Chatterjee}}, \citenamefont {{Breivik}},
  \citenamefont {{Rodriguez}}, \citenamefont {{Larson}},\ and\ \citenamefont
  {{Rasio}}}]{2018PhRvL.120s1103K}%
  \BibitemOpen
  \bibfield  {author} {\bibinfo {author} {\bibfnamefont {K.}~\bibnamefont
  {{Kremer}}}, \bibinfo {author} {\bibfnamefont {S.}~\bibnamefont
  {{Chatterjee}}}, \bibinfo {author} {\bibfnamefont {K.}~\bibnamefont
  {{Breivik}}}, \bibinfo {author} {\bibfnamefont {C.~L.}\ \bibnamefont
  {{Rodriguez}}}, \bibinfo {author} {\bibfnamefont {S.~L.}\ \bibnamefont
  {{Larson}}},\ and\ \bibinfo {author} {\bibfnamefont {F.~A.}\ \bibnamefont
  {{Rasio}}},\ }\href {https://doi.org/10.1103/PhysRevLett.120.191103}
  {\bibfield  {journal} {\bibinfo  {journal} {\prl}\ }\textbf {\bibinfo
  {volume} {120}},\ \bibinfo {eid} {191103} (\bibinfo {year} {2018})},\ \Eprint
  {https://arxiv.org/abs/1802.05661} {arXiv:1802.05661 [astro-ph.HE]}
  \BibitemShut {NoStop}%
\bibitem [{\citenamefont {{Kremer}}\ \emph {et~al.}(2019)\citenamefont
  {{Kremer}}, \citenamefont {{Rodriguez}}, \citenamefont {{Amaro-Seoane}},
  \citenamefont {{Breivik}}, \citenamefont {{Chatterjee}}, \citenamefont
  {{Katz}}, \citenamefont {{Larson}}, \citenamefont {{Rasio}}, \citenamefont
  {{Samsing}}, \citenamefont {{Ye}},\ and\ \citenamefont
  {{Zevin}}}]{2019PhRvD..99f3003K}%
  \BibitemOpen
  \bibfield  {author} {\bibinfo {author} {\bibfnamefont {K.}~\bibnamefont
  {{Kremer}}}, \bibinfo {author} {\bibfnamefont {C.~L.}\ \bibnamefont
  {{Rodriguez}}}, \bibinfo {author} {\bibfnamefont {P.}~\bibnamefont
  {{Amaro-Seoane}}}, \bibinfo {author} {\bibfnamefont {K.}~\bibnamefont
  {{Breivik}}}, \bibinfo {author} {\bibfnamefont {S.}~\bibnamefont
  {{Chatterjee}}}, \bibinfo {author} {\bibfnamefont {M.~L.}\ \bibnamefont
  {{Katz}}}, \bibinfo {author} {\bibfnamefont {S.~L.}\ \bibnamefont
  {{Larson}}}, \bibinfo {author} {\bibfnamefont {F.~A.}\ \bibnamefont
  {{Rasio}}}, \bibinfo {author} {\bibfnamefont {J.}~\bibnamefont {{Samsing}}},
  \bibinfo {author} {\bibfnamefont {C.~S.}\ \bibnamefont {{Ye}}},\ and\
  \bibinfo {author} {\bibfnamefont {M.}~\bibnamefont {{Zevin}}},\ }\href
  {https://doi.org/10.1103/PhysRevD.99.063003} {\bibfield  {journal} {\bibinfo
  {journal} {\prd}\ }\textbf {\bibinfo {volume} {99}},\ \bibinfo {eid} {063003}
  (\bibinfo {year} {2019})},\ \Eprint {https://arxiv.org/abs/1811.11812}
  {arXiv:1811.11812 [astro-ph.HE]} \BibitemShut {NoStop}%
\bibitem [{\citenamefont {{Fragione}}\ and\ \citenamefont
  {{Kocsis}}(2018)}]{2018PhRvL.121p1103F}%
  \BibitemOpen
  \bibfield  {author} {\bibinfo {author} {\bibfnamefont {G.}~\bibnamefont
  {{Fragione}}}\ and\ \bibinfo {author} {\bibfnamefont {B.}~\bibnamefont
  {{Kocsis}}},\ }\href {https://doi.org/10.1103/PhysRevLett.121.161103}
  {\bibfield  {journal} {\bibinfo  {journal} {\prl}\ }\textbf {\bibinfo
  {volume} {121}},\ \bibinfo {eid} {161103} (\bibinfo {year} {2018})},\ \Eprint
  {https://arxiv.org/abs/1806.02351} {arXiv:1806.02351 [astro-ph.GA]}
  \BibitemShut {NoStop}%
\bibitem [{\citenamefont {Licquia}\ and\ \citenamefont
  {Newman}(2015)}]{2015ApJ...806...96L}%
  \BibitemOpen
  \bibfield  {author} {\bibinfo {author} {\bibfnamefont {T.~C.}\ \bibnamefont
  {Licquia}}\ and\ \bibinfo {author} {\bibfnamefont {J.~A.}\ \bibnamefont
  {Newman}},\ }\href {https://doi.org/10.1088/0004-637x/806/1/96} {\bibfield
  {journal} {\bibinfo  {journal} {The Astrophysical Journal}\ }\textbf
  {\bibinfo {volume} {806}},\ \bibinfo {pages} {96} (\bibinfo {year}
  {2015})}\BibitemShut {NoStop}%
\bibitem [{\citenamefont {{S{\'e}rsic}}(1963)}]{1963BAAA....6...41S}%
  \BibitemOpen
  \bibfield  {author} {\bibinfo {author} {\bibfnamefont {J.~L.}\ \bibnamefont
  {{S{\'e}rsic}}},\ }\href@noop {} {\bibfield  {journal} {\bibinfo  {journal}
  {Boletin de la Asociacion Argentina de Astronomia La Plata Argentina}\
  }\textbf {\bibinfo {volume} {6}},\ \bibinfo {pages} {41} (\bibinfo {year}
  {1963})}\BibitemShut {NoStop}%
\bibitem [{\citenamefont {Navarro}\ \emph {et~al.}(1997)\citenamefont
  {Navarro}, \citenamefont {Frenk},\ and\ \citenamefont
  {White}}]{1997ApJ...490..493N}%
  \BibitemOpen
  \bibfield  {author} {\bibinfo {author} {\bibfnamefont {J.~F.}\ \bibnamefont
  {Navarro}}, \bibinfo {author} {\bibfnamefont {C.~S.}\ \bibnamefont {Frenk}},\
  and\ \bibinfo {author} {\bibfnamefont {S.~D.~M.}\ \bibnamefont {White}},\
  }\href {https://doi.org/10.1086/304888} {\bibfield  {journal} {\bibinfo
  {journal} {The Astrophysical Journal}\ }\textbf {\bibinfo {volume} {490}},\
  \bibinfo {pages} {493} (\bibinfo {year} {1997})}\BibitemShut {NoStop}%
\bibitem [{\citenamefont {{Chandrasekhar}}(1943)}]{1943ApJ....97..255C}%
  \BibitemOpen
  \bibfield  {author} {\bibinfo {author} {\bibfnamefont {S.}~\bibnamefont
  {{Chandrasekhar}}},\ }\href {https://doi.org/10.1086/144517} {\bibfield
  {journal} {\bibinfo  {journal} {\apj}\ }\textbf {\bibinfo {volume} {97}},\
  \bibinfo {pages} {255} (\bibinfo {year} {1943})}\BibitemShut {NoStop}%
\bibitem [{\citenamefont {{Binney}}\ and\ \citenamefont
  {{Tremaine}}(2008)}]{2008gady.book.....B}%
  \BibitemOpen
  \bibfield  {author} {\bibinfo {author} {\bibfnamefont {J.}~\bibnamefont
  {{Binney}}}\ and\ \bibinfo {author} {\bibfnamefont {S.}~\bibnamefont
  {{Tremaine}}},\ }\href@noop {} {\emph {\bibinfo {title} {{Galactic Dynamics:
  Second Edition}}}}\ (\bibinfo {year} {2008})\BibitemShut {NoStop}%
\bibitem [{\citenamefont {{Mapelli}}(2021)}]{2021arXiv210600699M}%
  \BibitemOpen
  \bibfield  {author} {\bibinfo {author} {\bibfnamefont {M.}~\bibnamefont
  {{Mapelli}}},\ }\href@noop {} {\bibfield  {journal} {\bibinfo  {journal}
  {arXiv e-prints}\ ,\ \bibinfo {eid} {arXiv:2106.00699}} (\bibinfo {year}
  {2021})},\ \Eprint {https://arxiv.org/abs/2106.00699} {arXiv:2106.00699
  [astro-ph.HE]} \BibitemShut {NoStop}%
\bibitem [{\citenamefont {{Baumgardt}}(2001)}]{2001MNRAS.325.1323B}%
  \BibitemOpen
  \bibfield  {author} {\bibinfo {author} {\bibfnamefont {H.}~\bibnamefont
  {{Baumgardt}}},\ }\href {https://doi.org/10.1046/j.1365-8711.2001.04272.x}
  {\bibfield  {journal} {\bibinfo  {journal} {\mnras}\ }\textbf {\bibinfo
  {volume} {325}},\ \bibinfo {pages} {1323} (\bibinfo {year} {2001})},\ \Eprint
  {https://arxiv.org/abs/astro-ph/0012330} {arXiv:astro-ph/0012330 [astro-ph]}
  \BibitemShut {NoStop}%
\bibitem [{\citenamefont {{Gieles}}\ and\ \citenamefont
  {{Baumgardt}}(2008)}]{2008MNRAS.389L..28G}%
  \BibitemOpen
  \bibfield  {author} {\bibinfo {author} {\bibfnamefont {M.}~\bibnamefont
  {{Gieles}}}\ and\ \bibinfo {author} {\bibfnamefont {H.}~\bibnamefont
  {{Baumgardt}}},\ }\href {https://doi.org/10.1111/j.1745-3933.2008.00515.x}
  {\bibfield  {journal} {\bibinfo  {journal} {\mnras}\ }\textbf {\bibinfo
  {volume} {389}},\ \bibinfo {pages} {L28} (\bibinfo {year} {2008})},\ \Eprint
  {https://arxiv.org/abs/0806.2327} {arXiv:0806.2327 [astro-ph]} \BibitemShut
  {NoStop}%
\bibitem [{\citenamefont {{Prieto}}\ and\ \citenamefont
  {{Gnedin}}(2008)}]{2008ApJ...689..919P}%
  \BibitemOpen
  \bibfield  {author} {\bibinfo {author} {\bibfnamefont {J.~L.}\ \bibnamefont
  {{Prieto}}}\ and\ \bibinfo {author} {\bibfnamefont {O.~Y.}\ \bibnamefont
  {{Gnedin}}},\ }\href {https://doi.org/10.1086/591777} {\bibfield  {journal}
  {\bibinfo  {journal} {\apj}\ }\textbf {\bibinfo {volume} {689}},\ \bibinfo
  {pages} {919} (\bibinfo {year} {2008})},\ \Eprint
  {https://arxiv.org/abs/astro-ph/0608069} {arXiv:astro-ph/0608069 [astro-ph]}
  \BibitemShut {NoStop}%
\bibitem [{\citenamefont {{Fall}}\ and\ \citenamefont
  {{Zhang}}(2001)}]{2001ApJ...561..751F}%
  \BibitemOpen
  \bibfield  {author} {\bibinfo {author} {\bibfnamefont {S.~M.}\ \bibnamefont
  {{Fall}}}\ and\ \bibinfo {author} {\bibfnamefont {Q.}~\bibnamefont
  {{Zhang}}},\ }\href {https://doi.org/10.1086/323358} {\bibfield  {journal}
  {\bibinfo  {journal} {\apj}\ }\textbf {\bibinfo {volume} {561}},\ \bibinfo
  {pages} {751} (\bibinfo {year} {2001})},\ \Eprint
  {https://arxiv.org/abs/astro-ph/0107298} {arXiv:astro-ph/0107298 [astro-ph]}
  \BibitemShut {NoStop}%
\bibitem [{\citenamefont {Fragione}\ \emph {et~al.}(2019)\citenamefont
  {Fragione}, \citenamefont {Antonini},\ and\ \citenamefont
  {Gnedin}}]{2019ApJ...871L...8F}%
  \BibitemOpen
  \bibfield  {author} {\bibinfo {author} {\bibfnamefont {G.}~\bibnamefont
  {Fragione}}, \bibinfo {author} {\bibfnamefont {F.}~\bibnamefont {Antonini}},\
  and\ \bibinfo {author} {\bibfnamefont {O.~Y.}\ \bibnamefont {Gnedin}},\
  }\href {https://doi.org/10.3847/2041-8213/aafc62} {\bibfield  {journal}
  {\bibinfo  {journal} {The Astrophysical Journal}\ }\textbf {\bibinfo {volume}
  {871}},\ \bibinfo {pages} {L8} (\bibinfo {year} {2019})}\BibitemShut
  {NoStop}%
\bibitem [{\citenamefont {{Dehnen}}\ \emph {et~al.}(2006)\citenamefont
  {{Dehnen}}, \citenamefont {{McLaughlin}},\ and\ \citenamefont
  {{Sachania}}}]{2006MNRAS.369.1688D}%
  \BibitemOpen
  \bibfield  {author} {\bibinfo {author} {\bibfnamefont {W.}~\bibnamefont
  {{Dehnen}}}, \bibinfo {author} {\bibfnamefont {D.~E.}\ \bibnamefont
  {{McLaughlin}}},\ and\ \bibinfo {author} {\bibfnamefont {J.}~\bibnamefont
  {{Sachania}}},\ }\href {https://doi.org/10.1111/j.1365-2966.2006.10404.x}
  {\bibfield  {journal} {\bibinfo  {journal} {\mnras}\ }\textbf {\bibinfo
  {volume} {369}},\ \bibinfo {pages} {1688} (\bibinfo {year} {2006})},\ \Eprint
  {https://arxiv.org/abs/astro-ph/0603825} {arXiv:astro-ph/0603825 [astro-ph]}
  \BibitemShut {NoStop}%
\bibitem [{\citenamefont {{Brown}}\ \emph {et~al.}(2010)\citenamefont
  {{Brown}}, \citenamefont {{Geller}}, \citenamefont {{Kenyon}},\ and\
  \citenamefont {{Diaferio}}}]{2010AJ....139...59B}%
  \BibitemOpen
  \bibfield  {author} {\bibinfo {author} {\bibfnamefont {W.~R.}\ \bibnamefont
  {{Brown}}}, \bibinfo {author} {\bibfnamefont {M.~J.}\ \bibnamefont
  {{Geller}}}, \bibinfo {author} {\bibfnamefont {S.~J.}\ \bibnamefont
  {{Kenyon}}},\ and\ \bibinfo {author} {\bibfnamefont {A.}~\bibnamefont
  {{Diaferio}}},\ }\href {https://doi.org/10.1088/0004-6256/139/1/59}
  {\bibfield  {journal} {\bibinfo  {journal} {\aj}\ }\textbf {\bibinfo {volume}
  {139}},\ \bibinfo {pages} {59} (\bibinfo {year} {2010})},\ \Eprint
  {https://arxiv.org/abs/0910.2242} {arXiv:0910.2242 [astro-ph.GA]}
  \BibitemShut {NoStop}%
\bibitem [{\citenamefont {{Kremer}}\ \emph
  {et~al.}(2020{\natexlab{b}})\citenamefont {{Kremer}}, \citenamefont {{Ye}},
  \citenamefont {{Rui}}, \citenamefont {{Weatherford}}, \citenamefont
  {{Chatterjee}}, \citenamefont {{Fragione}}, \citenamefont {{Rodriguez}},
  \citenamefont {{Spera}},\ and\ \citenamefont
  {{Rasio}}}]{2020ApJS..247...48K}%
  \BibitemOpen
  \bibfield  {author} {\bibinfo {author} {\bibfnamefont {K.}~\bibnamefont
  {{Kremer}}}, \bibinfo {author} {\bibfnamefont {C.~S.}\ \bibnamefont {{Ye}}},
  \bibinfo {author} {\bibfnamefont {N.~Z.}\ \bibnamefont {{Rui}}}, \bibinfo
  {author} {\bibfnamefont {N.~C.}\ \bibnamefont {{Weatherford}}}, \bibinfo
  {author} {\bibfnamefont {S.}~\bibnamefont {{Chatterjee}}}, \bibinfo {author}
  {\bibfnamefont {G.}~\bibnamefont {{Fragione}}}, \bibinfo {author}
  {\bibfnamefont {C.~L.}\ \bibnamefont {{Rodriguez}}}, \bibinfo {author}
  {\bibfnamefont {M.}~\bibnamefont {{Spera}}},\ and\ \bibinfo {author}
  {\bibfnamefont {F.~A.}\ \bibnamefont {{Rasio}}},\ }\href
  {https://doi.org/10.3847/1538-4365/ab7919} {\bibfield  {journal} {\bibinfo
  {journal} {\apjs}\ }\textbf {\bibinfo {volume} {247}},\ \bibinfo {eid} {48}
  (\bibinfo {year} {2020}{\natexlab{b}})},\ \Eprint
  {https://arxiv.org/abs/1911.00018} {arXiv:1911.00018 [astro-ph.HE]}
  \BibitemShut {NoStop}%
\bibitem [{\citenamefont {{Anderson}}\ and\ \citenamefont {{van der
  Marel}}(2010)}]{2010ApJ...710.1032A}%
  \BibitemOpen
  \bibfield  {author} {\bibinfo {author} {\bibfnamefont {J.}~\bibnamefont
  {{Anderson}}}\ and\ \bibinfo {author} {\bibfnamefont {R.~P.}\ \bibnamefont
  {{van der Marel}}},\ }\href {https://doi.org/10.1088/0004-637X/710/2/1032}
  {\bibfield  {journal} {\bibinfo  {journal} {\apj}\ }\textbf {\bibinfo
  {volume} {710}},\ \bibinfo {pages} {1032} (\bibinfo {year} {2010})},\ \Eprint
  {https://arxiv.org/abs/0905.0627} {arXiv:0905.0627 [astro-ph.GA]}
  \BibitemShut {NoStop}%
\bibitem [{\citenamefont {{Cheng}}\ \emph {et~al.}(2020)\citenamefont
  {{Cheng}}, \citenamefont {{Li}}, \citenamefont {{Wang}}, \citenamefont
  {{Li}},\ and\ \citenamefont {{Xu}}}]{2020ApJ...904..198C}%
  \BibitemOpen
  \bibfield  {author} {\bibinfo {author} {\bibfnamefont {Z.}~\bibnamefont
  {{Cheng}}}, \bibinfo {author} {\bibfnamefont {Z.}~\bibnamefont {{Li}}},
  \bibinfo {author} {\bibfnamefont {W.}~\bibnamefont {{Wang}}}, \bibinfo
  {author} {\bibfnamefont {X.}~\bibnamefont {{Li}}},\ and\ \bibinfo {author}
  {\bibfnamefont {X.}~\bibnamefont {{Xu}}},\ }\href
  {https://doi.org/10.3847/1538-4357/abbdfc} {\bibfield  {journal} {\bibinfo
  {journal} {\apj}\ }\textbf {\bibinfo {volume} {904}},\ \bibinfo {eid} {198}
  (\bibinfo {year} {2020})},\ \Eprint {https://arxiv.org/abs/2010.00908}
  {arXiv:2010.00908 [astro-ph.HE]} \BibitemShut {NoStop}%
\bibitem [{\citenamefont {{Tremou}}\ \emph {et~al.}(2018)\citenamefont
  {{Tremou}}, \citenamefont {{Strader}}, \citenamefont {{Chomiuk}},
  \citenamefont {{Shishkovsky}}, \citenamefont {{Maccarone}}, \citenamefont
  {{Miller-Jones}}, \citenamefont {{Tudor}}, \citenamefont {{Heinke}},
  \citenamefont {{Sivakoff}}, \citenamefont {{Seth}},\ and\ \citenamefont
  {{Noyola}}}]{2018ApJ...862...16T}%
  \BibitemOpen
  \bibfield  {author} {\bibinfo {author} {\bibfnamefont {E.}~\bibnamefont
  {{Tremou}}}, \bibinfo {author} {\bibfnamefont {J.}~\bibnamefont {{Strader}}},
  \bibinfo {author} {\bibfnamefont {L.}~\bibnamefont {{Chomiuk}}}, \bibinfo
  {author} {\bibfnamefont {L.}~\bibnamefont {{Shishkovsky}}}, \bibinfo {author}
  {\bibfnamefont {T.~J.}\ \bibnamefont {{Maccarone}}}, \bibinfo {author}
  {\bibfnamefont {J.~C.~A.}\ \bibnamefont {{Miller-Jones}}}, \bibinfo {author}
  {\bibfnamefont {V.}~\bibnamefont {{Tudor}}}, \bibinfo {author} {\bibfnamefont
  {C.~O.}\ \bibnamefont {{Heinke}}}, \bibinfo {author} {\bibfnamefont {G.~R.}\
  \bibnamefont {{Sivakoff}}}, \bibinfo {author} {\bibfnamefont {A.~C.}\
  \bibnamefont {{Seth}}},\ and\ \bibinfo {author} {\bibfnamefont
  {E.}~\bibnamefont {{Noyola}}},\ }\href
  {https://doi.org/10.3847/1538-4357/aac9b9} {\bibfield  {journal} {\bibinfo
  {journal} {\apj}\ }\textbf {\bibinfo {volume} {862}},\ \bibinfo {eid} {16}
  (\bibinfo {year} {2018})},\ \Eprint {https://arxiv.org/abs/1806.00259}
  {arXiv:1806.00259 [astro-ph.HE]} \BibitemShut {NoStop}%
\bibitem [{\citenamefont {{Baumgardt}}\ \emph {et~al.}(2018)\citenamefont
  {{Baumgardt}}, \citenamefont {{Amaro-Seoane}},\ and\ \citenamefont
  {{Sch{\"o}del}}}]{2018A&A...609A..28B}%
  \BibitemOpen
  \bibfield  {author} {\bibinfo {author} {\bibfnamefont {H.}~\bibnamefont
  {{Baumgardt}}}, \bibinfo {author} {\bibfnamefont {P.}~\bibnamefont
  {{Amaro-Seoane}}},\ and\ \bibinfo {author} {\bibfnamefont {R.}~\bibnamefont
  {{Sch{\"o}del}}},\ }\href {https://doi.org/10.1051/0004-6361/201730462}
  {\bibfield  {journal} {\bibinfo  {journal} {\aap}\ }\textbf {\bibinfo
  {volume} {609}},\ \bibinfo {eid} {A28} (\bibinfo {year} {2018})},\ \Eprint
  {https://arxiv.org/abs/1701.03818} {arXiv:1701.03818 [astro-ph.GA]}
  \BibitemShut {NoStop}%
\bibitem [{\citenamefont {{Alexander}}(2017)}]{2017ARA&A..55...17A}%
  \BibitemOpen
  \bibfield  {author} {\bibinfo {author} {\bibfnamefont {T.}~\bibnamefont
  {{Alexander}}},\ }\href {https://doi.org/10.1146/annurev-astro-091916-055306}
  {\bibfield  {journal} {\bibinfo  {journal} {\araa}\ }\textbf {\bibinfo
  {volume} {55}},\ \bibinfo {pages} {17} (\bibinfo {year} {2017})},\ \Eprint
  {https://arxiv.org/abs/1701.04762} {arXiv:1701.04762 [astro-ph.GA]}
  \BibitemShut {NoStop}%
\bibitem [{\citenamefont {{Peters}}\ and\ \citenamefont
  {{Mathews}}(1963)}]{1963PhRv..131..435P}%
  \BibitemOpen
  \bibfield  {author} {\bibinfo {author} {\bibfnamefont {P.~C.}\ \bibnamefont
  {{Peters}}}\ and\ \bibinfo {author} {\bibfnamefont {J.}~\bibnamefont
  {{Mathews}}},\ }\href {https://doi.org/10.1103/PhysRev.131.435} {\bibfield
  {journal} {\bibinfo  {journal} {Physical Review}\ }\textbf {\bibinfo {volume}
  {131}},\ \bibinfo {pages} {435} (\bibinfo {year} {1963})}\BibitemShut
  {NoStop}%
\bibitem [{\citenamefont {{Hoang}}\ \emph {et~al.}(2019)\citenamefont
  {{Hoang}}, \citenamefont {{Naoz}}, \citenamefont {{Kocsis}}, \citenamefont
  {{Farr}},\ and\ \citenamefont {{McIver}}}]{2019ApJ...875L..31H}%
  \BibitemOpen
  \bibfield  {author} {\bibinfo {author} {\bibfnamefont {B.-M.}\ \bibnamefont
  {{Hoang}}}, \bibinfo {author} {\bibfnamefont {S.}~\bibnamefont {{Naoz}}},
  \bibinfo {author} {\bibfnamefont {B.}~\bibnamefont {{Kocsis}}}, \bibinfo
  {author} {\bibfnamefont {W.~M.}\ \bibnamefont {{Farr}}},\ and\ \bibinfo
  {author} {\bibfnamefont {J.}~\bibnamefont {{McIver}}},\ }\href
  {https://doi.org/10.3847/2041-8213/ab14f7} {\bibfield  {journal} {\bibinfo
  {journal} {\apjl}\ }\textbf {\bibinfo {volume} {875}},\ \bibinfo {eid} {L31}
  (\bibinfo {year} {2019})},\ \Eprint {https://arxiv.org/abs/1903.00134}
  {arXiv:1903.00134 [astro-ph.HE]} \BibitemShut {NoStop}%
\bibitem [{\citenamefont {{Deme}}\ \emph {et~al.}(2020)\citenamefont {{Deme}},
  \citenamefont {{Hoang}}, \citenamefont {{Naoz}},\ and\ \citenamefont
  {{Kocsis}}}]{2020ApJ...901..125D}%
  \BibitemOpen
  \bibfield  {author} {\bibinfo {author} {\bibfnamefont {B.}~\bibnamefont
  {{Deme}}}, \bibinfo {author} {\bibfnamefont {B.-M.}\ \bibnamefont {{Hoang}}},
  \bibinfo {author} {\bibfnamefont {S.}~\bibnamefont {{Naoz}}},\ and\ \bibinfo
  {author} {\bibfnamefont {B.}~\bibnamefont {{Kocsis}}},\ }\href
  {https://doi.org/10.3847/1538-4357/abafa3} {\bibfield  {journal} {\bibinfo
  {journal} {\apj}\ }\textbf {\bibinfo {volume} {901}},\ \bibinfo {eid} {125}
  (\bibinfo {year} {2020})},\ \Eprint {https://arxiv.org/abs/2005.03677}
  {arXiv:2005.03677 [astro-ph.HE]} \BibitemShut {NoStop}%
\bibitem [{\citenamefont {{Fragione}}\ \emph
  {et~al.}(2018{\natexlab{c}})\citenamefont {{Fragione}}, \citenamefont
  {{Pavl{\'\i}k}},\ and\ \citenamefont {{Banerjee}}}]{2018MNRAS.480.4955F}%
  \BibitemOpen
  \bibfield  {author} {\bibinfo {author} {\bibfnamefont {G.}~\bibnamefont
  {{Fragione}}}, \bibinfo {author} {\bibfnamefont {V.}~\bibnamefont
  {{Pavl{\'\i}k}}},\ and\ \bibinfo {author} {\bibfnamefont {S.}~\bibnamefont
  {{Banerjee}}},\ }\href {https://doi.org/10.1093/mnras/sty2234} {\bibfield
  {journal} {\bibinfo  {journal} {\mnras}\ }\textbf {\bibinfo {volume} {480}},\
  \bibinfo {pages} {4955} (\bibinfo {year} {2018}{\natexlab{c}})},\ \Eprint
  {https://arxiv.org/abs/1804.04856} {arXiv:1804.04856 [astro-ph.GA]}
  \BibitemShut {NoStop}%
\bibitem [{\citenamefont {{Ye}}\ \emph {et~al.}(2020)\citenamefont {{Ye}},
  \citenamefont {{Fong}}, \citenamefont {{Kremer}}, \citenamefont
  {{Rodriguez}}, \citenamefont {{Chatterjee}}, \citenamefont {{Fragione}},\
  and\ \citenamefont {{Rasio}}}]{2020ApJ...888L..10Y}%
  \BibitemOpen
  \bibfield  {author} {\bibinfo {author} {\bibfnamefont {C.~S.}\ \bibnamefont
  {{Ye}}}, \bibinfo {author} {\bibfnamefont {W.-f.}\ \bibnamefont {{Fong}}},
  \bibinfo {author} {\bibfnamefont {K.}~\bibnamefont {{Kremer}}}, \bibinfo
  {author} {\bibfnamefont {C.~L.}\ \bibnamefont {{Rodriguez}}}, \bibinfo
  {author} {\bibfnamefont {S.}~\bibnamefont {{Chatterjee}}}, \bibinfo {author}
  {\bibfnamefont {G.}~\bibnamefont {{Fragione}}},\ and\ \bibinfo {author}
  {\bibfnamefont {F.~A.}\ \bibnamefont {{Rasio}}},\ }\href
  {https://doi.org/10.3847/2041-8213/ab5dc5} {\bibfield  {journal} {\bibinfo
  {journal} {\apjl}\ }\textbf {\bibinfo {volume} {888}},\ \bibinfo {eid} {L10}
  (\bibinfo {year} {2020})},\ \Eprint {https://arxiv.org/abs/1910.10740}
  {arXiv:1910.10740 [astro-ph.HE]} \BibitemShut {NoStop}%
\bibitem [{\citenamefont {{Fragione}}\ and\ \citenamefont
  {{Banerjee}}(2020)}]{2020ApJ...901L..16F}%
  \BibitemOpen
  \bibfield  {author} {\bibinfo {author} {\bibfnamefont {G.}~\bibnamefont
  {{Fragione}}}\ and\ \bibinfo {author} {\bibfnamefont {S.}~\bibnamefont
  {{Banerjee}}},\ }\href {https://doi.org/10.3847/2041-8213/abb671} {\bibfield
  {journal} {\bibinfo  {journal} {\apjl}\ }\textbf {\bibinfo {volume} {901}},\
  \bibinfo {eid} {L16} (\bibinfo {year} {2020})},\ \Eprint
  {https://arxiv.org/abs/2006.06702} {arXiv:2006.06702 [astro-ph.GA]}
  \BibitemShut {NoStop}%
\bibitem [{\citenamefont {Stanzione}\ \emph {et~al.}(2020)\citenamefont
  {Stanzione}, \citenamefont {West}, \citenamefont {Evans}, \citenamefont
  {Minyard}, \citenamefont {Ghattas},\ and\ \citenamefont
  {Panda}}]{10.1145/3311790.3396656}%
  \BibitemOpen
  \bibfield  {author} {\bibinfo {author} {\bibfnamefont {D.}~\bibnamefont
  {Stanzione}}, \bibinfo {author} {\bibfnamefont {J.}~\bibnamefont {West}},
  \bibinfo {author} {\bibfnamefont {R.~T.}\ \bibnamefont {Evans}}, \bibinfo
  {author} {\bibfnamefont {T.}~\bibnamefont {Minyard}}, \bibinfo {author}
  {\bibfnamefont {O.}~\bibnamefont {Ghattas}},\ and\ \bibinfo {author}
  {\bibfnamefont {D.~K.}\ \bibnamefont {Panda}},\ }in\ \href
  {https://doi.org/10.1145/3311790.3396656} {\emph {\bibinfo {booktitle}
  {Practice and Experience in Advanced Research Computing}}},\ \bibinfo {series
  and number} {PEARC '20}\ (\bibinfo  {publisher} {Association for Computing
  Machinery},\ \bibinfo {address} {New York, NY, USA},\ \bibinfo {year}
  {2020})\ p.\ \bibinfo {pages} {106–111}\BibitemShut {NoStop}%
\bibitem [{\citenamefont {{Perez}}\ and\ \citenamefont
  {{Granger}}(2007)}]{2007CSE.....9c..21P}%
  \BibitemOpen
  \bibfield  {author} {\bibinfo {author} {\bibfnamefont {F.}~\bibnamefont
  {{Perez}}}\ and\ \bibinfo {author} {\bibfnamefont {B.~E.}\ \bibnamefont
  {{Granger}}},\ }\href {https://doi.org/10.1109/MCSE.2007.53} {\bibfield
  {journal} {\bibinfo  {journal} {Computing in Science and Engineering}\
  }\textbf {\bibinfo {volume} {9}},\ \bibinfo {pages} {21} (\bibinfo {year}
  {2007})}\BibitemShut {NoStop}%
\bibitem [{\citenamefont {{Virtanen}}\ \emph {et~al.}(2020)\citenamefont
  {{Virtanen}}, \citenamefont {{Gommers}}, \citenamefont {{Oliphant}},
  \citenamefont {{Haberland}}, \citenamefont {{Reddy}}, \citenamefont
  {{Cournapeau}}, \citenamefont {{Burovski}}, \citenamefont {{Peterson}},
  \citenamefont {{Weckesser}}, \citenamefont {{Bright}}, \citenamefont {{van
  der Walt}}, \citenamefont {{Brett}}, \citenamefont {{Wilson}}, \citenamefont
  {{Millman}}, \citenamefont {{Mayorov}}, \citenamefont {{Nelson}},
  \citenamefont {{Jones}}, \citenamefont {{Kern}}, \citenamefont {{Larson}},
  \citenamefont {{Carey}}, \citenamefont {{Polat}}, \citenamefont {{Feng}},
  \citenamefont {{Moore}}, \citenamefont {{VanderPlas}}, \citenamefont
  {{Laxalde}}, \citenamefont {{Perktold}}, \citenamefont {{Cimrman}},
  \citenamefont {{Henriksen}}, \citenamefont {{Quintero}}, \citenamefont
  {{Harris}}, \citenamefont {{Archibald}}, \citenamefont {{Ribeiro}},
  \citenamefont {{Pedregosa}}, \citenamefont {{van Mulbregt}},\ and\
  \citenamefont {{SciPy 1. 0 Contributors}}}]{2020NatMe..17..261V}%
  \BibitemOpen
  \bibfield  {author} {\bibinfo {author} {\bibfnamefont {P.}~\bibnamefont
  {{Virtanen}}}, \bibinfo {author} {\bibfnamefont {R.}~\bibnamefont
  {{Gommers}}}, \bibinfo {author} {\bibfnamefont {T.~E.}\ \bibnamefont
  {{Oliphant}}}, \bibinfo {author} {\bibfnamefont {M.}~\bibnamefont
  {{Haberland}}}, \bibinfo {author} {\bibfnamefont {T.}~\bibnamefont
  {{Reddy}}}, \bibinfo {author} {\bibfnamefont {D.}~\bibnamefont
  {{Cournapeau}}}, \bibinfo {author} {\bibfnamefont {E.}~\bibnamefont
  {{Burovski}}}, \bibinfo {author} {\bibfnamefont {P.}~\bibnamefont
  {{Peterson}}}, \bibinfo {author} {\bibfnamefont {W.}~\bibnamefont
  {{Weckesser}}}, \bibinfo {author} {\bibfnamefont {J.}~\bibnamefont
  {{Bright}}}, \bibinfo {author} {\bibfnamefont {S.~J.}\ \bibnamefont {{van der
  Walt}}}, \bibinfo {author} {\bibfnamefont {M.}~\bibnamefont {{Brett}}},
  \bibinfo {author} {\bibfnamefont {J.}~\bibnamefont {{Wilson}}}, \bibinfo
  {author} {\bibfnamefont {K.~J.}\ \bibnamefont {{Millman}}}, \bibinfo {author}
  {\bibfnamefont {N.}~\bibnamefont {{Mayorov}}}, \bibinfo {author}
  {\bibfnamefont {A.~R.~J.}\ \bibnamefont {{Nelson}}}, \bibinfo {author}
  {\bibfnamefont {E.}~\bibnamefont {{Jones}}}, \bibinfo {author} {\bibfnamefont
  {R.}~\bibnamefont {{Kern}}}, \bibinfo {author} {\bibfnamefont
  {E.}~\bibnamefont {{Larson}}}, \bibinfo {author} {\bibfnamefont {C.~J.}\
  \bibnamefont {{Carey}}}, \bibinfo {author} {\bibfnamefont
  {{\.I}.}~\bibnamefont {{Polat}}}, \bibinfo {author} {\bibfnamefont
  {Y.}~\bibnamefont {{Feng}}}, \bibinfo {author} {\bibfnamefont {E.~W.}\
  \bibnamefont {{Moore}}}, \bibinfo {author} {\bibfnamefont {J.}~\bibnamefont
  {{VanderPlas}}}, \bibinfo {author} {\bibfnamefont {D.}~\bibnamefont
  {{Laxalde}}}, \bibinfo {author} {\bibfnamefont {J.}~\bibnamefont
  {{Perktold}}}, \bibinfo {author} {\bibfnamefont {R.}~\bibnamefont
  {{Cimrman}}}, \bibinfo {author} {\bibfnamefont {I.}~\bibnamefont
  {{Henriksen}}}, \bibinfo {author} {\bibfnamefont {E.~A.}\ \bibnamefont
  {{Quintero}}}, \bibinfo {author} {\bibfnamefont {C.~R.}\ \bibnamefont
  {{Harris}}}, \bibinfo {author} {\bibfnamefont {A.~M.}\ \bibnamefont
  {{Archibald}}}, \bibinfo {author} {\bibfnamefont {A.~H.}\ \bibnamefont
  {{Ribeiro}}}, \bibinfo {author} {\bibfnamefont {F.}~\bibnamefont
  {{Pedregosa}}}, \bibinfo {author} {\bibfnamefont {P.}~\bibnamefont {{van
  Mulbregt}}},\ and\ \bibinfo {author} {\bibnamefont {{SciPy 1. 0
  Contributors}}},\ }\href {https://doi.org/10.1038/s41592-019-0686-2}
  {\bibfield  {journal} {\bibinfo  {journal} {Nature Methods}\ }\textbf
  {\bibinfo {volume} {17}},\ \bibinfo {pages} {261} (\bibinfo {year} {2020})},\
  \Eprint {https://arxiv.org/abs/1907.10121} {arXiv:1907.10121 [cs.MS]}
  \BibitemShut {NoStop}%
\bibitem [{\citenamefont {{Hunter}}(2007)}]{2007CSE.....9...90H}%
  \BibitemOpen
  \bibfield  {author} {\bibinfo {author} {\bibfnamefont {J.~D.}\ \bibnamefont
  {{Hunter}}},\ }\href {https://doi.org/10.1109/MCSE.2007.55} {\bibfield
  {journal} {\bibinfo  {journal} {Computing in Science and Engineering}\
  }\textbf {\bibinfo {volume} {9}},\ \bibinfo {pages} {90} (\bibinfo {year}
  {2007})}\BibitemShut {NoStop}%
\bibitem [{\citenamefont {{van der Walt}}\ \emph {et~al.}(2011)\citenamefont
  {{van der Walt}}, \citenamefont {{Colbert}},\ and\ \citenamefont
  {{Varoquaux}}}]{2011CSE....13b..22V}%
  \BibitemOpen
  \bibfield  {author} {\bibinfo {author} {\bibfnamefont {S.}~\bibnamefont {{van
  der Walt}}}, \bibinfo {author} {\bibfnamefont {S.~C.}\ \bibnamefont
  {{Colbert}}},\ and\ \bibinfo {author} {\bibfnamefont {G.}~\bibnamefont
  {{Varoquaux}}},\ }\href {https://doi.org/10.1109/MCSE.2011.37} {\bibfield
  {journal} {\bibinfo  {journal} {Computing in Science and Engineering}\
  }\textbf {\bibinfo {volume} {13}},\ \bibinfo {pages} {22} (\bibinfo {year}
  {2011})},\ \Eprint {https://arxiv.org/abs/1102.1523} {arXiv:1102.1523
  [cs.MS]} \BibitemShut {NoStop}%
\bibitem [{\citenamefont {Meurer}\ \emph {et~al.}(2017)\citenamefont {Meurer}
  \emph {et~al.}}]{Meurer:2017yhf}%
  \BibitemOpen
  \bibfield  {author} {\bibinfo {author} {\bibfnamefont {A.}~\bibnamefont
  {Meurer}} \emph {et~al.},\ }\href {https://doi.org/10.7717/peerj-cs.103}
  {\bibfield  {journal} {\bibinfo  {journal} {PeerJ Comput. Sci.}\ }\textbf
  {\bibinfo {volume} {3}},\ \bibinfo {pages} {e103} (\bibinfo {year}
  {2017})}\BibitemShut {NoStop}%
\bibitem [{\citenamefont {{Pedregosa}}\ \emph {et~al.}(2012)\citenamefont
  {{Pedregosa}}, \citenamefont {{Varoquaux}}, \citenamefont {{Gramfort}},
  \citenamefont {{Michel}}, \citenamefont {{Thirion}}, \citenamefont
  {{Grisel}}, \citenamefont {{Blondel}}, \citenamefont {{M{\"u}ller}},
  \citenamefont {{Nothman}}, \citenamefont {{Louppe}}, \citenamefont
  {{Prettenhofer}}, \citenamefont {{Weiss}}, \citenamefont {{Dubourg}},
  \citenamefont {{Vanderplas}}, \citenamefont {{Passos}}, \citenamefont
  {{Cournapeau}}, \citenamefont {{Brucher}}, \citenamefont {{Perrot}},\ and\
  \citenamefont {{Duchesnay}}}]{2012arXiv1201.0490P}%
  \BibitemOpen
  \bibfield  {author} {\bibinfo {author} {\bibfnamefont {F.}~\bibnamefont
  {{Pedregosa}}}, \bibinfo {author} {\bibfnamefont {G.}~\bibnamefont
  {{Varoquaux}}}, \bibinfo {author} {\bibfnamefont {A.}~\bibnamefont
  {{Gramfort}}}, \bibinfo {author} {\bibfnamefont {V.}~\bibnamefont
  {{Michel}}}, \bibinfo {author} {\bibfnamefont {B.}~\bibnamefont {{Thirion}}},
  \bibinfo {author} {\bibfnamefont {O.}~\bibnamefont {{Grisel}}}, \bibinfo
  {author} {\bibfnamefont {M.}~\bibnamefont {{Blondel}}}, \bibinfo {author}
  {\bibfnamefont {A.}~\bibnamefont {{M{\"u}ller}}}, \bibinfo {author}
  {\bibfnamefont {J.}~\bibnamefont {{Nothman}}}, \bibinfo {author}
  {\bibfnamefont {G.}~\bibnamefont {{Louppe}}}, \bibinfo {author}
  {\bibfnamefont {P.}~\bibnamefont {{Prettenhofer}}}, \bibinfo {author}
  {\bibfnamefont {R.}~\bibnamefont {{Weiss}}}, \bibinfo {author} {\bibfnamefont
  {V.}~\bibnamefont {{Dubourg}}}, \bibinfo {author} {\bibfnamefont
  {J.}~\bibnamefont {{Vanderplas}}}, \bibinfo {author} {\bibfnamefont
  {A.}~\bibnamefont {{Passos}}}, \bibinfo {author} {\bibfnamefont
  {D.}~\bibnamefont {{Cournapeau}}}, \bibinfo {author} {\bibfnamefont
  {M.}~\bibnamefont {{Brucher}}}, \bibinfo {author} {\bibfnamefont
  {M.}~\bibnamefont {{Perrot}}},\ and\ \bibinfo {author} {\bibfnamefont
  {{\'E}.}~\bibnamefont {{Duchesnay}}},\ }\href@noop {} {\bibfield  {journal}
  {\bibinfo  {journal} {arXiv e-prints}\ ,\ \bibinfo {eid} {arXiv:1201.0490}}
  (\bibinfo {year} {2012})},\ \Eprint {https://arxiv.org/abs/1201.0490}
  {arXiv:1201.0490 [cs.LG]} \BibitemShut {NoStop}%
\bibitem [{\citenamefont {{Gerosa}}\ and\ \citenamefont
  {{Vallisneri}}(2017)}]{2017JOSS....2..222G}%
  \BibitemOpen
  \bibfield  {author} {\bibinfo {author} {\bibfnamefont {D.}~\bibnamefont
  {{Gerosa}}}\ and\ \bibinfo {author} {\bibfnamefont {M.}~\bibnamefont
  {{Vallisneri}}},\ }\href {https://doi.org/10.21105/joss.00222} {\bibfield
  {journal} {\bibinfo  {journal} {The Journal of Open Source Software}\
  }\textbf {\bibinfo {volume} {2}},\ \bibinfo {pages} {222} (\bibinfo {year}
  {2017})}\BibitemShut {NoStop}%
\bibitem [{\citenamefont {{Apostolatos}}\ \emph {et~al.}(1994)\citenamefont
  {{Apostolatos}}, \citenamefont {{Cutler}}, \citenamefont {{Sussman}},\ and\
  \citenamefont {{Thorne}}}]{1994PhRvD..49.6274A}%
  \BibitemOpen
  \bibfield  {author} {\bibinfo {author} {\bibfnamefont {T.~A.}\ \bibnamefont
  {{Apostolatos}}}, \bibinfo {author} {\bibfnamefont {C.}~\bibnamefont
  {{Cutler}}}, \bibinfo {author} {\bibfnamefont {G.~J.}\ \bibnamefont
  {{Sussman}}},\ and\ \bibinfo {author} {\bibfnamefont {K.~S.}\ \bibnamefont
  {{Thorne}}},\ }\href {https://doi.org/10.1103/PhysRevD.49.6274} {\bibfield
  {journal} {\bibinfo  {journal} {\prd}\ }\textbf {\bibinfo {volume} {49}},\
  \bibinfo {pages} {6274} (\bibinfo {year} {1994})}\BibitemShut {NoStop}%
\bibitem [{\citenamefont {{Kidder}}(1995)}]{1995PhRvD..52..821K}%
  \BibitemOpen
  \bibfield  {author} {\bibinfo {author} {\bibfnamefont {L.~E.}\ \bibnamefont
  {{Kidder}}},\ }\href {https://doi.org/10.1103/PhysRevD.52.821} {\bibfield
  {journal} {\bibinfo  {journal} {\prd}\ }\textbf {\bibinfo {volume} {52}},\
  \bibinfo {pages} {821} (\bibinfo {year} {1995})},\ \Eprint
  {https://arxiv.org/abs/gr-qc/9506022} {arXiv:gr-qc/9506022 [gr-qc]}
  \BibitemShut {NoStop}%
\bibitem [{\citenamefont {{Buonanno}}\ \emph {et~al.}(2003)\citenamefont
  {{Buonanno}}, \citenamefont {{Chen}},\ and\ \citenamefont
  {{Vallisneri}}}]{2003PhRvD..67j4025B}%
  \BibitemOpen
  \bibfield  {author} {\bibinfo {author} {\bibfnamefont {A.}~\bibnamefont
  {{Buonanno}}}, \bibinfo {author} {\bibfnamefont {Y.}~\bibnamefont {{Chen}}},\
  and\ \bibinfo {author} {\bibfnamefont {M.}~\bibnamefont {{Vallisneri}}},\
  }\href {https://doi.org/10.1103/PhysRevD.67.104025} {\bibfield  {journal}
  {\bibinfo  {journal} {\prd}\ }\textbf {\bibinfo {volume} {67}},\ \bibinfo
  {eid} {104025} (\bibinfo {year} {2003})},\ \Eprint
  {https://arxiv.org/abs/gr-qc/0211087} {arXiv:gr-qc/0211087 [gr-qc]}
  \BibitemShut {NoStop}%
\bibitem [{\citenamefont {{von Zeipel}}(1910)}]{1910AN....183..345V}%
  \BibitemOpen
  \bibfield  {author} {\bibinfo {author} {\bibfnamefont {H.}~\bibnamefont {{von
  Zeipel}}},\ }\href {https://doi.org/10.1002/asna.19091832202} {\bibfield
  {journal} {\bibinfo  {journal} {Astronomische Nachrichten}\ }\textbf
  {\bibinfo {volume} {183}},\ \bibinfo {pages} {345} (\bibinfo {year}
  {1910})}\BibitemShut {NoStop}%
\bibitem [{\citenamefont {{Lidov}}(1962)}]{1962P&SS....9..719L}%
  \BibitemOpen
  \bibfield  {author} {\bibinfo {author} {\bibfnamefont {M.~L.}\ \bibnamefont
  {{Lidov}}},\ }\href {https://doi.org/10.1016/0032-0633(62)90129-0} {\bibfield
   {journal} {\bibinfo  {journal} {\planss}\ }\textbf {\bibinfo {volume} {9}},\
  \bibinfo {pages} {719} (\bibinfo {year} {1962})}\BibitemShut {NoStop}%
\bibitem [{\citenamefont {{Kozai}}(1962)}]{1962AJ.....67..591K}%
  \BibitemOpen
  \bibfield  {author} {\bibinfo {author} {\bibfnamefont {Y.}~\bibnamefont
  {{Kozai}}},\ }\href {https://doi.org/10.1086/108790} {\bibfield  {journal}
  {\bibinfo  {journal} {\aj}\ }\textbf {\bibinfo {volume} {67}},\ \bibinfo
  {pages} {591} (\bibinfo {year} {1962})}\BibitemShut {NoStop}%
\bibitem [{\citenamefont {{Ito}}\ and\ \citenamefont
  {{Ohtsuka}}(2019)}]{2019MEEP....7....1I}%
  \BibitemOpen
  \bibfield  {author} {\bibinfo {author} {\bibfnamefont {T.}~\bibnamefont
  {{Ito}}}\ and\ \bibinfo {author} {\bibfnamefont {K.}~\bibnamefont
  {{Ohtsuka}}},\ }\href {https://doi.org/10.5047/meep.2019.00701.0001}
  {\bibfield  {journal} {\bibinfo  {journal} {Monographs on Environment, Earth
  and Planets}\ }\textbf {\bibinfo {volume} {7}},\ \bibinfo {pages} {1}
  (\bibinfo {year} {2019})},\ \Eprint {https://arxiv.org/abs/1911.03984}
  {arXiv:1911.03984 [astro-ph.EP]} \BibitemShut {NoStop}%
\bibitem [{\citenamefont {{Holman}}\ \emph {et~al.}(1997)\citenamefont
  {{Holman}}, \citenamefont {{Touma}},\ and\ \citenamefont
  {{Tremaine}}}]{1997Natur.386..254H}%
  \BibitemOpen
  \bibfield  {author} {\bibinfo {author} {\bibfnamefont {M.}~\bibnamefont
  {{Holman}}}, \bibinfo {author} {\bibfnamefont {J.}~\bibnamefont {{Touma}}},\
  and\ \bibinfo {author} {\bibfnamefont {S.}~\bibnamefont {{Tremaine}}},\
  }\href {https://doi.org/10.1038/386254a0} {\bibfield  {journal} {\bibinfo
  {journal} {\nat}\ }\textbf {\bibinfo {volume} {386}},\ \bibinfo {pages} {254}
  (\bibinfo {year} {1997})}\BibitemShut {NoStop}%
\bibitem [{\citenamefont {{Blaes}}\ \emph {et~al.}(2002)\citenamefont
  {{Blaes}}, \citenamefont {{Lee}},\ and\ \citenamefont
  {{Socrates}}}]{2002ApJ...578..775B}%
  \BibitemOpen
  \bibfield  {author} {\bibinfo {author} {\bibfnamefont {O.}~\bibnamefont
  {{Blaes}}}, \bibinfo {author} {\bibfnamefont {M.~H.}\ \bibnamefont {{Lee}}},\
  and\ \bibinfo {author} {\bibfnamefont {A.}~\bibnamefont {{Socrates}}},\
  }\href {https://doi.org/10.1086/342655} {\bibfield  {journal} {\bibinfo
  {journal} {\apj}\ }\textbf {\bibinfo {volume} {578}},\ \bibinfo {pages} {775}
  (\bibinfo {year} {2002})},\ \Eprint {https://arxiv.org/abs/astro-ph/0203370}
  {arXiv:astro-ph/0203370 [astro-ph]} \BibitemShut {NoStop}%
\bibitem [{\citenamefont {{Spitzer}}(1987)}]{1987degc.book.....S}%
  \BibitemOpen
  \bibfield  {author} {\bibinfo {author} {\bibfnamefont {L.}~\bibnamefont
  {{Spitzer}}},\ }\href@noop {} {\emph {\bibinfo {title} {{Dynamical evolution
  of globular clusters}}}}\ (\bibinfo {year} {1987})\BibitemShut {NoStop}%
\end{thebibliography}%

\end{document}